\documentclass[a4paper,11pt]{article}
\pdfoutput=1 

\usepackage{jcappub} 

\usepackage[T1]{fontenc} 
\usepackage{xspace}

\usepackage{cellspace}
\usepackage{xcolor}
\definecolor{darkblue}{rgb}{0.0, 0.0, 0.55}
\definecolor{darkgreen}{rgb}{0.0, 0.55, 0.2}
\definecolor{darkred}{rgb}{0.55, 0.0, 0}

\newcommand{\omegaEFT}{\Omega_0^\mathrm{EFT}}
\newcommand{\lcdm}{$\Lambda$CDM\xspace}
\newcommand{\pto}{P_{21}(k)}

\title{\boldmath Constraining beyond \lcdm models with 21cm intensity mapping forecast observations combined with latest CMB data}

\author[a,b,c]{M. Berti,}
\author[d,c,b]{M. Spinelli,}
\author[a,b,c]{B. S. Haridasu,}
\author[a,b,c,d]{M. Viel,}
\author[e]{A. Silvestri}

\affiliation[a]{SISSA- International School for Advanced Studies, Via Bonomea 265, 34136 Trieste, Italy}
\affiliation[b]{INFN – National Institute for Nuclear Physics, Via Valerio 2, I-34127 Trieste, Italy}
\affiliation[c]{IFPU, Institute for Fundamental Physics of the Universe, via Beirut2, 34151 Trieste, Italy}
\affiliation[d]{INAF, Osservatorio Astronomico di Trieste, Via G. B. Tiepolo 11, I-34131 Trieste, Italy}
\affiliation[e]{Institute  Lorentz,  Leiden  University,  PO  Box  9506,  Leiden  2300  RA,  The  Netherlands}

\emailAdd{mberti@sissa.it}
\emailAdd{marta.spinelli@inaf.it}
\emailAdd{sharidas@sissa.it}
\emailAdd{viel@sissa.it}
\emailAdd{silvestri@lorentz.leidenuniv.nl}

\abstract{We explore constraints on dark energy and modified gravity with forecast 21cm intensity mapping measurements using the Effective Field Theory approach. We construct a realistic mock data set forecasting a low redshift 21cm signal power spectrum $P_{21}(z,k)$ measurement from the MeerKAT radio-telescope. We compute constraints on cosmological and model parameters through Monte Carlo Markov chain techniques, testing both the constraining power of $P_{21}(k)$ alone and its effect when combined with the latest Planck 2018 CMB data.
We complement our analysis by testing the effects of tomography from an ideal mock data set of observations in multiple redshift bins. We conduct our analysis numerically with the codes \texttt{EFTCAMB}/\texttt{EFTCosmoMC}, which we extend by implementing a likelihood module fully integrated with original codes. We find that adding $\pto$ to CMB data provides significantly tighter constraints on $\Omega_ch^2$ and $H_0$, with a reduction of the error with respect to Planck results at the level of more than $60\%$. 
For the parameters describing beyond \lcdm theories, we observe a reduction in the error with respect to the Planck constraints at the level of $\lesssim 10\%$. The improvement increases up to $\sim 35\%$ when we constrain the parameters using ideal, tomographic mock observations. We conclude that the power spectrum of the 21cm signal is sensitive to variations of the parameters describing the examined beyond \lcdm models and, thus, $\pto$ observations could help to constrain dark energy. The constraining power on such theories is improved significantly by tomography.}

\begin{document}
\maketitle
\flushbottom

\section{Introduction}
\label{sec:intro}
 In modern Cosmology, the source of cosmic acceleration is assigned to an enigmatic constituent of the Universe, called Dark Energy (DE). The standard cosmological model is the $\Lambda$ - Cold Dark Matter (\lcdm) model, in which DE constitutes almost 70\% of the total content of the Universe and is consistently modeled as a Cosmological Constant~($\Lambda$) term. The \lcdm model has been accurately tested in the past couple of decades and, at the state-of-the-art, it is in remarkable agreement with several independent cosmological and astrophysical observations, such as the Cosmic Microwave Background (CMB), the Supernovae observations and measurements from the large scale structure~\cite{planck:2018,abbott:2018,Hildebrandt:2017,riess:2018}.

Despite the success of the standard cosmological model, current tensions on the values of the \lcdm parameters measured by different observations~\cite{riess:2019,wong:2019,freedman:2019,Verde:2019} could imply new physics. Moreover, the Cosmological Constant raises many questions that are still unresolved, such as the nature of $\Lambda$ and why it is forced to assume the small value that is consistent with observations~\cite{carroll:2000,weinberg:1988,bull:2015}. To this extent, a plethora of alternative models have been proposed and studied~\cite{ferreira:2019,ishak:2019}, ranging from dynamical DE to modifications of gravity (MG). In the former an additional field, typically a scalar, is added to the matter content. In the latter, the gravity sector of the action is modified, normally resulting in additional gravitational degrees of freedom (DOFs). In practice, the distinction between DE and MG models is not always a strict one, and one can rather focus on the additional, propagating DOFs and employ a unifying framework to study deviations from  \lcdm. 

One such framework is offered by the Effective Field Theory (EFT) of DE~\cite{Gubitosi:2013,bloomfield:2013}, which allows to survey  large samples of DE/MG models for an optimal exploitation of the wealth of  upcoming,  high-precision data.
In this work, we adopt such framework and, in order to interface with the data and forecast observations, we adopt the \texttt{EFTCAMB} and \texttt{EFTCosmoMC}~\citep{hu:2014,Raveri:2014,num_notes_v3_EFTCAMB} codes, which implement EFT of DE in the popular Einstein-Boltzmann solver \texttt{CAMB}~\cite{Lewis:2000} and in the Monte Carlo Markov Chain (MCMC)~\cite{gilks:1995} sampler \texttt{CosmoMC}~\cite{Lewis:2002,Lewis:2013}, respectively. 
Latest constraints for EFT models produced with \texttt{EFTCAMB}/\texttt{EFTCosmoMC} can be found in ref.~\cite{Frusciante:2019}. 

In the last decade, DE-MG models were extensively tested. For comprehensive reviews on the state-of-the-art, we refer to refs.~\cite{ferreira:2019,ishak:2019}. Although no significant proof supporting a specific beyond \lcdm model has yet been found, the upcoming observations will allow high precision tests of gravity on the large cosmological scales. In this work, we focus on the 21cm signal detection through line intensity mapping techniques, which allows to sample the neutral hydrogen (HI) distribution in a wide redshift range~\cite{Bharadwaj:2000,Bharadwaj:2001,Battye:2004,McQuinn:2005,Kovetz:2017,Villaescusa-Navarro:2018}. These observations will open an important window  on DE/MG theories~\cite{Chang:2007,Hall:2012, Lima:2016,Brax:2012,Zhao:2015,Heneka:2018,Dash:2020}. 
Currently, several purpose-built radio-telescopes, such as the Canadian Hydrogen Intensity Mapping Experiment (CHIME),\footnote{See \url{https://chime-experiment.ca/en}.} the Hydrogen Intensity and Real-Time Analysis experiment (HIRAX),\footnote{See \url{https://hirax.ukzn.ac.za/}.} the Five-hundred-meter Aperture Spherical Telescope (FAST),\footnote{See \url{https://fast.bao.ac.cn/}.} are already taking data or are under construction. Moreover, radio cosmology and the study of DE/MG theories is one of the main science goals of the world largest interferometer, the Square Kilometer Array Observatory (SKAO),\footnote{See \url{https://www.skatelescope.org/}.} in particular with intensity mapping techniques~\cite{Bacon:2018}. 
The MeerKAT telescope,\footnote{See \url{https://www.sarao.ac.za/gallery/meerkat/}.} the SKAO precursor in South Africa, can be successfully exploit for 21cm intensity mapping~\cite{Santos:2015,Santos:2017} and has attained encouraging results~\cite{Wang:2021}.
In this work, we produce constraints on model parameters from forecast 21cm intensity mapping observations. The new observable we use is $P_{21}(z,k)$, the power spectrum of the 21cm signal. We construct the likelihood function for $P_{21}(z,k)$ by building a mock data set of realistic, upcoming MeerKAT observations of $P_{21}(z,k)$ at redshift $z=0.39$. We test both the effect of the 21cm signal observations alone and combined with the latest Planck 2018 CMB data~\cite{planck:2018}. We explore both a \lcdm scenario and several EFT models beyond \lcdm. We equip our likelihood also with additional information on background quantities, derived from further 21cm intensity mapping forecasts at higher redshift~\cite{Obuljen:2017}. Moreover, to test the effect of tomography on beyond \lcdm theories, we further extend the likelihood to include four additional, ideal mock data sets, constructed to mimic $P_{21}(z,k)$ observations at different redshifts. To perform our analysis, we expand the codes \texttt{EFTCAMB}/\texttt{EFTCosmoMC} and include a new likelihood module to compute constraints through MCMC techniques.

Compared to similar studies~\cite{Hall:2012,Pourtsidou:2015,Bull:2014,Obuljen:2017,Heneka:2018}, this work includes the following important new aspects: $i)$ we start by mocking realistic 21cm observations that are about to be achieved by present day experiments, once the foregrounds will be removed from the data; $ii)$ we employ a full MCMC Bayesian analysis, rather than adopting a Fisher Matrix approach; $iii)$ we start by considering the present day full CMB data from Planck, rather than assuming priors on cosmological parameters measured; $iv)$ we combine 21cm observations only with CMB data, without considering other probes, to disentangle precisely the constraining power of such future observations; $v)$ we include a conservative treatment of the astrophysical quantities that are used to mock the 21cm power spectrum, like the brightness temperature and the atomic hydrogen bias.

The structure of this paper is the following. In the first part, i.e. section~\ref{sec:method}, we delineate our methodology, describe the modeling of the 21cm signal power spectrum, and present the theoretical models we consider. Then, in section~\ref{sec:results} we present and discuss the results we obtain in different scenarios. We begin by examining the effects of the new $P_{21}$ likelihood on the \lcdm model in section~\ref{sec:res_lcdm}. In section~\ref{sec:res_EFT_lcdm}, we discuss results for models beyond \lcdm, which we study assuming a \lcdm background evolution. Then, we investigate the effect of implementing the HI bias and the brightness temperature as nuisance parameters in section~\ref{sec:res_nuisance}. We show results for models beyond \lcdm on a $w$CDM background evolution in section~\ref{sec:res_EFT_wCDM}. We explore a more ideal case, where we combine mock observations at five different redshifts, in section~\ref{sec:res_tomography}. Adding to the $\pto$ likelihood measures in different redshift bins, which we refer to as tomography, should improve the constraints on the considered beyond \lcdm models. We conclude by outlining the main results of this work in section~\ref{sec:res_summary}.

\section{Methods}
\label{sec:method}
Our main goal is to investigate whether future intensity mapping observations could play a role in improving constraints on cosmological parameters. To this end, we conduct a Monte-Carlo Markov-Chain (MCMC) analysis~\cite{gilks:1995} computing a likelihood function for the 21cm power spectrum, the $P_{21}$ likelihood, from a mock data set of observations. In section~\ref{sec:P21_model} we describe the model for the 21cm power spectrum, while in section~\ref{sec:likelihood_implementation} we provide details on the likelihood implementation. In section~\ref{sec:theoretical_framework}, instead, we summarize the main aspects of EFT for cosmic acceleration, outline the theoretical models that we test, and list the most recent constraints available in literature. Finally, we list other data sets used to test the $P_{21}$ likelihood in section~\ref{sec:data_sets}.

We adopt the Planck 2018 best fit results~\citep{planck:2018}, i.e. we consider as fiducial cosmology a $\Lambda$CDM model described by the parameters:
\begin{table}[tbh]
\centering
\begin{tabular}{ll}
$\Omega_b h^2=0.022383$,  & $\tau =0.0543$,\\
$\Omega_c h^2=0.12011$,  &  $\ln(10^{10} A_s)=3.0448$,\\
$ 100\theta_{\mathrm{MC}}= 1.040909$, & $n_s = 0.96605$ ,\\
\end{tabular}
\end{table}

\noindent obtained from the TT, TE, EE+lensing+lowE likelihood data sets. 

\subsection{Model for the 21cm signal}
\label{sec:P21_model}

We consider the following model for the 21cm signal linear power spectrum~\cite{kaiser1987,Villaescusa-Navarro:2018, Bacon:2018}:
\begin{equation}
    \label{eq:P_21_full}
     P_{21}(z,\,k,\, \mu) = \Bar{T}_b^2(z) \left[( b_{\mathrm{HI}}(z) + f(z)\, \mu)^2 P_{\rm m}(z,k) + P_{\mathrm{SN}}\right],
\end{equation}
where $\Bar{T}_b(z)$ is the mean brightness temperature, $b_{\mathrm{HI}}(z)$ is the HI linear bias, $f(z)$ is the linear growth rate, $\mu= \hat{k} \cdot \hat{z}$, $P_{\rm m}(z,k)$ is the total matter power spectrum and $P_{\mathrm{SN}}$ is the HI shot noise. 
For the time being, we restrict ourselves to the case $\mu=1$ and neglect the effect of the shot noise term~\cite{Pourtsidou:2016,Villaescusa-Navarro:2018}, leaving the study of different values of $\mu$ and the introduction of the shot noise for future work. I.e., we use:
\begin{equation}
    \label{eq:P_21}
     P_{21}(z,\,k) = \Bar{T}_b^2(z) ( b_{\mathrm{HI}}(z) + f(z))^2 P_{\rm m}(z,k).
\end{equation}

For the mean brightness temperature we rely on the following parametrization from ref.~\cite{Furlanetto:2006}:
\begin{equation}
    \label{eq:T_b}
    T_b(z) = 23.88 \left(\frac{\Omega_b h^2}{0.02}\right) \sqrt{\frac{0.15}{\Omega_m h^2} \frac{(1+z)}{10}}\; x_{\mathrm{HI}}\;\mathrm{mK},
\end{equation}
 where $\Omega_b$ and $\Omega_m$ are the baryon and total matter density parameters today and $x_{\mathrm{HI}}= \frac{\Omega_{\mathrm{HI}}}{\Omega_{\mathrm{H}}}$ is the fraction of neutral atomic hydrogen. Following results in literature we assume the hydrogen fraction to be $\Omega_{\mathrm{H}} = 0.74 \Omega_b$~\cite{Spinelli:2019} and consider a constant value of $\Omega_{\mathrm{HI}} \sim 10^{-3}$. This choice of $\Omega_{\mathrm{HI}}$ is justified by observations~\cite{Crighton:2015} and has been already adopted in other works~\cite{Villaescusa-Navarro:2018,Bacon:2018}.

We model the linear bias $ b_{\mathrm{HI}}(z)$ from numerical simulation results. In ref.~\cite{Villaescusa-Navarro:2018} $ b_{\mathrm{HI}}$ is estimated for a discrete set of redshifts $z=0,\, 1,\, 2,\, 3,\, 4,\, 5$ in a $\Lambda$CDM framework. We interpolate linearly these points to estimate $ b_{\mathrm{HI}}(z)$ also at intermediate redshift values. Notice that, at the scales we operate, we can safely consider the bias to be scale-independent. 

The linear growth rate is computed as $f(z)=f\sigma_8(z)/\sigma_8(z)$, where $f\sigma_8(z)$ is numerically computed from the density-velocity correlation, i.e. $f\sigma_8(z)=\sigma_8^{(vd)}(z)^2/\sigma_8^{(dd)}(z)$, as defined in ref.~\cite{planck:2015cosmo}. This way of obtaining the growth rate is model dependent and allows us to estimate it also for models beyond $\Lambda$CDM, where the $f(z)$ behavior may not be trivial. This method, on the other hand, cannot account for any possible scale dependence of the growth rate, a known feature of several MG models. However, the growth rate for the subset of models we study results to be scale-independent in the range of scales we investigate and, thus, we are allowed to use this method.

\begin{figure}[tbp]
    \centering
    \includegraphics[width=1\textwidth]{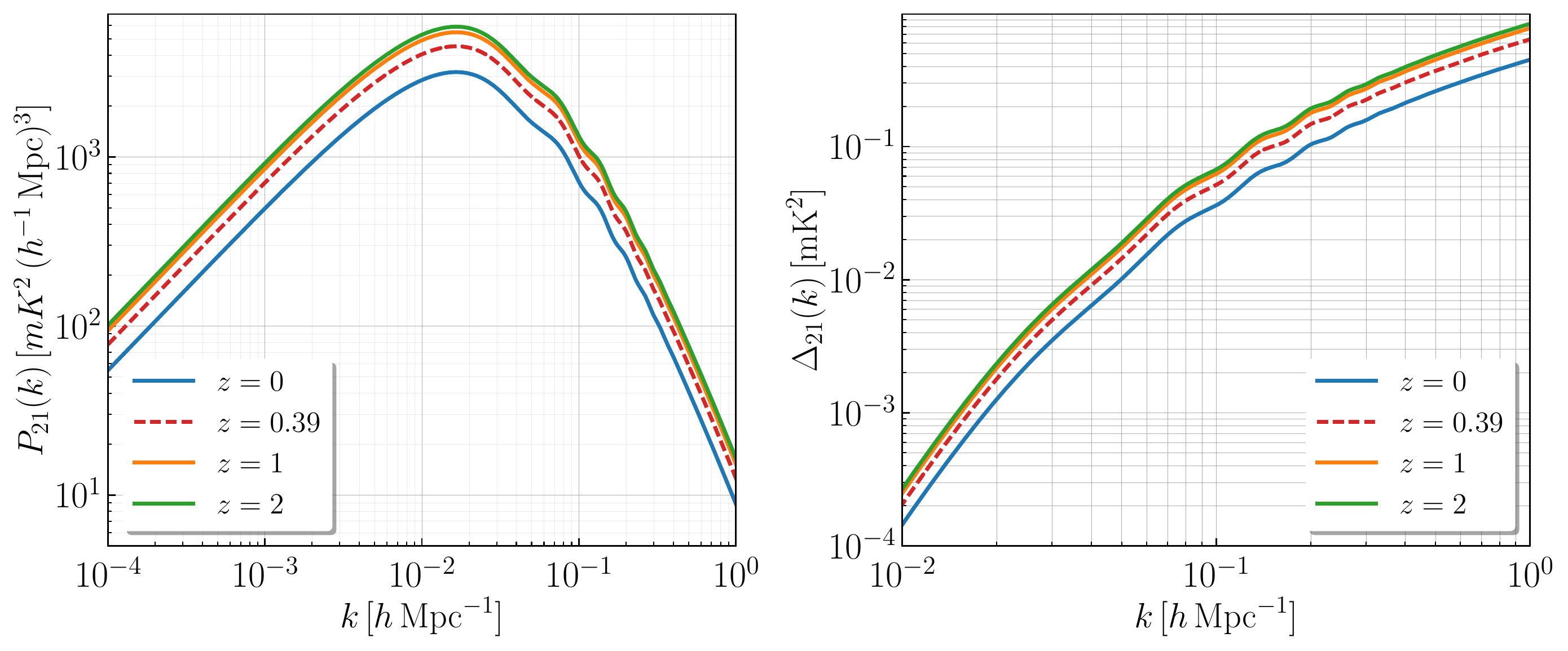}
    \caption{\label{fig:P21_delta_redshift} Linear power spectra for the 21cm signal at different redshifts. Results are computed with the model in eq.~\eqref{eq:P_21} for the $\Lambda$CDM fiducial cosmology, i.e. Planck 2018 best fit results~\citep{planck:2018}.  We show both the full power spectrum $P_{21}(k)$ (left) and the adimensional one  $\Delta_{21} (k) \equiv k^3\, P_{21}(k)/2\pi^2$ (right). }
\end{figure}

Finally, as well as $f\sigma_8(z)$ and $\sigma_8(z)$, the matter power spectrum $P_{\rm m}(z,k)$ is computed with the Eistein-Boltzmann solvers \texttt{CAMB}~\cite{Lewis:2000} and \texttt{EFTCAMB}~\cite{hu:2014,num_notes_v3_EFTCAMB}, the \texttt{CAMB} extension to test MG/DE models in an Effective Field Theory approach.\footnote{See \url{https://camb.info/} and \url{http://eftcamb.org/}.} We highlight that in our construction of $P_{21}(z,k)$ model dependent quantities are the matter power spectrum $P_{\rm m}(z,k)$ and the growth rate $f$, which we are able to compute for \lcdm and beyond \lcdm models. The other factors in eq.~\eqref{eq:P_21}, i.e. the brightness temperature $T_b$ and the bias $b_{\mathrm{HI}}$, are fixed for different models. This is because it is not clear how to compute these quantities for models beyond \lcdm. We further discuss this topic in section~\ref{sec:res_nuisance}. The theoretical predictions of $P_{21}(z,k)$ in the fiducial \lcdm cosmology at different redshifts are shown in figure~\ref{fig:P21_delta_redshift}.

\subsection{21cm signal likelihood implementation}
\label{sec:likelihood_implementation}
Given a set of observed $P_{21}(z,k)$ values, $\vec{P}_{21}^{obs}$, at a fixed redshift, we compute theoretical predictions of the 21cm signal, i.e. $\vec{P}_{21}^{th}$, at the same redshift and $k$. As described in section~\ref{sec:P21_model}, $P_{21}^{th}(z,k)$ is derived from the brightness temperature, the HI bias, the growth rate and the matter power spectrum, as in eq.~\eqref{eq:P_21}. At each MCMC step values of $P_{\rm m}(k)$ and $f$ are provided by \texttt{CAMB}/\texttt{EFTCAMB}, whereas $T_b$ and $b_{\mathrm{HI}}$ are kept fixed to their values from the fiducial cosmology. 

Thus, the  $\pto$ logarithmic likelihood function $\mathcal{L}$ is computed as:
\begin{equation}\label{eq:likelihood}
       - \ln\left[\mathcal{L}\left(\vec{P}_{21}^{obs}|\vec{P}_{21}^{th}\right)\right] =   \frac{1}{2} (\vec{P}_{21}^{th} - \vec{P}_{21}^{obs})^{\mathrm{T}} \,  \mathbf{C}^{-1} \, (\vec{P}_{21}^{th} - \vec{P}_{21}^{obs}) ,
\end{equation}
where $\mathbf{C}$ is the covariance matrix, with dimensions $n \times n$, being $n$ the number of data points. We assume $\mathbf{C}$ to be diagonal, given that we rely only on linear scales, and compute it from the observational errors $\vec{\sigma}_{P_{21}}$ as:
\begin{equation}
\mathbf{C}_{ij} = \sigma^2_{P_{21},i}\, \delta_{ij},
\end{equation}
with $\delta_{ij}$ being the usual Kronecker delta and $i,\,j = 1, \dots, n$.

We consider also the possibility to compute the likelihood from different redshift bins. If that is the case, observations at each redshift are assumed independent. Each contribution to the logarithmic likelihood is calculated separately and then added together. Moreover, it is also possible to choose an alternative parametrization for the brightness temperature $T_b$. From the definition in eq.~\eqref{eq:T_b} we can extrapolate the redshift dependence and rewrite $T_b$ in a more general way:
\begin{equation}\label{eq:T_b_}
    T_b = a_{T_b} \sqrt{1 + z},
\end{equation}
where in this case all the contributions from cosmological parameters and $\Omega_{\mathrm{HI}}$ absorbed in the value of the amplitude $a_{T_b}$. 

It is then possible to treat $a_{T_b}$, along with the bias $b_{HI}$, as nuisance parameters of the $\pto$ likelihood function, as we discuss in section~\ref{sec:res_nuisance}. 
We assume this to take into account the effect of possible variations in the brightness temperature and the bias, for which we lack of a theoretical model, in particular for beyond \lcdm theories.

We numerically compute the constraints by means of the latest public available version of the MCMC codes \texttt{CosmoMC}~\cite{Lewis:2002} and \texttt{EFTCosmoMC}~\cite{Raveri:2014}, the \texttt{CosmoMC} extension for studying MG/DE models in an EFT framework.\footnote{See also \url{https://cosmologist.info/cosmomc/}.} \texttt{CosmoMC}/\texttt{EFTCosmoMC} allow the exploration of the parameter space by computing the likelihood from several different observational data. We expand this set by implementing the computation of the likelihood for $\pto$ from upcoming intensity mapping observations as described above.\footnote{To this end we write a new module, \texttt{twentyonepk.f90}, to be added to the Fortran version of \texttt{CosmoMC}/\texttt{EFTCosmoMC} .}

\subsubsection{A realistic mock data set}
\label{sec:mock_data}
\begin{figure}[tbp]
    \centering
    \includegraphics[width=0.7\textwidth]{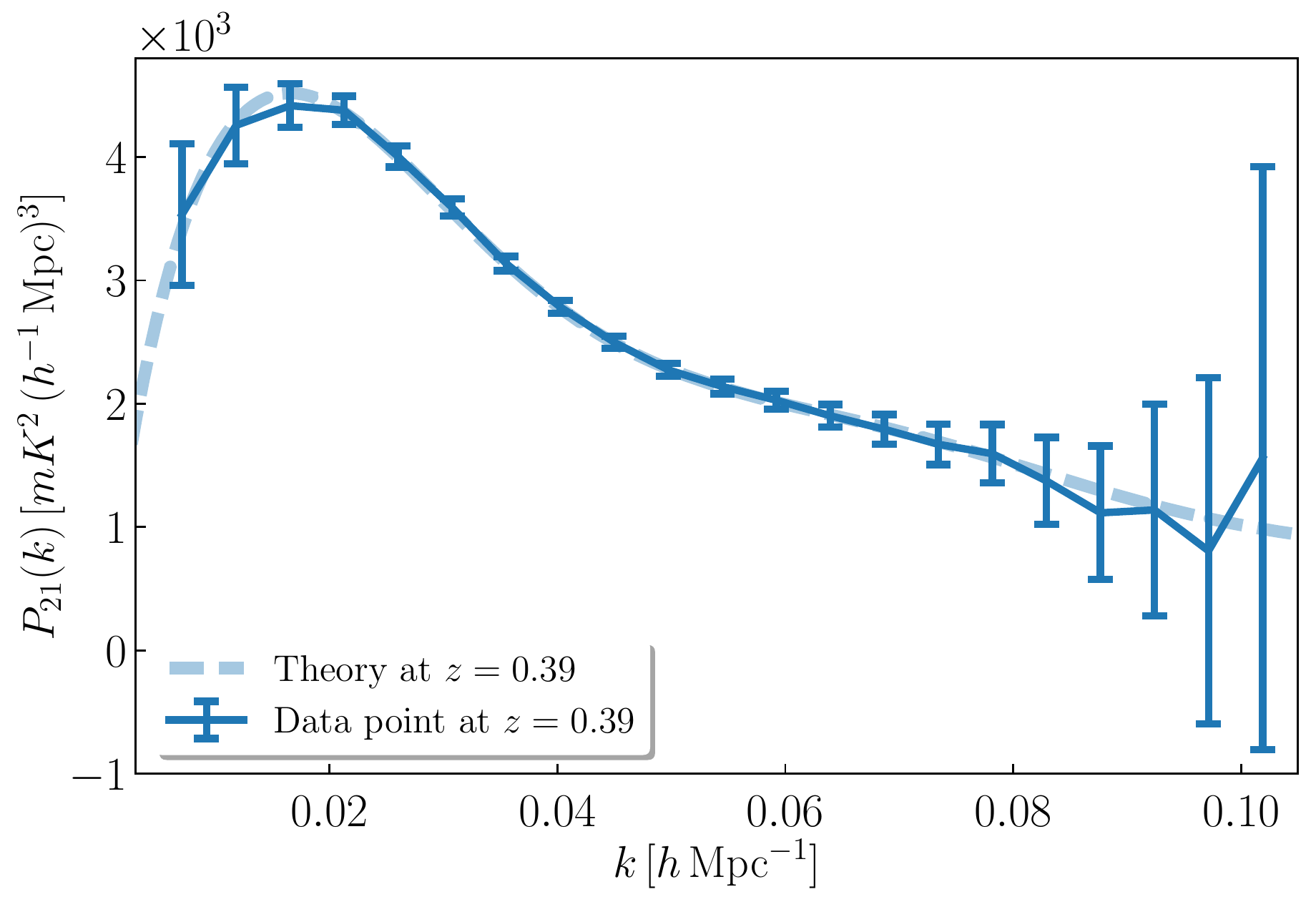}
        \caption{\label{fig:mock_data_points} Mock data set observations and theory prediction at $z=0.39$ for the 21cm signal power spectrum. Error bars are generated considering a realistic intensity mapping survey with the MeerKAT telescope. Central values are instead obtained from points predicted by the theory for $\Lambda$CDM and the fiducial cosmology we assume. Each point from the theory is then displaced randomly according to a gaussian distribution with a fifth of the error on that point as standard deviation.}
\end{figure}

A central ingredient for the likelihood of eq.~\eqref{eq:likelihood} is the measured 21cm power spectrum $P_{21}^{obs}$ and its observational error $\sigma^2_{P_{21}}$ as a function of the $k$ in each of the $i^{th}$ available bin. With the aim of obtaining realistic constraints for both the standard and non-standard models explored in this work, we adopt the following approach to generate mock data. We assume we are using the MeerKAT telescope as a collection of 64 single 13.5 m dishes, as for the preferred practice in intensity mapping analysis~\cite{Santos:2015, Santos:2017, Wang:2021}, scanning a total sky area of 2000~deg$^2$. We consider L-band observations and thus a central redshift of $z=0.39$. We assume for the signal the same modeling as for section~\ref{sec:P21_model} and consider the effect of both the instrumental noise and the telescope beam. We consider a thermal noise level given by the radiometry equation with a system temperature taken directly from MeerKAT technical available documentation, a frequency resolution of 1~MHz, and 2.400~h of total observing time. The beam is assumed to be Gaussian and to scale proportionally to $\lambda /D_{\rm eff}$, where $\lambda$ is the observed frequency and $D_{\rm eff}$ is the effective dish size. We then compute the error as:
\begin{equation}\label{eq:P21obserr}
\sigma^2_{P_{21},i} = \frac{(2\pi)^3}{V_{\rm sur}}\frac{1}{4\pi k_i^2 \Delta k} \left( P_{21}(k_i)+P_{\rm noise} \right),
\end{equation}
where $V_{\rm sur}$ is the survey total volume, $\Delta k = 2 \pi/V_{\rm sur}^{1/3}$ is the $k-$bin width and $P_{\rm noise}$ encodes both the effect of the beam and the thermal noise. 
Note that we create our mock data only once, assuming a standard \lcdm cosmology for $\pto$. Given the fiducial cosmology and the errors of eq.~\eqref{eq:P21obserr} the final data set is constructed generating, for each $k-$bin, a new data point Gaussian distributed around the theory with a standard deviation of $\sigma^2_{P_{21},i}$, as depicted in figure~\ref{fig:mock_data_points}. Intensity mapping experiments have yet to completely solve additional challenges, such as the cleaning of the foreground emissions, e.g. see ref.~\cite{Switzer:2013}, beam modeling uncertainties~\cite{Asad:2021,Matshawule:2021}, and possible residual artifacts in the data~\cite{Harper:2018,Li:2021}. In this work, we do not model these and leave them to further studies.

\subsubsection{Additional external intensity mapping forecasts on background quantities}
\label{sec:background}

We also test the efficacy of the $P_{21}(k)$ likelihood combined with additional intensity mapping information on background and growth of structure quantities provided by other experiments at higher redshift. As described in ref.~\cite{Obuljen:2017}, 21cm intensity mapping observations in the redshift range $2.5< z< 5$ are expected to provide powerful constraints on cosmological parameters. In ref.~\cite{Obuljen:2017}, constraints on the quantities $f\sigma_8(z)$, the angular diameter distance $D_A(z)$, and the Hubble parameter $H(z)$ are computed for a hypothetical observation of the telescope HIRAX~\cite{Newburgh:2016} in this redshift range.  Our aim is to equip the $\pto$ likelihood with additional information on the background evolution, coming from external 21cm intensity mapping observations. To do so we construct an additional likelihood and mock data sets for $f\sigma_8(z)$, $D_A(z)$ and $H(z)$. In the following, we refer to this likelihood as the \textit{background} likelihood.
\begin{table}[tbp]
\centering
\begin{tabular}{lccc}
\hline
\hline
                                  &$\sigma_x/x(z=2.5)$  &  $x(z=0.39)$    & $\sigma_x(z=0.39)$ \\
\hline
$f\sigma_8$                       & $0.075$             & $0.478$         & $0.036$\\
$H$ [km s$^{-1}$ Mpc$^{-1}$]      & $0.014$             & $83.3$          & $1.2$\\
$D_A$ $[\mathrm{Mpc}]$            & $0.053$             & $1130$          & $60$ \\
\hline
\hline
\end{tabular}
\caption{\label{tab:background} Data points and errors on $f\sigma_8$, the angular diameter distance $D_A$ and the Hubble parameter $H$ at redshift $z=0.39$. In the first column we show the forecast relative errors on these quantities at $z=2.5$ computed in ref.~\cite{Obuljen:2017}. In the second column we show theoretical expectations for $f\sigma_8$, $D_A$ and $H$ obtained with \texttt{CAMB} for our fiducial cosmology at redshift $z=0.39$. In the third column there are the absolute errors on each quantity, which we derived from the relative errors at redshift $z=2.5$ as $\sigma_x(z=0.39)= \sigma_x/x(z=2.5) \cdot  x(z=0.39)$. I.e., we set $\sigma_x/x(z=0.39) = \sigma_x/x(z=2.5)$.}
\end{table}

The mock data set build for $f\sigma_8(z)$, $D_A(z)$ and $H(z)$ at redshift $z=0.39$, is shown in table~\ref{tab:background}. To construct it, we start from the forecast relative errors on these quantities at redshift $z=2.5$~\cite{Obuljen:2017}. Following the trend found in ref.~\cite{Obuljen:2017}, we assume the relative errors on $f\sigma_8$, $D_A$ and $H$ to be increasing with the redshift. Then, in order to be conservative, we set the relative error at $z=0.39$ to the upper limit given by the relative errors at $z=2.5$. We compute absolute errors and central points from theoretical estimate of $f\sigma_8(z)$, $D_A(z)$ and $H(z)$, as explained in table~\ref{tab:background}.

From this mock data set we implement an additional likelihood term for each $x=f\sigma_8$, $H$, $D_A$ at $z=0.39$:
$$ -\ln\left[\mathcal{L}\left(x^{obs}|x^{th}\right)\right] =   \frac{1}{2} (x^{th} - x^{obs})\,  \frac{1}{\sigma_x^2} \, (x^{th} - x^{obs})\,.$$
We then sum each piece to the $\pto$ likelihood, defined in eq.~\eqref{eq:likelihood}. 
We stress that the 21cm power spectrum data and the background data obtained from 21cm measurements are considered to be independent. This is of course an assumption that deserves to be investigated with dedicated mock data sets.

\subsection{Theoretical framework}
\label{sec:theoretical_framework}

In this section we review general aspects of the beyond \lcdm models that we consider for our analysis. As anticipated, we work within the Effective Field Theory (EFT) framework. In the following,  we first review basic aspects of EFT of DE. Then, we describe in more detail the set of EFT models that we consider, and, finally, we  briefly summarize the latest available constraints. We stress that in our work we follow mainly the notation of refs.~\cite{hu:2014,num_notes_v3_EFTCAMB}. 

\subsubsection{Effective Field Theory formalism}
Upcoming, high precision cosmological observations are expected to shed light on the nature of gravity on large scales and the physics of DE. Given the plethora of candidate DE/MG, it is important to adopt a unifying framework that allows for an efficient comparison with data. To this extent, the Effective Field Theory (EFT) approach is a powerful tool that provides us with a general, unified description of the late-time DE/MG at the level of the action~\cite{Frusciante:2019}. 
The EFT framework has been originally introduced in Cosmology in the context of inflation~\cite{Creminelli:2006,Cheung:2007,Weinberg:2008}. Subsequently, it was applied to quintessence~\cite{Creminelli:2008wc} and eventually to scalar-tensor models of DE/MG ~\cite{Gubitosi:2013,bloomfield:2013}. The EFT of DE is formulated in terms of a  unifying action that allows for an efficient and broad sampling of the theory space under general conditions of stability.  It can be used in an agnostic way, with a set of free functions of time parametrizing theoretically allowed deviations from $\Lambda$CDM; or in a model-specific way, where the functions are customized to specific models through a mapping procedure.  For a detailed description of this, as well as a complete review of the state-of-the-art constraints on EFT models we refer the reader to ref.~\cite{Frusciante:2019}. 

The EFT formalism allows sampling a large number of different theoretical scenarios, of increasing complexity. For a first thorough forecast of 21cm intensity mapping constraints on DE/MG, we start from the subset representing Generalized Brans Dicke (GBD) theories, which includes f(R) and, more generally, chameleon type theories. In the EFT language, this class of theories can be explored simply by varying two functions of time, which determine both background and dynamics of linear perturbations. In our analysis, we opt for the Hubble parameter, $H(a)$ and the conformal coupling $\Omega(a)$ (see ref.~\cite{Frusciante:2016}
for further details on the mapping of these theories into EFT of DE and \texttt{EFTCAMB}.). Hereafter, we will use $\Omega^{\rm EFT}(a)$ to indicate this function, in order to avoid confusion with matter parameters that we vary in our analysis.  This corresponds to assuming a designer approach, in which we fix the expansion history and explore the space of non-minimally coupled DE/MG models reproducing it by means of varying the coupling $\Omega^{\rm EFT}(a)$\footnote{In the sub-case of $f(R)$ gravity, $\Omega^{\rm EFT}=df/dR$ in our convention.}. Let us recall that a non-zero conformal coupling implies a running of the Planck mass which can be alternatively explored with the function $\alpha_M\equiv H^{-1} d\ln{M_P^2}/dt$. Within the context of GBD , we simply have $\alpha_M=H^{-1} d \ln{(1+ \Omega^{\rm EFT})}/dt$. We will present constraints on both $\Omega^{\rm EFT}$ and $\alpha_M$.

GBD models represent a large set of DE/MG models which survive the latest constraints from the direct detection of gravitational waves, while still displaying an interesting phenomenology at the level of large scale structure \cite{Espejo:2018hxa}. While being relatively simple, they are still subject to the so-called stability constraints, which ensure that the theory is free from pathologies, such as ghost and gradient instabilities~\cite{Zhao:2008,Salvatelli:2016,Peirone:2017}. Imposing such viability conditions correctly is of crucial importance when studying EFT models in an MCMC framework, in order to select the correct parameter space volume. In the \texttt{EFTCAMB}/\texttt{EFTCosmoMC} codes routines to check the stability of a model are already implemented. In the following, we refer to these as viability conditions~\cite{hu:2014,num_notes_v3_EFTCAMB}.

\subsubsection{Selected models and their latest constraints}
\label{sec:tested_model}
\begin{figure}[tbp]
    \centering
    \includegraphics[width=1\textwidth]{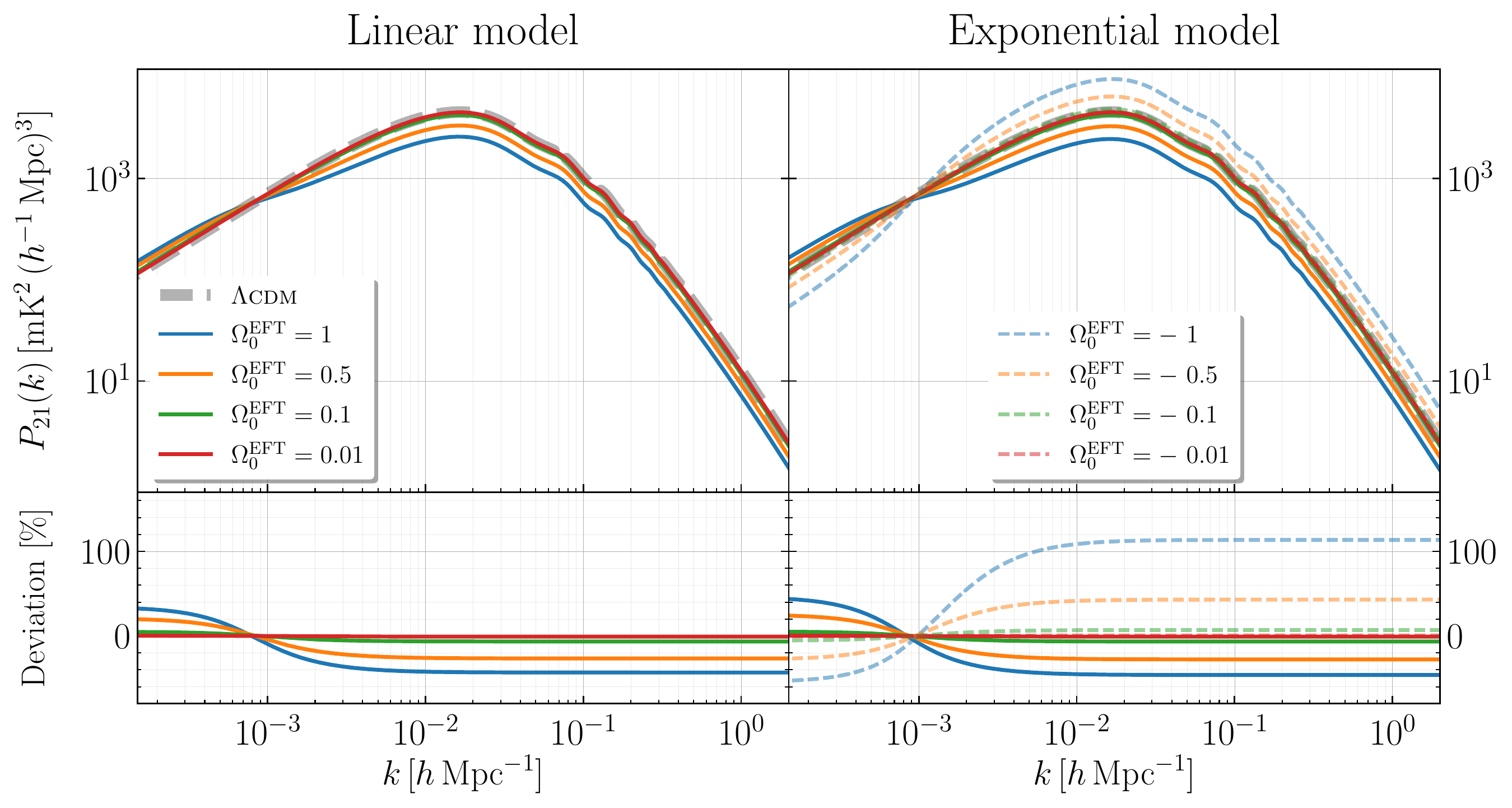}
    \caption{\label{fig:P21_linear} Upper panels: linear power spectra for the 21cm signal for $\Lambda$CDM and several \textit{pure} EFT models, described by different values of the parameter $\Omega_0^{\mathrm{EFT}}$. Lower panels: percentage deviations from $\Lambda$CDM predictions, computed as $(P_{21}^{\mathrm{EFT}} - P_{21}^{\Lambda \mathrm{CDM}})\times 100/P_{21}^{\Lambda \mathrm{CDM}} $.  We show theoretical predictions at redshift $z=0.39$ for a \textit{pure} linear EFT model (left panel) and an exponential one with fixed $\beta = 1$ (right panel). For the linear case we consider $\Omega_0^{\mathrm{EFT}} = 0.01,\, 0.1,\, 0.5,\, 1$, while for the exponential $\Omega_0^{\mathrm{EFT}} = \pm 0.01,\, \pm 0.1,\, \pm 0.5,\, \pm 1$. The other cosmological parameters are set to the assumed fiducial cosmology values.}
\end{figure}
\begin{figure}[tbp]
    \centering
    \includegraphics[width=1\textwidth]{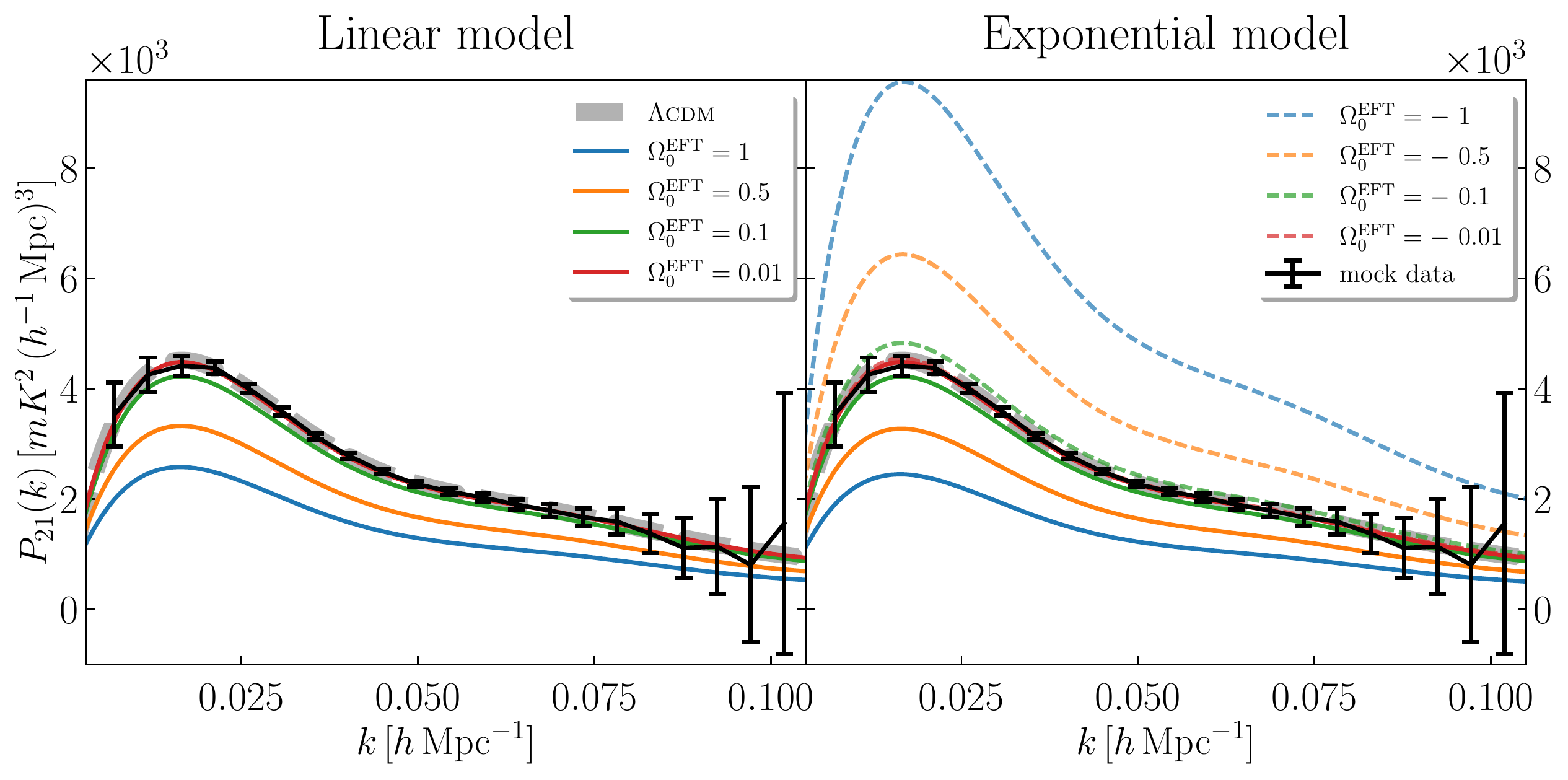}
    \caption{\label{fig:P21_vs_data} Linear power spectra for the 21cm signal for $\Lambda$CDM and several \textit{pure} EFT models, described by different values of the parameter $\Omega_0^{\mathrm{EFT}}$. We show theoretical predictions at redshift $z=0.39$ for a \textit{pure} linear EFT model (left panel) and an exponential one with fixed $\beta = 1$ (right panel). For the linear case we consider $\Omega_0^{\mathrm{EFT}} = 0.01,\, 0.1,\, 0.5,\, 1$, while for the exponential $\Omega_0^{\mathrm{EFT}} = \pm 0.01,\, \pm 0.1,\, \pm 0.5,\, \pm 1$. The other cosmological parameters are set to the assumed fiducial cosmology values. Black points with errorbars are the mock data set we construct in this work, as described in section~\ref{sec:mock_data}.}
\end{figure}

Let us outline the models that we consider for our analysis with the $P_{21}(k)$ likelihood while reviewing the most recent constraints on them. Our results relative to each scenario are discussed in section~\ref{sec:results}.  

We start by analyzing the $\Lambda$CDM case, for which theoretical predictions of the 21cm power spectrum at different redshifts are shown in Figure~\ref{fig:P21_delta_redshift}. We also consider the $w$CDM model, one of the simplest extensions to \lcdm, where the DE equation of state parameter $w$ is constant but different from $-1$, as it is the case for $\Lambda$CDM. No conformal coupling in this case. These two cosmologies are our starting point, on top of which we then turn on the conformal coupling, exploring different choices. Since we work in the designer approach, where we have fixed $H(a)$, we can focus on the effects of the coupling on perturbations. 

We explore different choices for $\Omega^{\rm EFT}(a)$, adopting parametrizations that have been already explored in the context of different observables. This allows us to draw a more meaningful comparison of the constraining power of $\pto$ with other probes. We start with the linear case, for which $\Omega^{\mathrm{EFT}}(a)$ is:
\begin{equation}
    \Omega^{\mathrm{EFT}}(a) = \omegaEFT a,
\end{equation}
where $\omegaEFT$ is a constant and all other second order EFT functions are set to zero. As it was shown in ref.~\cite{Raveri:2014}, the viability conditions require a non-negative $\omegaEFT$, i.e. $\omegaEFT \geq 0$, both in the case of a \lcdm and $w$CDM background.  We recover the \lcdm limit when $\omegaEFT \to 0$. We refer to this model as \textit{pure} linear EFT. The parameter $\omegaEFT$ has been constrained to be $\omegaEFT < 0.043$ $(95\%\,\,\text{CL})$ on a \lcdm background, from Planck 2015 TT, TE, EE power spectra combined with  BAO, Supernovae and $H_0$ priors data sets~\cite{planck:2015de-mg}. On a $w$CDM background, instead, the constraint $\omegaEFT < 0.058$ $(95\%\,\,\text{CL})$ was found from Planck 2013 TT power spectrum and lensing data combined with BAO observations~\cite{Raveri:2014}.
Both the matter power spectrum and the growth rate depend on $\omegaEFT$, thus the 21cm power spectrum is expected to be sensitive to this parameter as well. 

Another interesting parametrization is the exponential one, i.e:
\begin{equation}\label{eq:exponential}
    \Omega^{\mathrm{EFT}}(a) =  \exp ( \omegaEFT a^\beta ) -1\,.
\end{equation}
It is possible to see that this model has a simple mapping into the alternative parametrization referred to as the $\alpha$-basis~\cite{Bellini:2014}, of which we discussed the first function, $\alpha_M$.
Namely, it corresponds to a non-minimally coupled model in which the kinetic braiding and conformal coupling are related in a simple way, i.e. $\alpha_B=-\alpha_M$, and $\alpha_K$ is determined by the background, i.e. it is a function of $H$ and $\alpha_M$~\cite{planck:2015de-mg}. The function $\alpha_M$ evolves as:
\begin{equation}
    \alpha_M = \alpha_M^0 a^\beta,
\end{equation}
where we identify $\alpha_M^0 = \omegaEFT \beta$. We refer to this model as \textit{pure} exponential EFT. 

The latest, most stringent constraints on the exponential model are $\alpha_M^0 < 0.062$ $(95\%\,\,\text{CL})$, and $\beta = 0.92^{+0.53}_{-0.24}$ $(68\%\,\,\text{CL})$,
from Planck 2015 data combined with BAO and Supernovae datasets on a \lcdm background~\cite{planck:2015de-mg}. From Planck 2018 data, instead, a mild ($1.6\,\sigma$) preference for a negative running of the Planck mass has been observed~\cite{planck:2018}.\footnote{This preference is reduced to $<1\,\sigma$, with the inclusion of the BAO and weak lensing datasets.} Constraints computed from Planck 2018 TT, TE, EE power spectra, lensing and low polarization CMB data are $\omegaEFT = -0.049^{+0.037}_{-0.024}$ $ (68\%\,\,\text{CL})$, $\alpha_M^0 = -0.040^{+0.041}_{-0.016}$ $(68\%\,\,\text{CL})$, $\beta = 0.72^{+0.38}_{-0.14}$ $(68\%\,\,\text{CL})$.

Our predictions of $P_{21}(k)$ for a wide range of $\omegaEFT$ values are shown in figure~\ref{fig:P21_linear}, where we observe that the 21cm power spectrum is sensitive to variations in the EFT parameters. At small scales and redshift $z=0.39$, the amplitude of $\pto$ is rescaled for different values of $\omegaEFT$. This is the range of scales we investigate, as described in section~\ref{sec:mock_data}. Variations in this regime are at the level of $1 - 40\%$ (continuous lines), up to more than $100\%$ for negative values of $\omegaEFT$ (dashed lines). At large scales, instead, we observe a transfer of power, but it is not possible to probe this regime with current experiments. The behavior of $\pto$ for EFT models is induced mainly by the matter power spectrum $P_\mathrm{m}(z,k)$. In the $\pto$ model of eq.~\eqref{eq:P_21}, the only quantities sensitive to EFT modifications are $P_\mathrm{m}(z,k)$ and $f(z)$, which origins an overall rescaling. Thus, if we have a good estimate for the astrophysical quantities that enters in the $\pto$ model, e.g. the brightness temperature, we expect that the $\pto$ may be useful to constrain these kind of EFT models. 

\begin{figure}[tbp]
    \centering
    \includegraphics[width=0.9\textwidth]{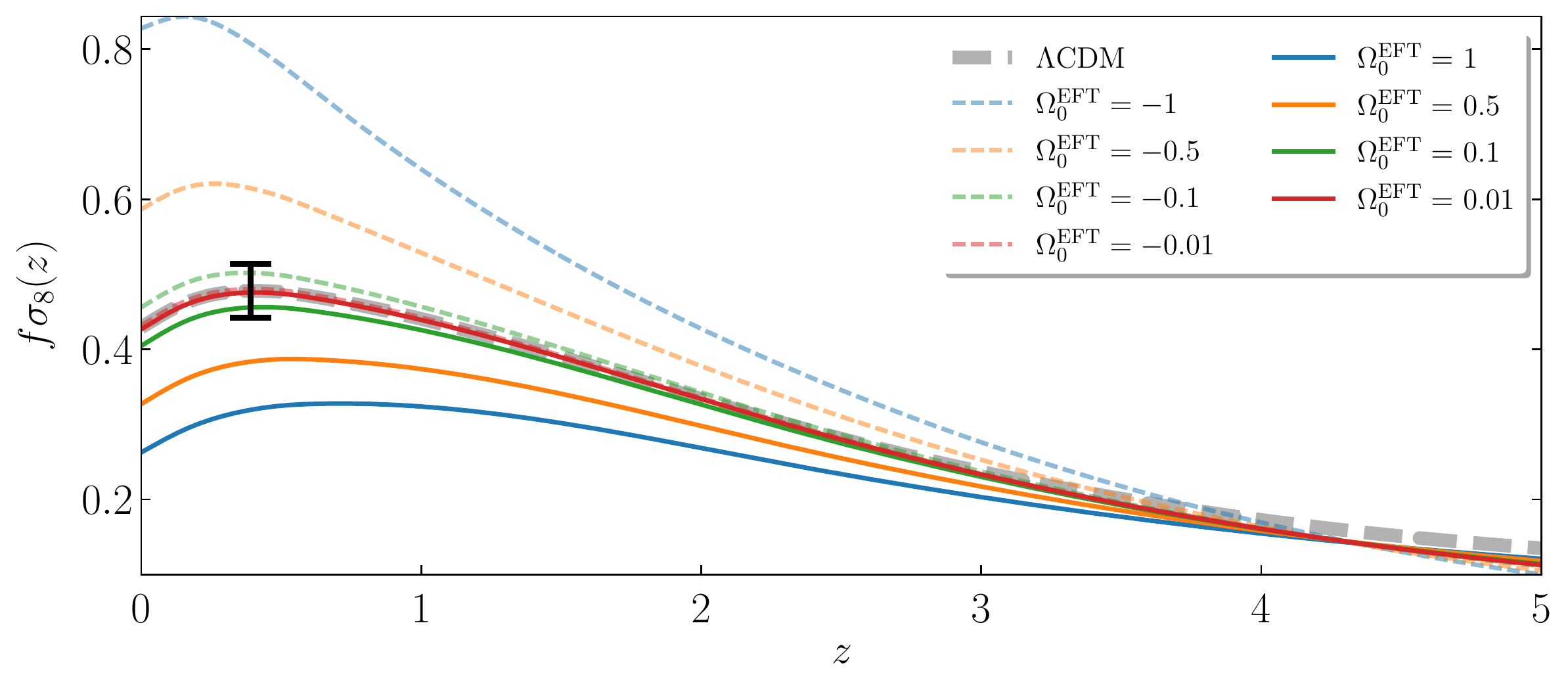}
    \caption{\label{fig:fsigma8} Theoretical predictions for the quantity $f\sigma_8$. We show the redshift evolution of $f\sigma_8$ for different models, i.e. $\Lambda$CDM, a \textit{pure} linear EFT model described by $\Omega_0^{\mathrm{EFT}} = 0.01,\, 0.1,\, 0.5,\, 1$ (solid lines), a \textit{pure} exponential EFT model with $\Omega_0^{\mathrm{EFT}} =- 0.01,\, -0.1,\, -0.5,\, -1$ and $\beta=1$. The black data point with error bars at $z=0.39$ is the one given in table~\ref{tab:background}. The other cosmological parameters are set to the assumed fiducial cosmology values.}
\end{figure}
In figures~\ref{fig:P21_vs_data} and~\ref{fig:fsigma8}, we compare EFT modifications to $\pto$ and $f\sigma_8$ with the mock data sets we construct in sections~\ref{sec:mock_data} and~\ref{sec:background} respectively. With the errorbars shown we expect the $\pto$ likelihood to be sensitive to EFT modifications. We stress that the current constraints from CMB data on EFT parameters limit $\omegaEFT$ to be $\omegaEFT \leq 0.1$ (green lines). EFT modifications for $\omegaEFT \leq 0.1$ fall within the errorbars of the mock data. We discuss the implications of this feature in section~\ref{sec:res_EFT_lcdm}. 

When performing MCMC analyses for the models outlined above, we vary the six cosmological parameters $\{\Omega_b h^2$, $\Omega_c h^2$, $100\theta_{MC}$, $\tau$, $A_s$, $n_s\}$ and the EFT parameters describing each EFT models. Results for all other quantities, such as $H_0$ and $\sigma_8$, are derived from constraints on the free parameters. If not explicitly stated, we consider the flat priors listed in table~\ref{tab:flat_priors}.
\begin{table}[tbp]
    \centering\small
    \begin{tabular}{lSc|lSc|lSc} \hline\hline
{\textsc{Parameter}} & {\textsc{Prior}} & {\textsc{Parameter}} & {\textsc{Prior}} & {\textsc{Parameter}} & {\textsc{Prior}}  \\\hline
    $ \Omega_b h^2 $ \dotfill & $[5\times 10^{-3}, 0.1]$ & $\tau$ \dotfill &$[0.01, 0.8]$
    & $w_0 $ \dotfill &$[-1.5,0]$\\
$ \Omega_c h^2 $ \dotfill  & $[0.1\times 10^{-3}, 0.99]$ & $\ln(A_s)$ \dotfill &$[1.61, 3.91]$
& $\omegaEFT $ \dotfill &$[-1,1 ]$\\
$100\theta_{MC}$ \dotfill&  $[0.5, 100]$&$ n_s $ \dotfill &$[0.8,1.2]$ 
& $\beta $ \dotfill &$[0,3]$\\ \hline\hline
    \end{tabular}
    \caption{Adopted flat priors on cosmological and \textit{pure} EFT parameters.}
    \label{tab:flat_priors}
\end{table}

\subsection{CMB data sets}
\label{sec:data_sets}

For a complete analysis of the efficacy of the $\pto$ likelihood, we also investigate the implications of using it combined with CMB observations. We work with the Planck measurements from the 2018 data release~\cite{planck:2018}. In particular, we include the high-$\ell$ TT, TE, EE lite likelihood in the interval of multipoles $30\leq\ell\leq 2508$ for TT and $30\leq\ell\leq 19696$ for TE, EE. Lite likelihoods are  calculated with the \texttt{Plik lite} likelihood~\cite{planck:2018like}. Instead for the low-$\ell$ TT power spectrum we use data from the \texttt{Commander} component-separation algorithm in the range $2 \leq \ell \leq 29$. We stress that we always include the low-$\ell$ likelihood, even when not explicitly stated. We adopt also the Planck CMB lensing likelihood and the low EE polarization power spectrum, referred to as lowE, in the range $2 \leq \ell\leq 29$, calculated from the likelihood code \texttt{SimAll}~\citep{planck:2018_maps}.

\section{Results}
\label{sec:results}
In this section, we illustrate the results obtained for the models outlined in section~\ref{sec:tested_model}. We present constraints on cosmological and EFT parameters from several different data sets. On the one hand, we test the effect of 21cm data alone. On the other, we analyze the impact of adding $\pto$ to CMB observations. We refer to appendix~\ref{app:constr_tables} for tables showing the full results of our analyses, i.e. the complete sets of constraints on all the model parameters.

\subsection{\lcdm reference results}
\label{sec:res_lcdm}
\begin{figure}[tpb]
    \centering
    \includegraphics[width=0.9\textwidth]{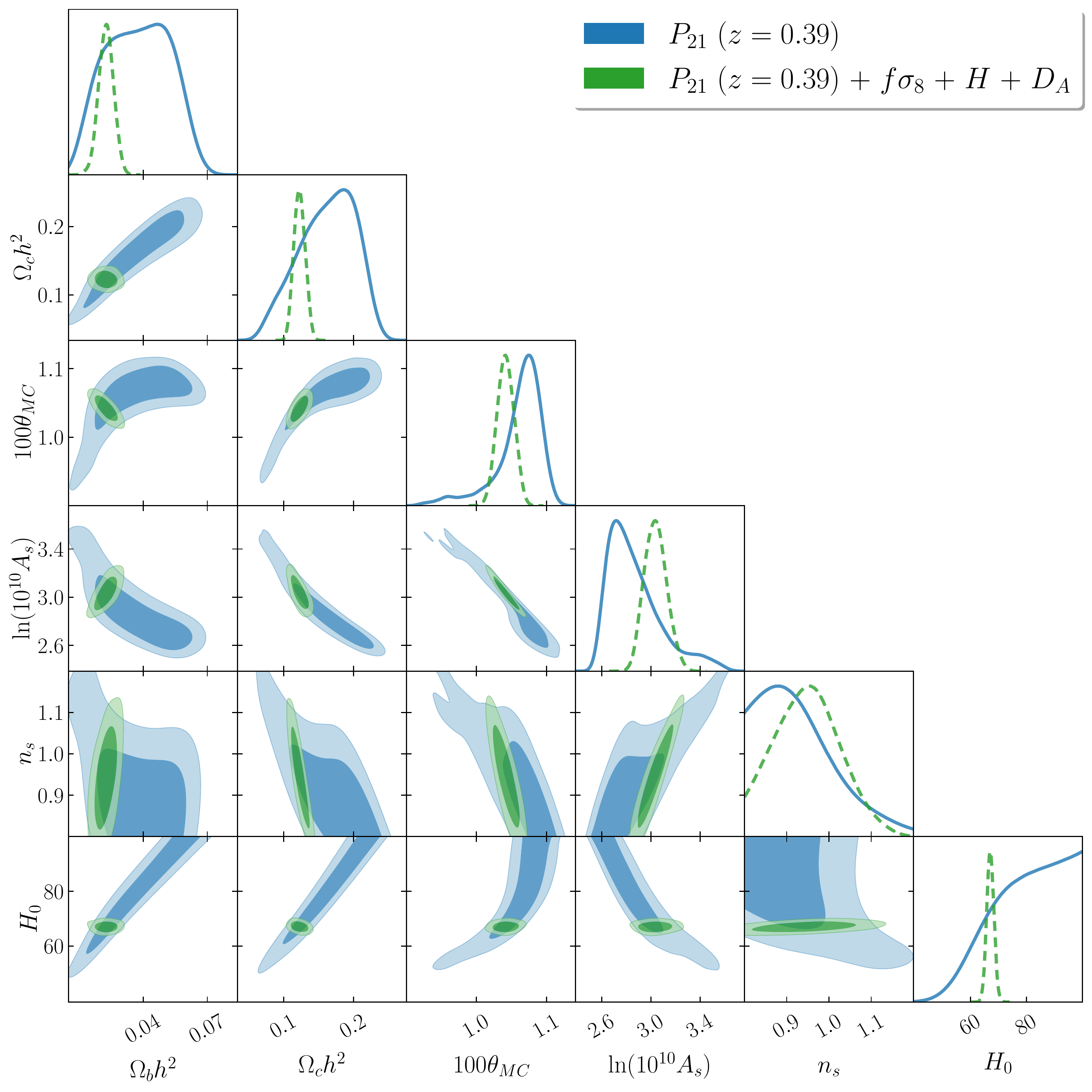}
    \caption{\label{fig:res_LCDM_P21_back} Joint constraints (68\% and 95\% confidence regions) and marginalized posterior distributions on cosmological parameters for a $\Lambda$CDM model. Here the label $P_{21} (z=0.39)$ stands for the 21cm power spectrum likelihood at redshift $z=0.39$, while $f\sigma_8 + H + D_A$ represents the additional \textit{background} likelihood described in section~\ref{sec:background} and computed at the same redshift $z=0.39$. We consider the parameter $\tau$ fixed to the fiducial cosmology value, i.e. $\tau = 0.0543$.}
\end{figure}

We begin by testing the constraining power of the mock data set we constructed. We constrain cosmological parameters for \lcdm using the $\pto$ likelihood alone and combined with the \textit{background} likelihood (see section~\ref{sec:background}). Results of this first analysis are shown in figure~\ref{fig:res_LCDM_P21_back} and table~\ref{tab:lcdm_testo}. As one could expect, $\pto$ alone is not able to constrain all the six cosmological parameters at once and some of them remain completely unconstrained. However, fixing $\tau$ to its fiducial cosmology value, resulted to be enough to reach numerical convergence and obtain broad constraints. Looking at the 2D contour in figure~\ref{fig:res_LCDM_P21_back}, we can observe a clear positive correlations between $\Omega_c h^2$ and $H_0$. We anticipate that this feature will be pivotal when combining $\pto$ with CMB data. 
\begin{figure}[tbp]
    \centering
    \includegraphics[width=0.9\textwidth]{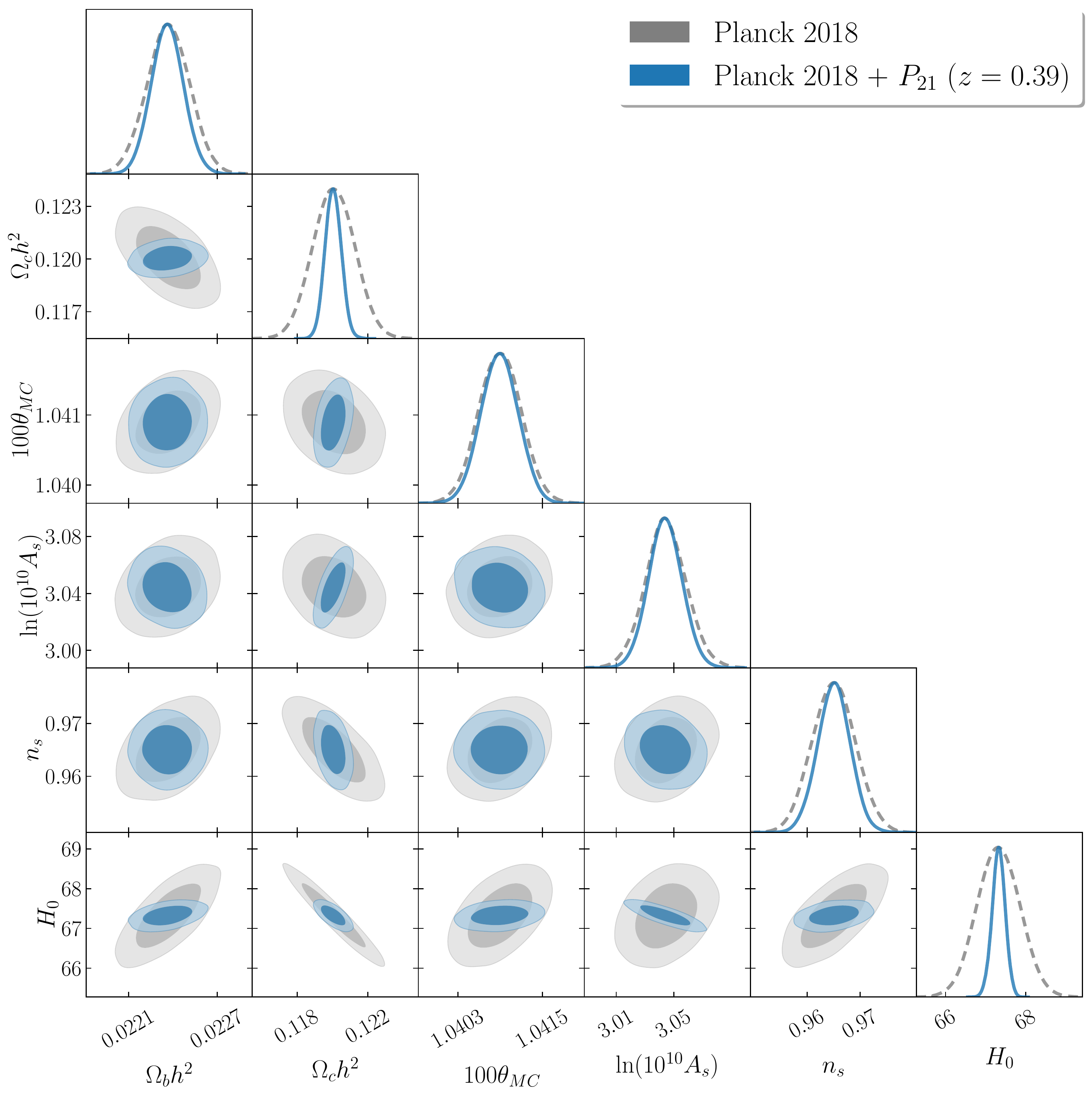}
    \caption{\label{fig:res_LCDM_Planck_P21} Joint constraints (68\% and 95\% confidence regions) and marginalized posterior distributions on cosmological parameters for a $\Lambda$CDM model. Here the label Planck 2018 stands for TT, TE, EE + lowE + lensing while the label $P_{21} (z=0.39)$ stands for the 21cm power spectrum likelihood at redshift $z=0.39$.  }
\end{figure}

This correlation is almost completely removed when we combine $\pto$ with the \textit{background} likelihood, i.e. including the three more data points on $f\sigma_8$, $H$ and $D_A$. In this case we obtain significantly tighter and smoother posterior distributions, as shown in figure~\ref{fig:res_LCDM_P21_back} and we obtain a competitive error on the estimate of $H_0$, i.e.
\begin{equation}
    H_0 = 67.1 \pm 1.3 \text{ km s}^{-1} \text{Mpc}^{-1}.
\end{equation}
Thus, we observe that 21cm observables alone can constrain $H_0$ with an error comparable to measurements obtained from other probes~\cite{Verde:2019}. E.g. early time probes constrain $H_0$ with an error of  $\sim 1.2 \text{ km s}^{-1} \text{Mpc}^{-1}$, from the Dark Energy Survey observations combined with BAO and BBN data~\cite{abbott:2018}.
\begin{table}[tbp]
    \centering\small
    \begin{tabular}{lScScScSc} \hline\hline
{\textsc{Parameter}} & Planck 2018 & Planck 2018 + $P_{21}$ & $P_{21}$ & $P_{21}$ + \textit{background}   \\\hline
    $ \Omega_b h^2 $ \dotfill & $0.02237\pm 0.00014$ & $0.02236\pm 0.00011$ $(\mathbf{-24}\%)$ & $0.038\pm 0.015$ & $0.0226\pm 0.0035$  \\
$ \Omega_c h^2 $ \dotfill & $0.1201\pm 0.0012$ & $0.12004\pm 0.00046$ $(\mathbf{-61\%})$ & $0.162^{+0.050}_{-0.033}$ & $0.1227\pm 0.0081$  \\
$ n_s $ \dotfill & $0.9650\pm 0.0041$ & $0.9651\pm 0.0031$ $(\mathbf{-25}\%)$ & $< 1.10^\dag$ & $0.951^{+0.072}_{-0.085}$ \\ \hline
$ H_0 $ \dotfill & $67.32\pm 0.53$ & $67.32\pm 0.16$ $(\mathbf{-69\%})$ &$> 57.8^\dag$ & $67.1\pm 1.3$ \\ \hline\hline
    \end{tabular}
    \caption{Marginalized constraints on cosmological parameters at the $68\%$ confidence level for a \lcdm model. $95\%$ confidence levels are marked with $^\dag$. Deviations in the error with respect to Planck 2018 results are shown in brackets. Here the label Planck 2018 stands for TT, TE, EE + lowE + lensing, while the label $P_{21} (z=0.39)$ stands for the 21cm power spectrum likelihood at redshift $z=0.39$. The label \textit{background} represents the additional \textit{background} likelihood described in section~\ref{sec:background}. Full constraints available in tables~\ref{tab:LCDM_fixed_aTb}.}
    \label{tab:lcdm_testo}
\end{table}
Once the impact of $\pto$ alone is established, we turn to combine $\pto$ with Planck 2018 CMB data. Marginalized posterior distributions are shown in figure~\ref{fig:res_LCDM_Planck_P21}, while constraints are listed in table~\ref{tab:lcdm_testo}. We find that adding $\pto$ significantly improves on the constraints produced by Planck data alone, reducing the error on cosmological parameters by at least $\sim 10 \%$ to $\sim 70\%$. The maximum effect is obtained for $\Omega_c h^2$ and $H_0$. With Planck data we obtain $\sigma_{\Omega_c h^2} = 1.2 \times 10^{-3}$,  and $\sigma_{H_0} = 0.53$, while adding the $\pto$ likelihood the errors reduce to $\sigma_{\Omega_c h^2} = 4.6 \times 10^{-4}$ and $\sigma_{H_0} = 0.16$. The percentage reduction is of the $61\%$ and the $69\%$, respectively. 

Comparing the 2D contour plots of figures~\ref{fig:res_LCDM_P21_back} and~\ref{fig:res_LCDM_Planck_P21}, we observe that $\Omega_c h^2$ and $H_0$ are correlated both using $\pto$ or Planck data alone. Nevertheless, the correlation appears to develop along two orthogonal directions for the two data sets. When using the data sets combined, the contour reduces to the intersection of these two regions, thus producing very tight constraints on $\Omega_c h^2$ and $H_0$. Therefore, the $\pto$ likelihood performs as expected, i.e. it provides complementary correlations to that of CMB and remarkably  improves on the constraints.

Although we do not show results here, we note that the effect of adding BAO data has also been tested. We observe no significant impact when we combine BAO both with $\pto$ data alone and with Planck + $\pto$. 

\subsection{\textit{Pure} EFT models on a \lcdm background}
\label{sec:res_EFT_lcdm}
\begin{figure}[tbp]
    \centering
    \includegraphics[width=0.65\textwidth]{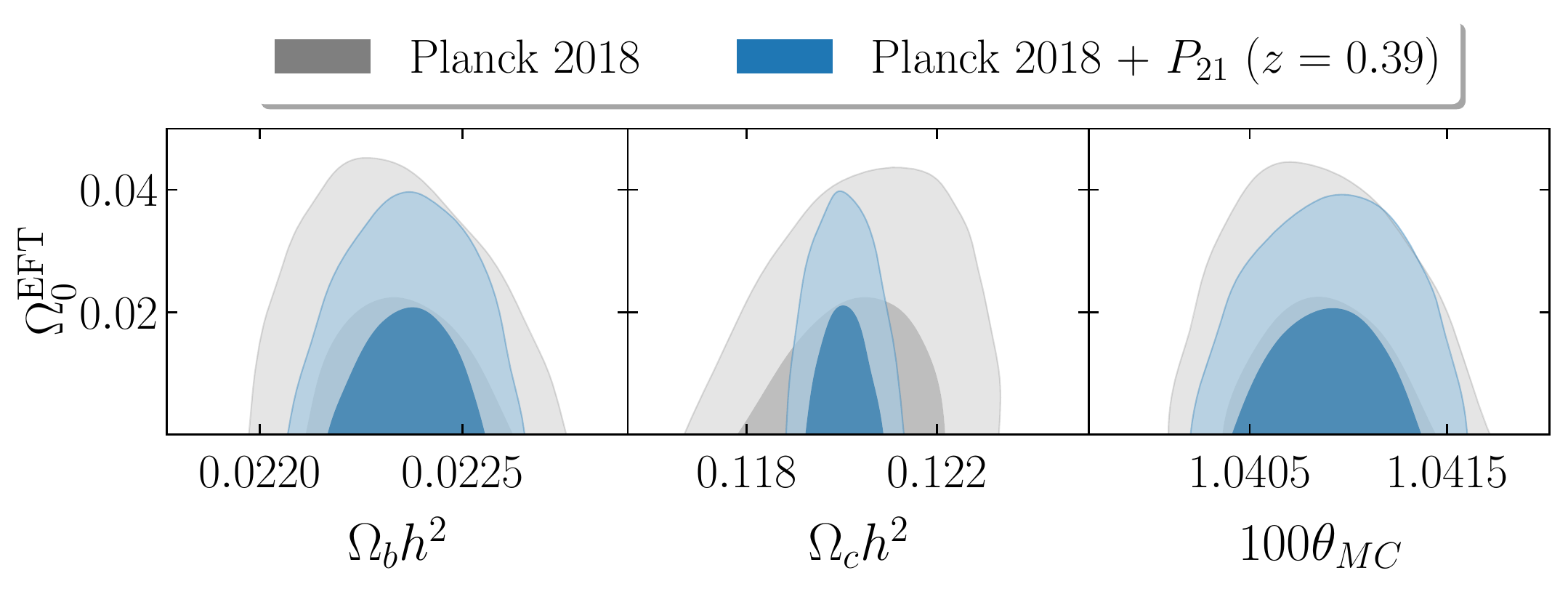}
    \includegraphics[width=0.65\textwidth]{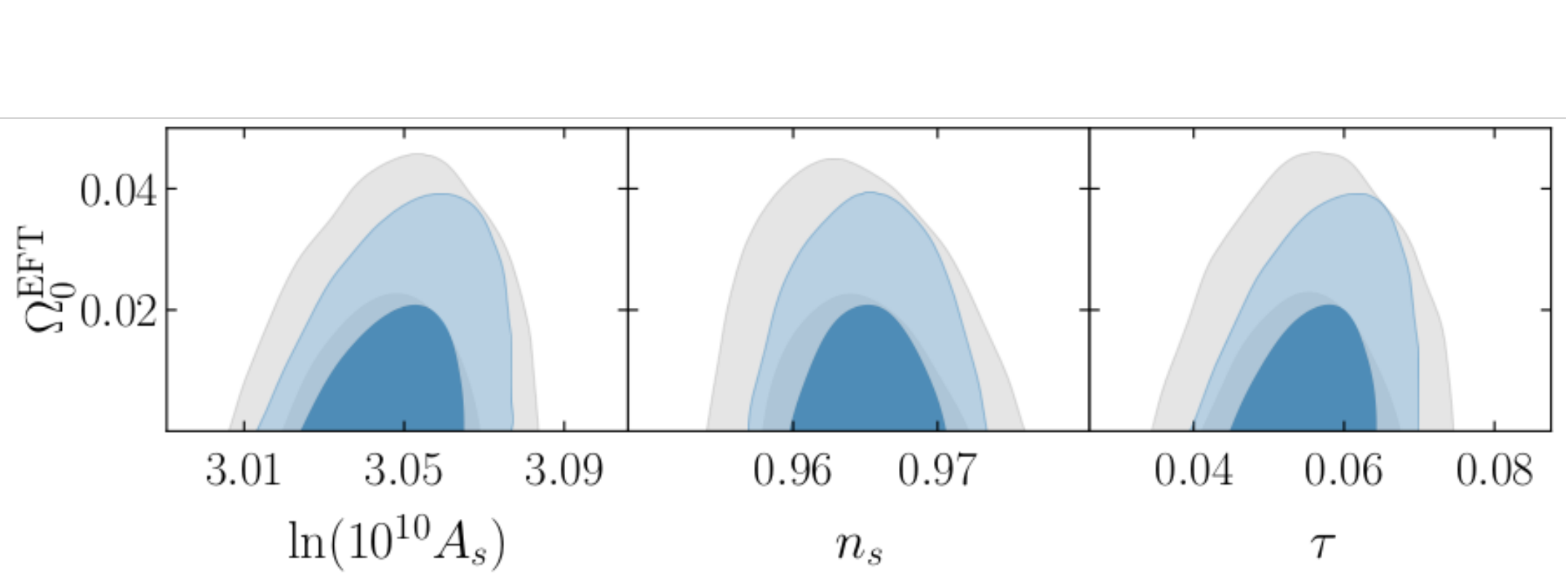}
    \caption{\label{fig:res_linear_Planck_P21} Joint constraints (68\% and 95\% confidence regions) on cosmological and EFT parameters for a \textit{pure} linear EFT model on a $\Lambda$CDM background. Here the label Planck 2018 stands for TT, TE, EE + lowE + lensing while the label $P_{21} (z=0.39)$ stands for the 21cm power spectrum likelihood at redshift $z=0.39$. }
\end{figure}

\begin{figure}[tbp]
    \centering
    \includegraphics[width=0.49\textwidth]{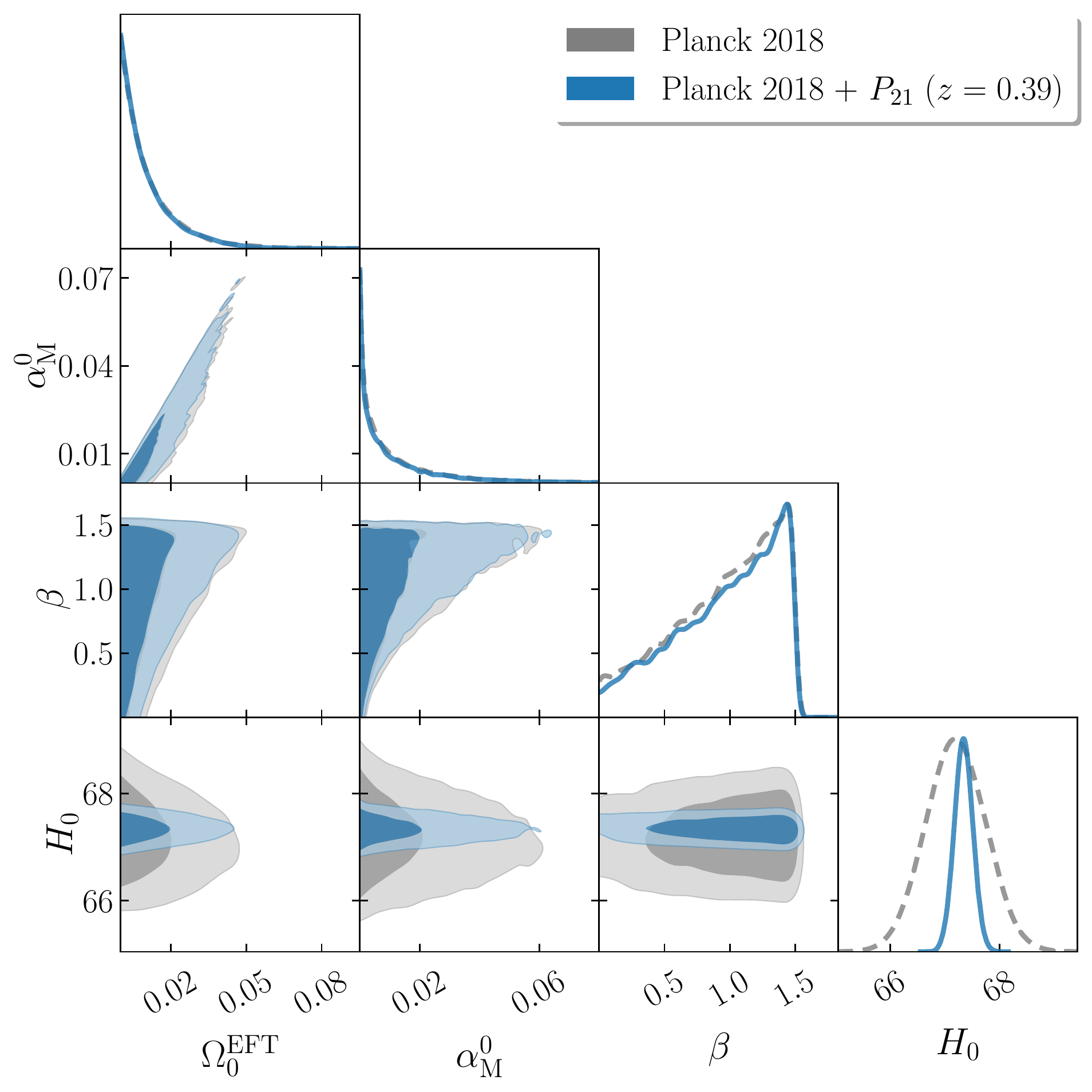}
    \includegraphics[width=0.49\textwidth]{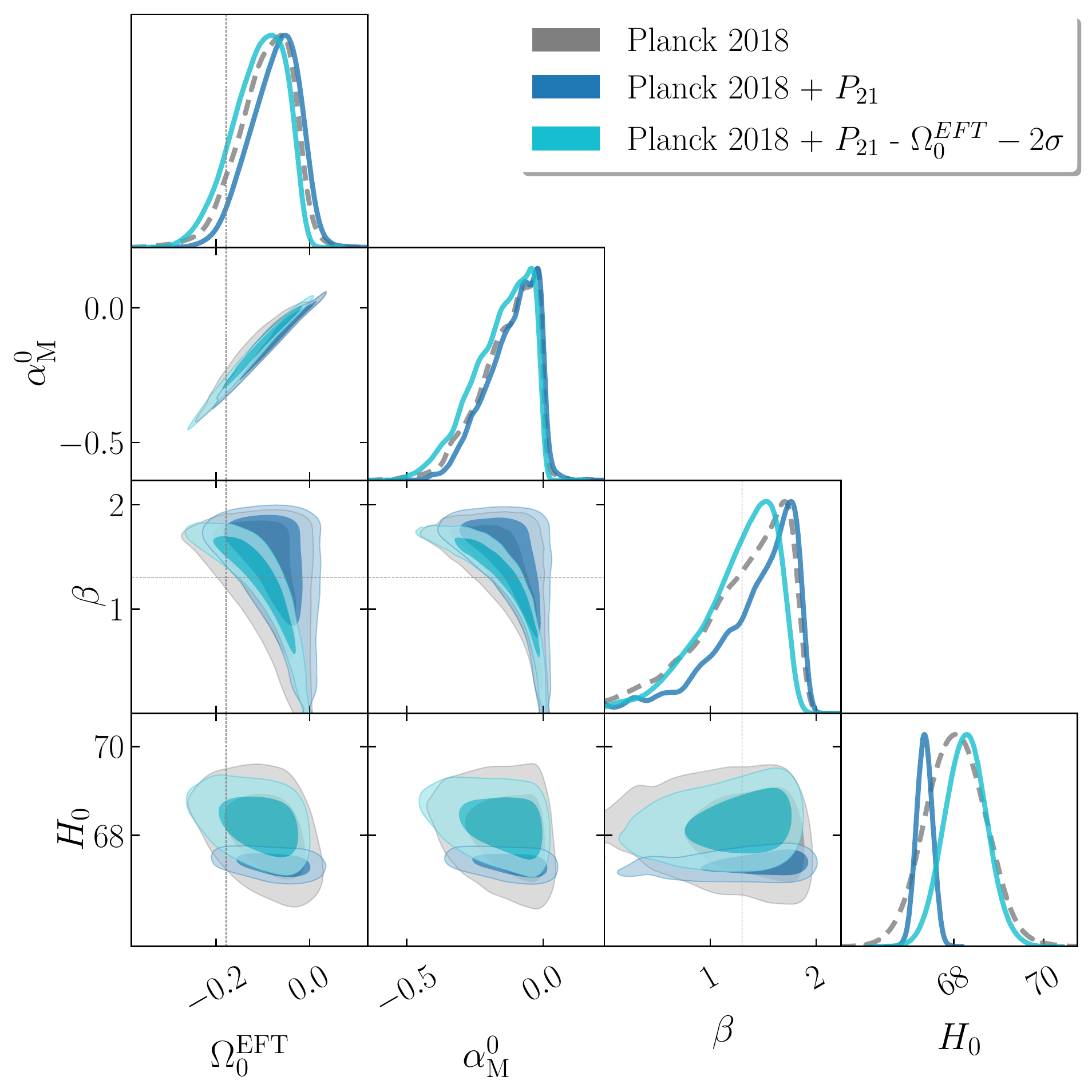}
    \caption{\label{fig:res_exponential_Planck_P21} Joint constraints (68\% and 95\% confidence regions) and marginalized posterior distributions on cosmological and EFT parameters for a \textit{pure} exponential EFT model on a $\Lambda$CDM background (left panel). We show also results for the same model for a negative running of the Planck mass (right panel). Here the label Planck 2018 stands for TT, TE, EE + lowE + lensing while the label $P_{21} (z=0.39)$ stands for the 21cm power spectrum likelihood at redshift $z=0.39$. $\omegaEFT - 2\sigma$ mark the constraints obtained from a new mock data set constructed setting $\omegaEFT$ $2\sigma$ away from the Planck estimate. I.e., we set $\omegaEFT=-0.178$, $2\sigma$ away from the Planck constraints, and $\beta = 1.28$, its Planck constraint. These values are marked with gray lines in the right panel plots.}
\end{figure}

After having established the constraining power of the 21cm signal, we study its impact on beyond \lcdm theories. We consider the \textit{pure} EFT models on a \lcdm background, which are described in section~\ref{sec:tested_model}. We test three different scenarios: $i)$ the \textit{pure} linear EFT model, $ii)$ the \textit{pure}  exponential EFT model, $iii)$ a \textit{pure} exponential EFT model for which we allow a negative running of the Planck mass. Results are shown in figures~\ref{fig:res_linear_Planck_P21},~\ref{fig:res_exponential_Planck_P21} and table~\ref{tab:EFT_lcdm_testo}.  

As a first check, we compute constraints from $\pto$ alone. As before, in this analysis, $\tau$ is kept fixed to its fiducial cosmology value. 
We observe that the state-of-the-art mock uncertainties still provide a weak constraining power for these models and 
the EFT parameters appear to be loosely constrained. As displayed in table~\ref{tab:EFT_lcdm_testo}, for both the \textit{pure} linear and exponential EFT models we find $\omegaEFT$ to be unconstrained. 
The best result is achieved on the parameter $\beta$ describing the exponential EFT models. $\pto$ alone is able to constrain $\beta$ with slightly broader error with respect to the one we obtain from Planck data. Moreover, we observe that adding the background data slightly improves the results on the EFT parameters, although with no significant impact. This is consistent with the fact that we assume a designer approach, where we parametrize independently the background and the EFT functions.  Consequently, adding information on the background does not have a direct impact on  EFT parameters constraints.
We conclude that the mock data set we constructed for the redshift bin $z=0.39$ exhibits a limited constraining power on the EFT parameters. However, we anticipate that future tomographic observations at multiple redshifts could significantly help to increase the constraining power of this observable alone on EFT parameters. We further discuss this claim in section~\ref{sec:res_tomography}.

We proceed by combining $\pto$ with the latest Planck 2018 CMB observations. Overall, we observe that constraints on cosmological parameters remain unaffected compared to the \lcdm case. Furthermore, adding $\pto$ produces an improvement on EFT parameters at the level of $1-18\%$. In the following we discuss results for each \textit{pure} EFT model individually. 

\begin{table}[tbp]
    \centering\small
    \begin{tabular}{lScScScSc} \hline\hline
{\textsc{Parameter}~~~~~~~~~~~~~} & Planck 2018 & Planck 2018 + $P_{21}$ & $P_{21}$ & $P_{21}$ + \textit{background} \\\hline
Linear EFT  \dotfill& \\
 $ \Omega_0^{\rm EFT}\dotfill $ & $< 0.035^\dag$ & $< 0.031^\dag$ $(\mathbf{-11}\%)$ & --- & ---   \\ 
  $H_0 $ \dotfill & $67.21\pm 0.55$ & $67.33\pm 0.17$ $(-70\%)$ & $> 57.7^\dag$ & $67.0\pm 1.3$   \\ \hline
Exponential EFT  \dotfill& \\
$ \Omega_0^{\rm EFT} $ \dotfill & $< 0.036^\dag$ & $< 0.034^\dag$ $(\mathbf{-5}\%)$ & ---   & --- \\
$ \beta $ \dotfill & $0.96^{+0.54}_{-0.19}$ & $0.97^{+0.53}_{-0.19}$ $(\mathbf{-1}\%)$ & $1.30\pm 0.57$  & $1.20^{+0.58}_{-0.47}$   \\
$ \alpha_M^0 $ \dotfill & $< 0.045^\dag$ & $< 0.044^\dag$ $(\mathbf{-3}\%)$ & $< 1.89^\dag$   &$< 1.72^\dag$ \\
$H_0$ \dotfill & $67.20\pm 0.55$ & $67.34\pm 0.17$ $(-69\%)$ & $> 64.3^\dag$  & $66.9\pm 1.3$\\\hline
Exponential EFT  &  \\
negative running \dotfill \\
 $ \Omega_0^{\rm EFT} $ \dotfill & $-0.094^{+0.071}_{-0.042}$ & $-0.075^{+0.063}_{-0.042}$ $(\mathbf{-7}\%)$ & --- & $0.19^{+0.69}_{-0.34}$   \\
$ \beta $ \dotfill & $1.28^{+0.56}_{-0.21}$ & $1.41^{+0.48}_{-0.15}$ $(\mathbf{-18}\%)$ & $1.21^{+0.57}_{-0.70}$ & $1.38\pm 0.55$   \\
$ \alpha_{\rm M}^0 $ \dotfill & $-0.133^{+0.13}_{-0.048}$ & $-0.117^{+0.12}_{-0.052}$ $(\mathbf{-3}\%)$ & $0.31^{+0.84}_{-0.70}$ & $0.36\pm 0.80$   \\
$ H_0 $ \dotfill & $68.03\pm 0.66$ & $67.36\pm 0.18$ $(-73\%)$ &$> 56.3^\dag$ & $67.0\pm 1.3$  \\
 \hline\hline
    \end{tabular}
    \caption{Marginalized constraints on cosmological parameters at the $68\%$ confidence level for \textit{pure} EFT models on a \lcdm background. $95\%$ confidence levels are marked with $^\dag$, while the symbol --- means that we do not find any constraint. Deviations in the error with respect to Planck 2018 results are shown in brackets. Here the label Planck 2018 stands for TT, TE, EE + lowE + lensing while the label $P_{21}$ stands for the 21cm power spectrum likelihood at redshift $z=0.39$. The label \textit{background} represents the additional \textit{background} likelihood described in section~\ref{sec:background}. Full constraints available in tables~\ref{tab:EFT_fixed_aTb},~\ref{tab:EFT_linear},~\ref{tab:EFT_exponential},~\ref{tab:EFT_exp_neg}.}
    \label{tab:EFT_lcdm_testo}
\end{table}
First, for the linear model from Planck data alone we constrain $\omegaEFT$ to be:
\begin{equation}
    \omegaEFT < 0.035\quad (95\%\,\,\text{CL}) ,
\end{equation}
as we can see in table~\ref{tab:EFT_lcdm_testo}. This upper limit improves on previous results in literature ($\omegaEFT < 0.043$), which were produced using Planck 2015 data sets and an older version of the code \texttt{EFTCAMB} (see section~\ref{sec:tested_model}). When we add the $\pto$ likelihood, we get: 
\begin{equation}
    \omegaEFT < 0.031 \quad (95\%\,\,\text{CL}) ,
\end{equation}
which improves the constraint from Planck alone at the level of $11\%$. Contour plots are shown in figure~\ref{fig:res_linear_Planck_P21}. We can observe that adding $\pto$ has a mild effect on removing the correlation between $\omegaEFT$ and some of the cosmological parameters, e.g. $\Omega_c h^2$. 

Second, as shown in table~\ref{tab:EFT_lcdm_testo} for the \textit{pure} exponential EFT model we obtain constraints from Planck 2018 data alone which are compatible with previous results in literature, produced with Planck 2015 data as before (see section~\ref{sec:tested_model}). Adding $\pto$ provides similar effects as for the linear model. With $\pto$ + Planck we obtain: 
\begin{align}
    \begin{split}
    \omegaEFT < 0.034 \quad (95\%\,\,\text{CL}), \\
        \beta = 0.97^{+0.53}_{-0.19} \quad (68\%\,\,\text{CL}), \\
        \alpha_M^0 < 0.044\quad (95\%\,\,\text{CL}) ,
    \end{split}
\end{align}
Note a $5\%$ reduction of the upper limit on $\omegaEFT$, and consequently on $\alpha_M^0$, with respect to Planck data alone. Instead, constraints on the parameter $\beta$ appear to be unaffected when we add $\pto$\footnote{Notice that the posterior distribution of the parameter $\beta$ is asymmetrical resulting in different confidence levels above and below the mean value. When we compute percentage deviations between different results, we compare the mean of the upper and lower errors. We do so for all asymmetrical constraints.}. Contour plots are shown in figure~\ref{fig:res_exponential_Planck_P21} (left panel). As for the linear case, we notice a small reduction of the correlation between $\omegaEFT$ and $H_0$. The characteristic shape of the $\beta - \alpha_M^0$ contour is mainly set by viability requirements, which sharply cut the samples at $\beta \sim 1.6$. For lower values of $\beta$, the MCMC samples converge towards $\omegaEFT=0$, resulting in an elongated shape of the contour. This is an attribute of the functional form of the EFT function $\Omega^{\mathrm{EFT}}(a)$. According to eq.~\eqref{eq:exponential}, when $\omegaEFT = 0$, i.e. in the \lcdm limit, also $\Omega^{\mathrm{EFT}}(a)=0$ for any value of $\beta$. Thus, near the \lcdm limit $\beta$ is unconstrained and the MCMC samples populate all the values allowed by viability conditions. We can observe this property for all the exponential models, i.e. in both the left and the right panels of figure~\ref{fig:res_exponential_Planck_P21}.

Lastly, we tested the case of a \textit{pure} exponential EFT model where we allow negative values of the function $\Omega^{\rm EFT}$. This scenario corresponds to a negative running of the Planck mass. From Planck 2018 data we find 
\begin{align}
    \begin{split}
    \omegaEFT = -0.094^{+0.071}_{-0.042} \quad (68\%\,\,\text{CL}) , \\
        \alpha_M^0 = -0.133^{+0.13}_{-0.048} \quad (68\%\,\,\text{CL}) ,\\
        \beta = 1.28^{+0.56}_{-0.21} \quad (68\%\,\,\text{CL}).
    \end{split}
\end{align}
Compared with state-of-the-art constraints for the same data sets reported in section~\ref{sec:tested_model}, our results present slightly different features, although we check that they carry the same statistical significance. $\chi^2$ deviations with respect to the \lcdm constraints are comparable. We find $\Delta \chi^2 = \chi^2_{\rm \Lambda CDM} - \chi^2_{\rm EFT}= -4.4$ to be compared with $\Delta \chi^2 = -4.3$ found in ref.~\cite{planck:2018}. Moreover, the significance of the deviation from $\omegaEFT = 0$, being $1.3\sigma$, is the same for our results and results in literature. The disparities that we find could be caused by a different choice of viability conditions. E.g., for $\omegaEFT$ it may be that we allow to sample a larger portion of the parameter space resulting in different but statistically equivalent constraints. 

As presented in table~\ref{tab:EFT_lcdm_testo}, when we add $\pto$ we find
\begin{align}
    \begin{split}
    \omegaEFT = -0.075^{+0.063}_{-0.042} \quad (68\%\,\,\text{CL}), \\
        \alpha_M^0 = -0.117^{+0.12}_{-0.052} \quad (68\%\,\,\text{CL}), \\
        \beta = 1.41^{+0.48}_{-0.15} \quad (68\%\,\,\text{CL}),
    \end{split}
\end{align}
with a reduction on the errors at the level of $7\%$ on $\omegaEFT$ and $18\%$ on $\beta$. In the contour plots of figure~\ref{fig:res_exponential_Planck_P21} (right panel), we observe the full shape of the $\beta - \alpha_M^0$ confidence region. 

Compared with the results in the left panel, the sharp cut-off at $\beta\sim 1.6$ here disappears, given that we relax some of the viability conditions. This allows us to reconstruct the full shape of the posterior distribution for higher values of $\beta$. For lower values of $\beta$, instead, we observe the clustering of the samples along the $\omegaEFT=0$ axis, as for the case with stronger viability conditions shown in the left panel. This feature impairs the contour on $\omegaEFT$, pushing it towards the \lcdm limit. Knowing or fine-tuning $\beta$ a priori would remove these artificial effects, and it could significantly help in better constraining $\omegaEFT$. We highlight that for the time being we do not introduce weak lensing data, which should move the constraints to the \lcdm limit, as one can observe in ref.~\cite{planck:2018}.

The significance of the deviation from zero for the data sets Planck 2018 + $\pto$ is $1.2\sigma$, slightly lower than the value obtained with Planck data alone. From this, we may infer that the 21cm signal seems to bring back $\omegaEFT$ to its \lcdm limit. However, this feature is caused by the choice we made for the central points of the mock data set for $\pto$, that we generate assuming a \lcdm cosmology.

In order to test this statement, we conduct a further consistency check for our mock data set for the \textit{pure}  exponential EFT model with negative values of $\omegaEFT$. We consider a new mock data set with the same errors, but different central points. To generate such points we use the predicted $\pto$ power spectrum for an exponential EFT model described by the parameter $\omegaEFT=-0.178$, which is $2\sigma$ away from the mean value of the Planck constraint, taken from table~\ref{tab:EFT_lcdm_testo}. We set $\beta$ and all the other cosmological parameter to their Planck constraint. Results are shown in the right panel of figure~\ref{fig:res_exponential_Planck_P21}. For Planck 2018 + $\pto$ likelihood computed on this new data set, we find $\omegaEFT$ constrained to be
\begin{equation}
    \omegaEFT =-0.109^{+0.071}_{-0.041} \quad (68\%\,\,\text{CL})\, .
\end{equation}
While the error on $\omegaEFT$ remains unchanged, this estimate appears to be away from $0$ at the level of $1.5\sigma$, to be compared with the previous $1.3\sigma$ from Planck 2018 data. 
The contour in the $\beta - \alpha_M^0$ confidence region (figure~\ref{fig:res_exponential_Planck_P21}, right panel, cyan contour) is significantly reduced with respect to the previous case (blue contour). Assuming that one could place a prior or fix $\beta$ and remove the artificial clustering around the $\omegaEFT=0$ axis, the significance of the deviation from the \lcdm limit would increase remarkably. Thus, if $\pto$ data would prefer an EFT Universe, we should be able to detect and distinguish it from \lcdm. The constraints on the EFT parameters ($\omegaEFT,\, \beta$) appropriately decrease when using the EFT mock dataset, while the cosmological parameters ($\Omega_ch^2,\, H_0$) show only mild improvement, remaining equivalent to that of Planck constraints. Essentially, the $\pto$ mock modelled around $\Lambda$CDM, aids to bring the constraints closer to Planck constraints for all the parameters. While the $\pto$ modelled around the EFT, as expected, provides better agreement with the Planck constraints on the cosmological parameters and additionally shows improvement for the EFT parameters. 

In conclusion, we notice that the conformal coupling characteristic of GBD models, explored here with the \textit{pure} EFT approach, is sensitive to 21cm signal observations. However, the constraining power of a realistic mock data set is still limited by the capability of state-of-the-art instruments. When used without other probes like the CMB, the $\pto$ likelihood shows a mild constraining power. When combined with CMB measurements, results are dominated by the more comprehensive Planck 2018 data sets. In this framework, the realistic, MeerKAT-like, single bin $\pto$ likelihood improves the constraints on EFT parameters at the level of $\sim 10\%$. We expect that adding tomographic observations of $\pto$ for multiple redshift bins would improve the constraining power of this new observable, as we examine in section~\ref{sec:res_tomography}.

\subsection{Adding likelihood nuisance parameters}
\label{sec:res_nuisance}
\begin{figure}[tbp]
    \centering
    \includegraphics[width=0.49\textwidth]{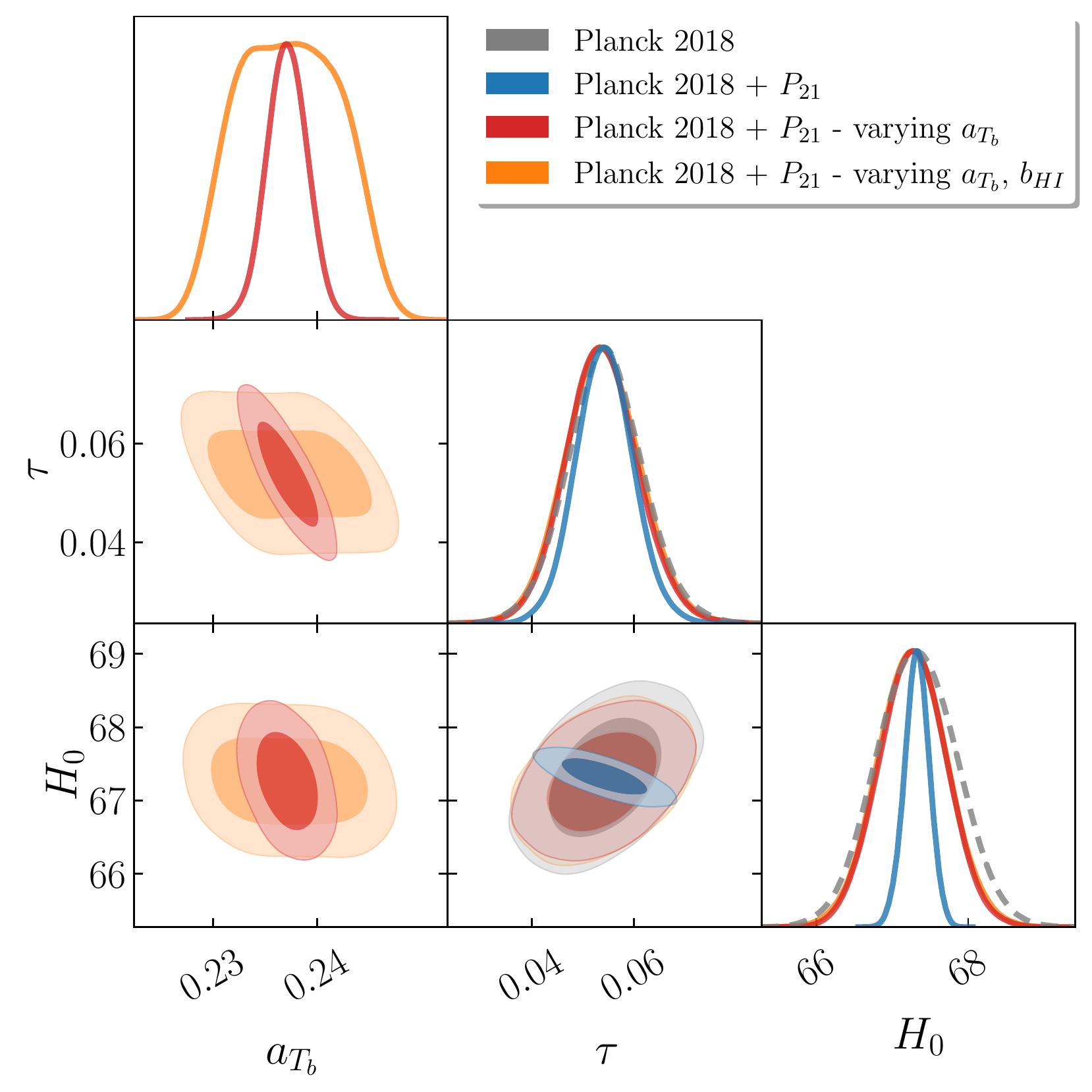}
    \includegraphics[width=0.49\textwidth]{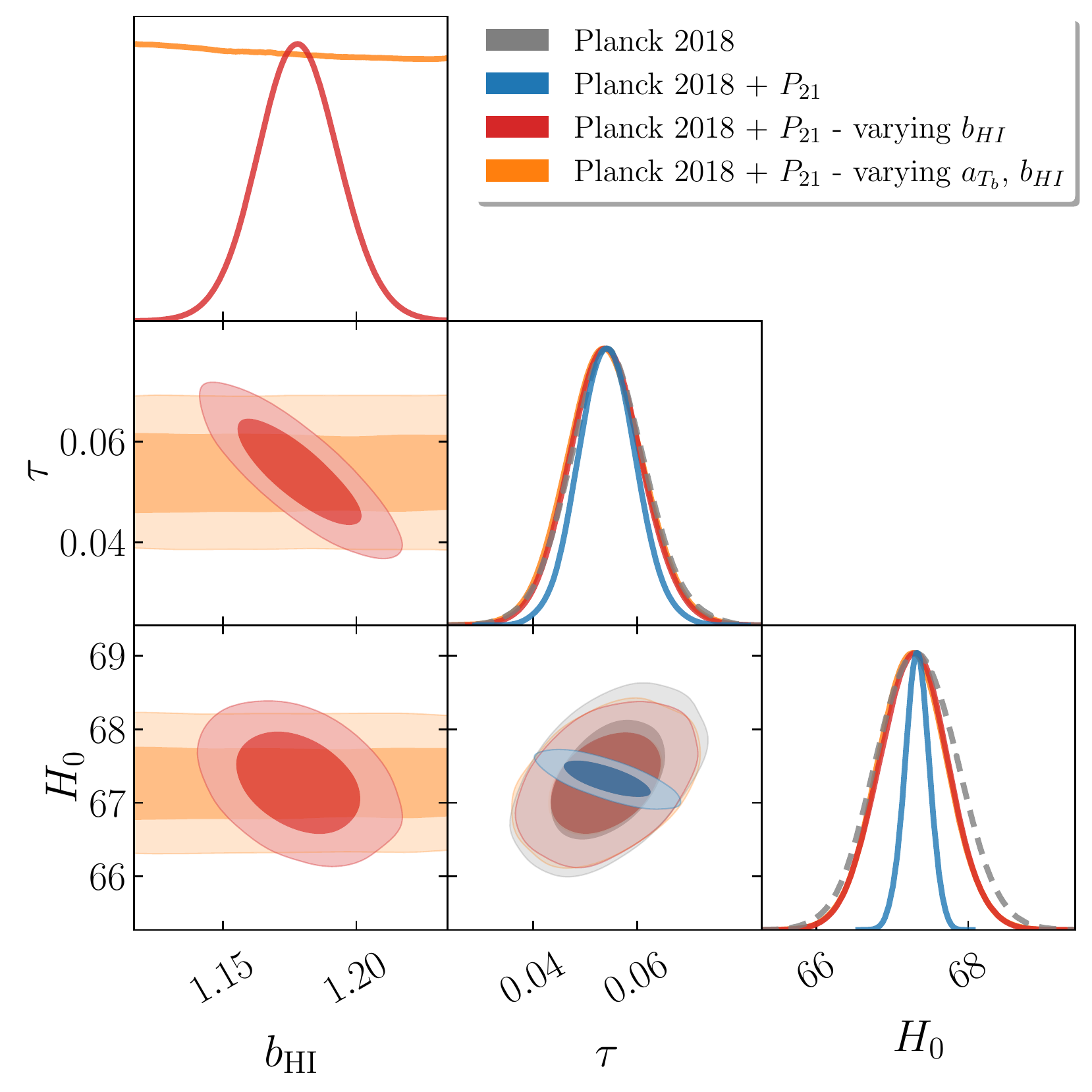}
    \caption{\label{fig:res_nuisance_par} Joint constraints (68\% and 95\% confidence regions) and marginalized posterior distributions on cosmological parameters for a $\Lambda$CDM model. We open the parameter space to the two nuisance parameters $a_{T_b}$ and $b_{HI}$, i.e. the amplitude of the brightness temperature and the HI bias. We study different configurations: both the parameters are fixed, one is free the other is fixed, they are both free. Here the label Planck 2018 stands for TT, TE, EE + lowE + lensing while the label $P_{21}$ stands for the 21cm power spectrum likelihood at redshift $z=0.39$. }
\end{figure}
When modeling $\pto$ we keep both the bias $b_{\rm HI}$ and the amplitude of the brightness temperature $a_{T_b}$ fixed to their \lcdm limit, lacking a theoretical prediction of how these two quantities might change for beyond \lcdm scenarios. See refs.~\cite{Hall:2012} for this type of computations. This assumption is accurate enough for our purposes since we expect to detect small variations from \lcdm.
                            
In this section we relax this assumption and implement the bias and the brightness temperature as nuisance likelihood parameters and test their effects on the parameter constraints. Note that varying these two quantities correspond to alter the amplitude of the power spectrum. 
We compare four different case studies, where: $i)$ we keep fixed both $a_{T_b}$ and $b_{\rm HI}$, $ii)$ we vary only $b_{\rm HI}$, $iii)$ we vary only $a_{T_b}$, $iv)$we vary both $a_{T_b}$ and $b_{\rm HI}$. Confidence regions and marginalized posteriors are shown in figure~\ref{fig:res_nuisance_par}, but we refer also to figure~\ref{fig:res_nuisance_EFT_2}, in appendix~\ref{app:nuisance_par}, and tables~\ref{tab:LCDM_$w$CDM}-\ref{tab:EFT_exp_neg}, in appendix~\ref{app:constr_tables}. We start studying the effect of nuisance parameters using Planck 2018 data + $\pto$ assuming a \lcdm universe.

We observe that adding separately $b_{\rm HI}$ or $a_{T_b}$ (case $ii)$ and $iii)$) produces comparable results. Red contours in figure~\ref{fig:res_nuisance_par} represent this scenario. We notice that, as expected, the addition of one nuisance parameter reduces the constraining power of $\pto$. With respect to Planck constraints, we observe a reduction of the error at the level of $\sim 15\%$, to be compared with the $\sim 65\%$ that we found with fixed $a_{T_b}$ and $b_{\rm HI}$.

When two nuisance parameters are varied (case iv)), we obtain a similar effect on cosmological parameters, while losing constraining power on $a_{T_b}$ and $b_{\rm HI}$. In figure~\ref{fig:res_nuisance_par} (orange contours), we see that the brightness temperature $a_{T_b}$ is loosely constrained while the bias $b_{\rm HI}$ remains completely unconstrained. 
In conclusion, although $b_{\rm HI}$ and $a_{T_b}$ are two very different physical quantities, they have similar effects on the parameter constraints. In the following, we fix the bias $b_{\rm HI}$ and leave the brightness temperature $a_{T_b}$ as nuisance parameter.
\begin{figure}[tbp]
    \centering
    \includegraphics[width=0.9\textwidth]{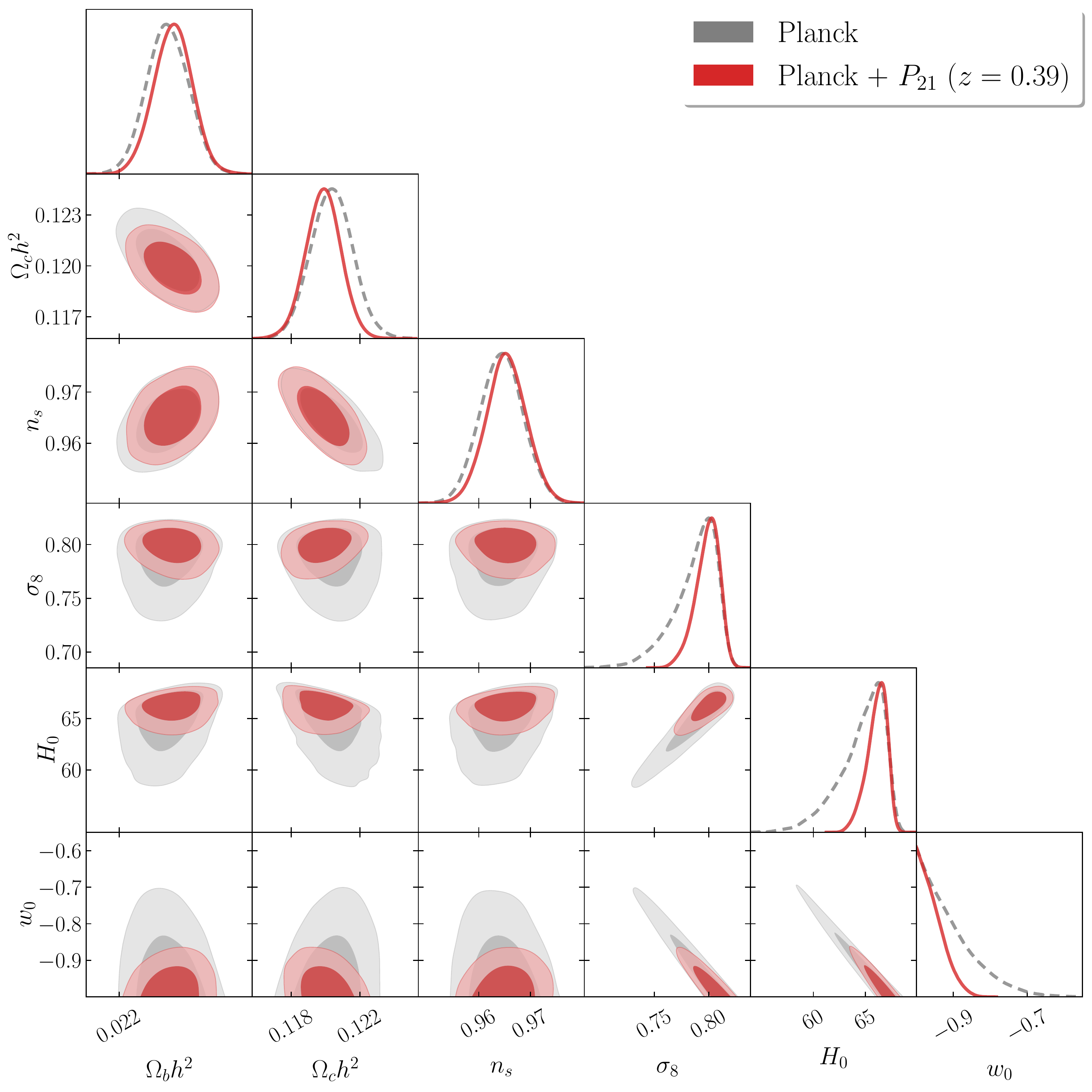}
    \caption{\label{fig:res_$w$CDM_Planck_P21}  Joint constraints (68\% and 95\% confidence regions) and marginalized posterior distributions on cosmological parameters for a $w$CDM model. Here the label Planck 2018 stands for TT, TE, EE + lowE + lensing while the label $P_{21} (z=0.39)$ stands for the 21cm power spectrum likelihood at redshift $z=0.39$. We vary the nuisance parameter $a_{T_b}$, i.e. the amplitude of the brightness temperature, keeping fixed the HI bias $b_{HI}$.}
\end{figure}

We have tested the effect of varying $a_{T_b}$ on EFT parameter constraints. We find that, as expected, adding a nuisance parameter reduces the constraining power on model parameters. When varying $a_{T_b}$, we observe it to be mildly correlated with EFT parameters (see appendix~\ref{app:nuisance_par}).

\subsection{Results for a $w$CDM background}
\label{sec:res_EFT_wCDM}
\begin{table}[tbp]
    \centering\small
    \begin{tabular}{lScSc} \hline\hline
    {\textsc{Parameter}~~~~~~~~~~~~~} & Planck 2018 & Planck 2018 + $P_{21}$ \\
$w$CDM \dotfill \\
$ w_0 $ \dotfill & $< -0.76^\dag$ & $< -0.89^\dag$ $(\mathbf{-53}\%)$ \\
$ H_0 $ \dotfill & $64.6^{+2.7}_{-1.1}$ & $66.1^{+1.2}_{-0.69}$ $(-50\%)$ \\
$ \sigma_8 $ \dotfill & $0.787^{+0.025}_{-0.011}$ & $0.799^{+0.013}_{-0.0085}$ $(-40\%)$ \\\hline
Linear EFT  \dotfill\\
$ w_0 $ \dotfill & $-0.925^{+0.020}_{-0.079}$ & $-0.960^{+0.015}_{-0.043}$ $(-41\%)$ \\
$ \omegaEFT$ \dotfill & $< 0.034^\dag$ & $< 0.034^\dag$   \\
$ H_0 $ \dotfill & $64.9^{+2.4}_{-0.97}$ & $66.1^{+1.2}_{-0.70}$ $(-44\%)$ \\\hline
Exponential EFT  \dotfill \\
$ w_0 $ \dotfill & $-0.911^{+0.028}_{-0.084}$ & $-0.953^{+0.018}_{-0.041}$ $(-47\%)$ \\
$ \Omega_0^{\rm EFT} $ \dotfill & $< 0.051^\dag$ & $< 0.047^\dag$ $(\mathbf{-7}\%)$ \\
$ \beta $ \dotfill & --- & $1.79^{+1.1}_{-0.47}$ $(-5\%)$ \\
    $ H_0 $ \dotfill & $64.6^{+2.5}_{-1.1}$ & $66.0^{+1.1}_{-0.73}$ $(\mathbf{-49}\%)$ \\
 \hline\hline
    \end{tabular}
    \caption{Marginalized constraints on cosmological parameters at the $68\%$ confidence level for \textit{pure} EFT models on a $w$CDM background. For $w$CDM the equation of state parameter is $w_0>-1$. $95\%$ confidence levels are marked with $^\dag$, while the symbol --- means that we do not find any constraint. Deviations in the error with respect to Planck 2018 results are shown in brackets. Here the label Planck 2018 stands for TT, TE, EE + lowE + lensing while the label $P_{21}$ stands for the $P_{21}$ likelihood at redshift $z=0.39$. Full constraints available in tables~\ref{tab:LCDM_$w$CDM}-\ref{tab:EFT_exp_neg}.}
    \label{tab:$w$CDM_testo}
\end{table}
\begin{figure}[tbp]
    \centering
    \includegraphics[width=0.49\textwidth]{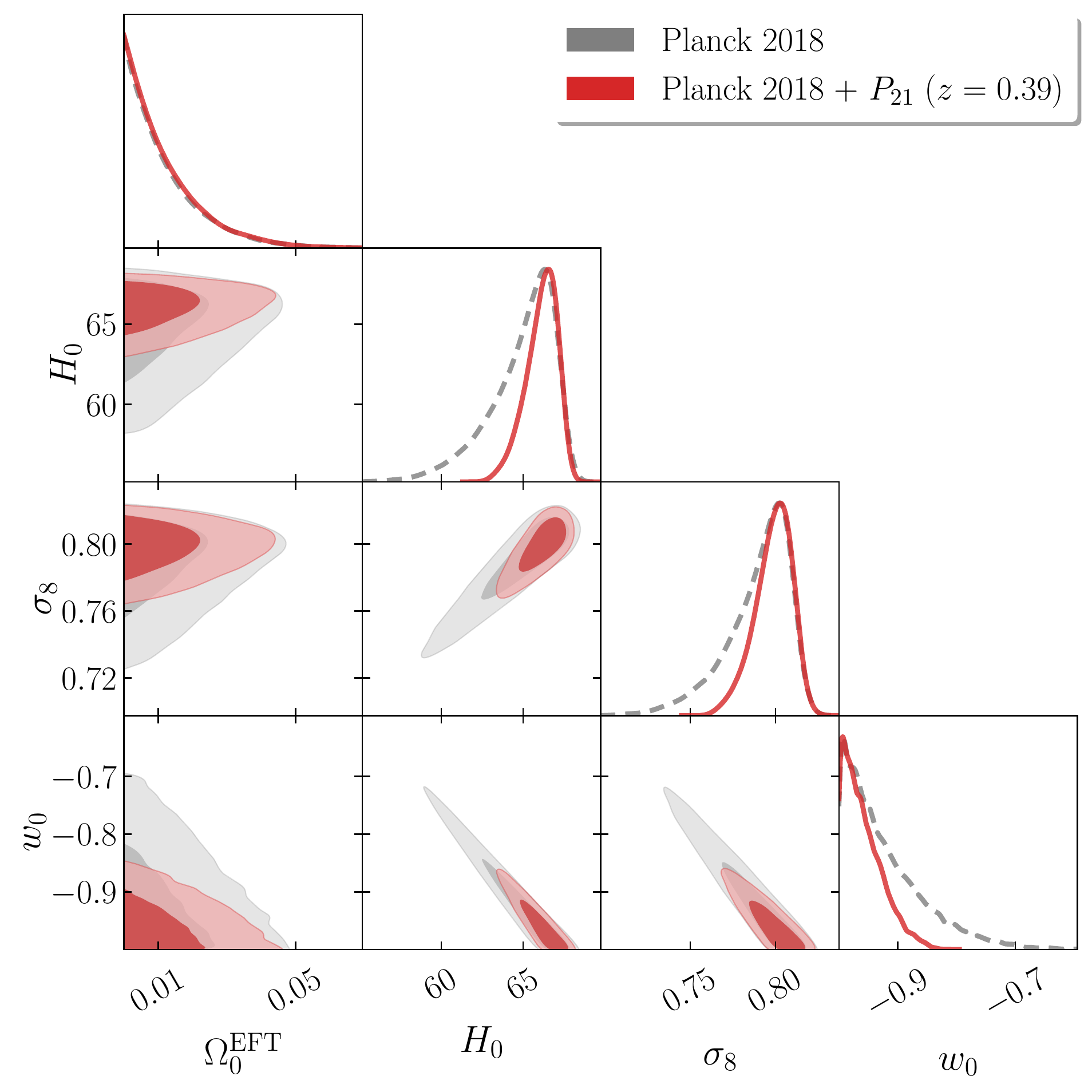}
    \includegraphics[width=0.49\textwidth]{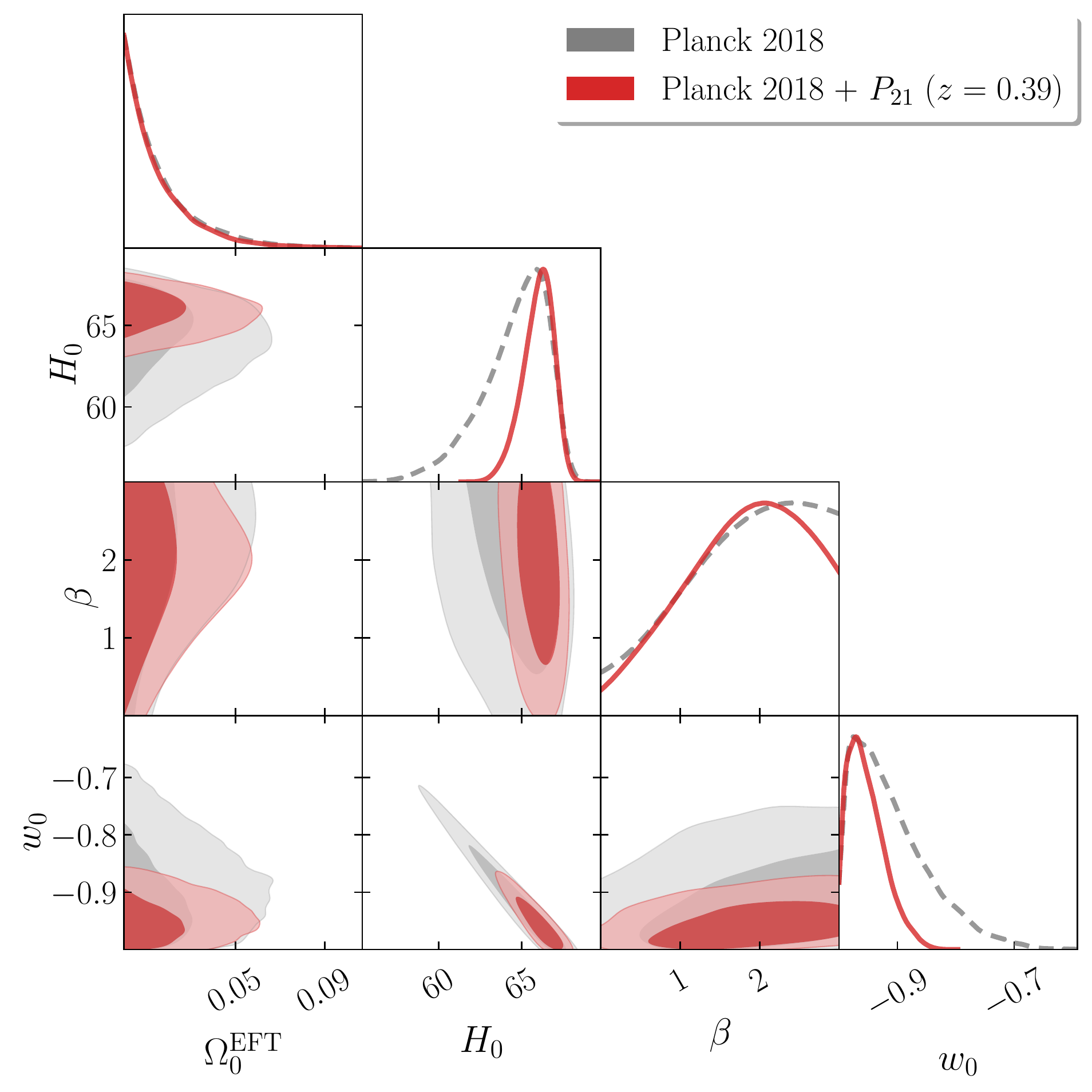}
    \caption{\label{fig:res_EFT_$w$CDM} Joint constraints (68\% and 95\% confidence regions) and marginalized posterior distributions on cosmological and EFT parameters for a \textit{pure}  linear EFT model on a $w$CDM background (left panel). We show also results for a \textit{pure} exponential EFT model (right panel). Here the label Planck 2018 stands for TT, TE, EE + lowE + lensing, while the label $P_{21} (z=0.39)$ stands for the 21cm power spectrum likelihood at redshift $z=0.39$. $f\sigma_8 + H + D_A$ represents the additional \textit{background} likelihood described in section~\ref{sec:background} and computed at the same redshift $z=0.39$. We vary the nuisance parameter $a_{T_b}$, i.e. the amplitude of the brightness temperature, keeping fixed the HI bias $b_{HI}$.}
\end{figure}
We now turn our attention to EFT models on a $w$CDM background. We would like to test the impact of the $\pto$ likelihood in a more complex framework, in which we open the parameter space to the DE equation of state parameter $w_0$. We highlight that throughout this analysis we always vary $a_{T_b}$, the amplitude of the brightness temperature along with the other cosmological parameters, as discussed in~\ref{sec:res_nuisance}. Using $\pto$ alone and with background data we find broad constraints, comparable with the ones described above in section~\ref{sec:res_EFT_lcdm}. In the following we focus on constraints obtained combining the $\pto$ likelihood with CMB. We refer to appendix~\ref{app:constr_tables} for the complete set of constraints. 

First, we test the effect of $\pto$ for a simple $w$CDM model, i.e. we only add a constant $w_0$ to the parameter space. 
As shown in figure~\ref{fig:res_$w$CDM_Planck_P21} and table~\ref{tab:$w$CDM_testo}, $\pto$ data  significantly reduces the error on $w_0$ and consequently on $H_0$ and $\sigma_8$. E.g., on $w_0$ we obtain an improvement of $53\%$ on the upper limit found from Planck data alone. Given that $w_0$ is degenerate with both $H_0$ and $\sigma_8$, this translates into a reduction on the errors on these parameters at the level of $50\%$ and $40\%$ respectively. Adding $\pto$ reduces the degeneracy between $w_0$ and the other parameters. Notice that, the improvement on cosmological parameter errors is compatible with the results we obtained for \lcdm with $a_{T_b}$ as nuisance.

We show results for EFT models in figure~\ref{fig:res_EFT_$w$CDM} and table~\ref{tab:$w$CDM_testo}. We test the linear and the \textit{pure} exponential EFT models. Overall, the improvement brought by $\pto$ on $w_0$ and derived parameters is compatible with results for the vanilla $w$CDM case. For the linear model, we obtain the upper limit:
\begin{align}
    \begin{split}
    \omegaEFT < 0.034 \quad (95\%\,\,\text{CL}) ,
    \end{split}
\end{align}
from both Planck 2018 data with or without the addition of $\pto$ Note that this constraint significantly improves over previous results in the literature($\omegaEFT < 0.058$), computed form Planck 2013~\cite{Planck:2013} and WMAP~\cite{WMAP:2012} data sets and without polarization data (see section~\ref{sec:tested_model}). Although adding $\pto$ does not impact on the upper limit on $\omegaEFT$, in the contours (left panel of figure~\ref{fig:res_EFT_$w$CDM}) we observe that it reduces the correlation between $\omegaEFT$ and other parameters.
For the exponential model, instead, adding $\pto$ reduces the errors on EFT parameters at the level of $5-7\%$, as presented in table~\ref{tab:$w$CDM_testo}. For Planck 2018 + $\pto$ we find:
\begin{align}
    \begin{split}
    \omegaEFT < 0.047 \quad (95\%\,\,\text{CL}) , \\
    \beta = 1.79^{+1.1}_{-0.47}\quad (68\%\,\,\text{CL}) .
    \end{split}
\end{align}
In the 2D contours (right panel of figure~\ref{fig:res_EFT_$w$CDM}) again we notice that $\pto$ helps removing the correlation between EFT and cosmological parameters. We highlight that the confidence regions in the plane $\omegaEFT-\beta$ here do not show the cut-off for $\beta\sim 1.6$, that we found for the same model on a \lcdm background. This means that when we add $w_0$ to the parameter space, a wider region is allowed by viability conditions and we are able to reconstruct the contour for higher values of $\beta$. Note that viability conditions are responsible also for the skewed elongated posteriors that we observe in figure~\ref{fig:res_EFT_$w$CDM}. As before (see figure~\ref{fig:res_exponential_Planck_P21}), the samples are clustered along the $\omegaEFT=0$, due to the modeling we consider.

On a $w$CDM background evolution the major impact of the $\pto$ likelihood is to reduce the estimate on the equation of state parameter $w_0$, and consequently on $H_0$ and $\sigma_8$. For EFT models, $\pto$ produces results similar to what we found for the \lcdm background. I.e., the $\pto$ likelihood has mild constraining power on the EFT parameters and it helps in reducing the degeneracy with the other model parameters.

\subsection{Tomography}
\label{sec:res_tomography}
In the results above we explored the constraining power of a realistic $\pto$ mock data set at the redshift $z=0.39$, with conservative error bars. In this section, instead, we would like to test a more ideal case, by exploiting the tomographic nature of 21cm intensity mapping observations. To this end, we construct new mock data sets with multiple redshifts and we investigate how this improves on the constraints for beyond \lcdm models. 
\begin{table}[tbp]
    \centering\footnotesize
    \begin{tabular}{lScScSc} \hline\hline
     & Planck 2018  & Planck 2018 & Planck 2018\\
 {\textsc{Parameter}}& + $P_{21}^{\rm EFT}(z=0.39)$ & + $P_{21}^{\rm EFT}$(bins) & + $P_{21}^{\rm EFT}$(bins) - halved errors\\\hline
$\Omega_ch^2$\dotfill  & $0.1194\pm 0.0011$ $(-22\%)$ & $0.12042\pm 0.00080$ $(-43\%)$ & $0.12046\pm 0.00055$ $(-61\%)$   \\
 $ \Omega_0^{\rm EFT} $ \dotfill & $-0.086^{+0.064}_{-0.038}$ $(-10\%)$ & $-0.079^{+0.047}_{-0.036}$ $(\mathbf{-26\%})$& $-0.103^{+0.042}_{-0.032}$ $(\mathbf{-35\%})$ \\
$ \beta $ \dotfill & $1.28^{+0.58}_{-0.22}$ $(+4\%)$ & $1.08^{+0.42}_{-0.25}$ $(\mathbf{-13\%})$& $1.06^{+0.30}_{-0.19}$  $(\mathbf{-36\%})$  \\
$ H_0 $ \dotfill & $67.63\pm 0.50$$(-24\%)$ & $67.15\pm 0.36$ $(-46\%)$ & $67.13\pm 0.24$ $(-65\%)$ \\
 \hline\hline
 \\
 \hline\hline
      & Planck 2018  & Planck 2018 & Planck 2018\\
{\textsc{Parameter}} & + $P_{21}^{\Lambda\rm CDM}$ $(z=0.39)$ & + $P_{21}^{\Lambda\rm CDM}$(bins)& + $P_{21}^{\Lambda\rm CDM}$(bins)- halved errors\\\hline
$ \Omega_c h^2 $ \dotfill & $0.1194\pm 0.0011$ $(-22\%)$ & $0.11957\pm 0.00082$ $(-41\%)$ & $0.11997\pm 0.00055$  $(-61\%)$\\
$ \Omega_0^{\rm EFT} $ \dotfill &  $-0.086^{+0.068}_{-0.039}$ $(-10\%)$ & $-0.066^{+0.055}_{-0.031}$ $(-\mathbf{24}\%)$ & $-0.047^{+0.047}_{-0.027}$ $(-\mathbf{35}\%)$ \\
$ \beta $ \dotfill  & $1.28^{+0.58}_{-0.22}$ $(+4\%)$ & $1.18^{+0.57}_{-0.26}$$(-\mathbf{8}\%)$ & $1.26^{+0.55}_{-0.18}$  $(-\mathbf{5}\%)$  \\
$ H_0 $ \dotfill  & $67.63\pm 0.51$ $(-23\%)$ & $67.54\pm 0.37$ $(-44\%)$ & $67.36\pm 0.24$ $(-65\%)$\   \\
 \hline\hline
 \\
 \hline\hline
{\textsc{Parameter}} &$P_{21}^{\Lambda\rm CDM}$(bins) & $P_{21}^{\rm EFT}$(bins) &$P_{21}^{\rm EFT}$(bins) - halved errors \\\hline
 $ \Omega_0^{\rm EFT} $ \dotfill & $0.053^{+0.075}_{-0.17}$ & $-0.14^{+0.13}_{-0.10}$   & $-0.131^{+0.074}_{-0.045}$  \\
$ \beta $ \dotfill & $1.26^{+0.55}_{-0.30}$ &  $1.10^{+0.49}_{-0.29}$    & $0.94^{+0.48}_{-0.32}$ \\
$ H_0 $ \dotfill &  $74.1^{+8.1}_{-11}$ & $70\pm 9$ & $67.32^{+0.77}_{-1.6}$  \\
 \hline\hline
    \end{tabular}
    \caption{Marginalized constraints on cosmological parameters at the $68\%$ confidence level. We test a \textit{pure} exponential EFT model on a \lcdm background, with negative running of the Planck mass. We test the effects of tomography by means of two new 21cm power spectrum mock data sets: $P_{21}^{\rm EFT}$, constructed around an EFT theory described by $\omegaEFT=-0.178$ and $\beta=1.28$, and $P_{21}^{\Lambda\rm CDM}$, constructed around our fiducial cosmology. We compare the constraining power of a single bin $(z=0.39)$ with the five bins at $z=0,\,0.39,\,0.53,\,0.67,\,2.5$ combined (bins) and this same set with halved error bars. We use the 21cm power spectrum likelihood combined with Planck observations. The label Planck 2018 stands for TT, TE, EE + lowE + lensing. Deviations in the error with respect to Planck 2018 results are shown in brackets.
    }
    \label{tab:tomography}
\end{table}

We add to the mock data at $z=0.39$ four new mock $\pto$ measurements at redshifts $z = 0,\,0.53,\,0.67,\,2.5$, that we add to the one at $z=0.39$ in the likelihood. We consider observations in different redshift bins to be independent. The errors for the bins $z=0$, $z=0.53$ and $z=0.67$ forecast plausible MeerKAT-like observations and they are constructed with the same technique of section~\ref{sec:mock_data}. For the bin at $z=2.5$ we assume to have an observation of $\pto$ from the extended HIRAX experiment, as in ref.~\cite{Obuljen:2017}. For this bin we use the errorbars adopted for the data set at redshift $z=0.67$. For all bins, we consider also a more optimistic case, i.e. we halve the errors on each point. 
\begin{figure}[tbp]
    \centering
    \includegraphics[width=0.45\textwidth]{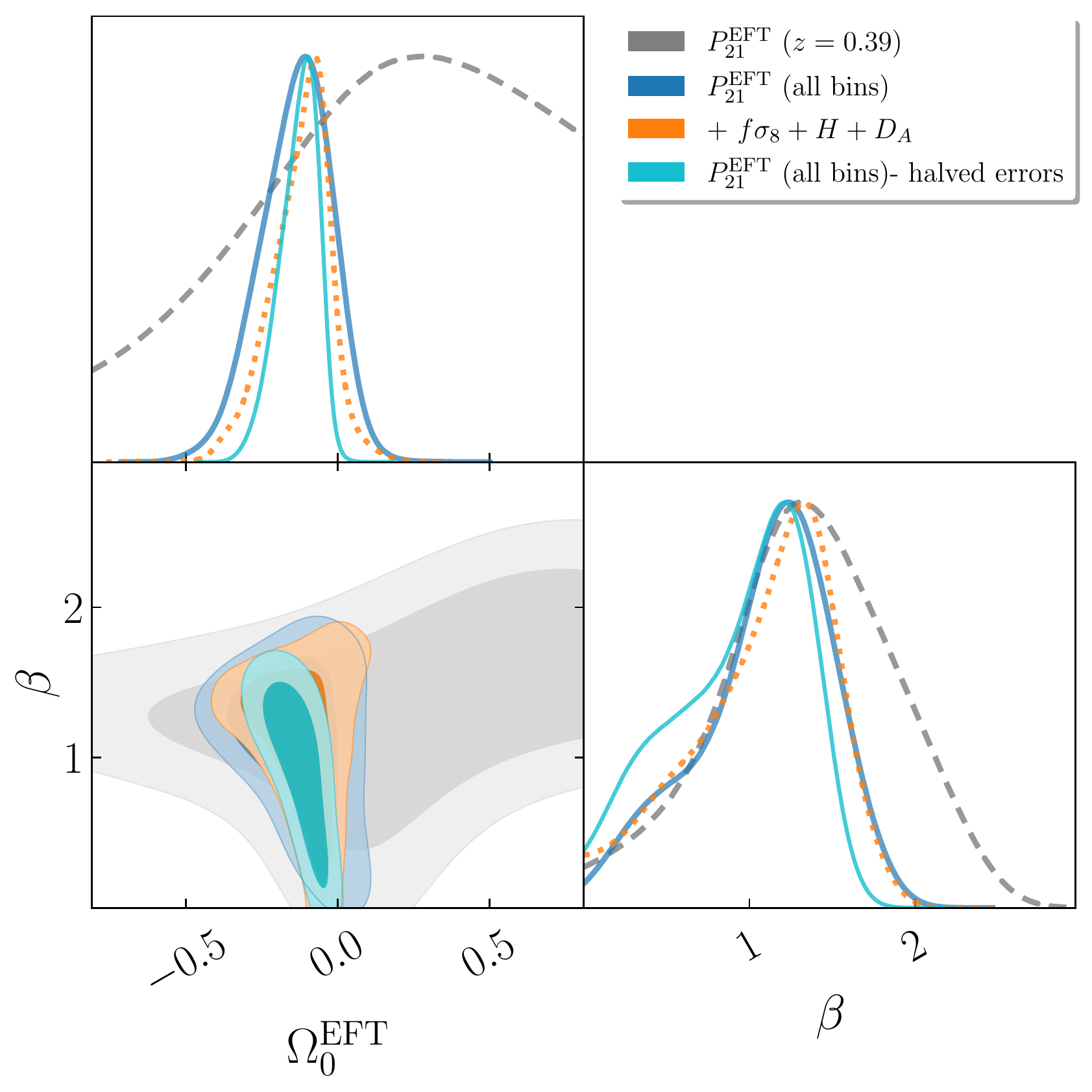}
    \includegraphics[width=0.45\textwidth]{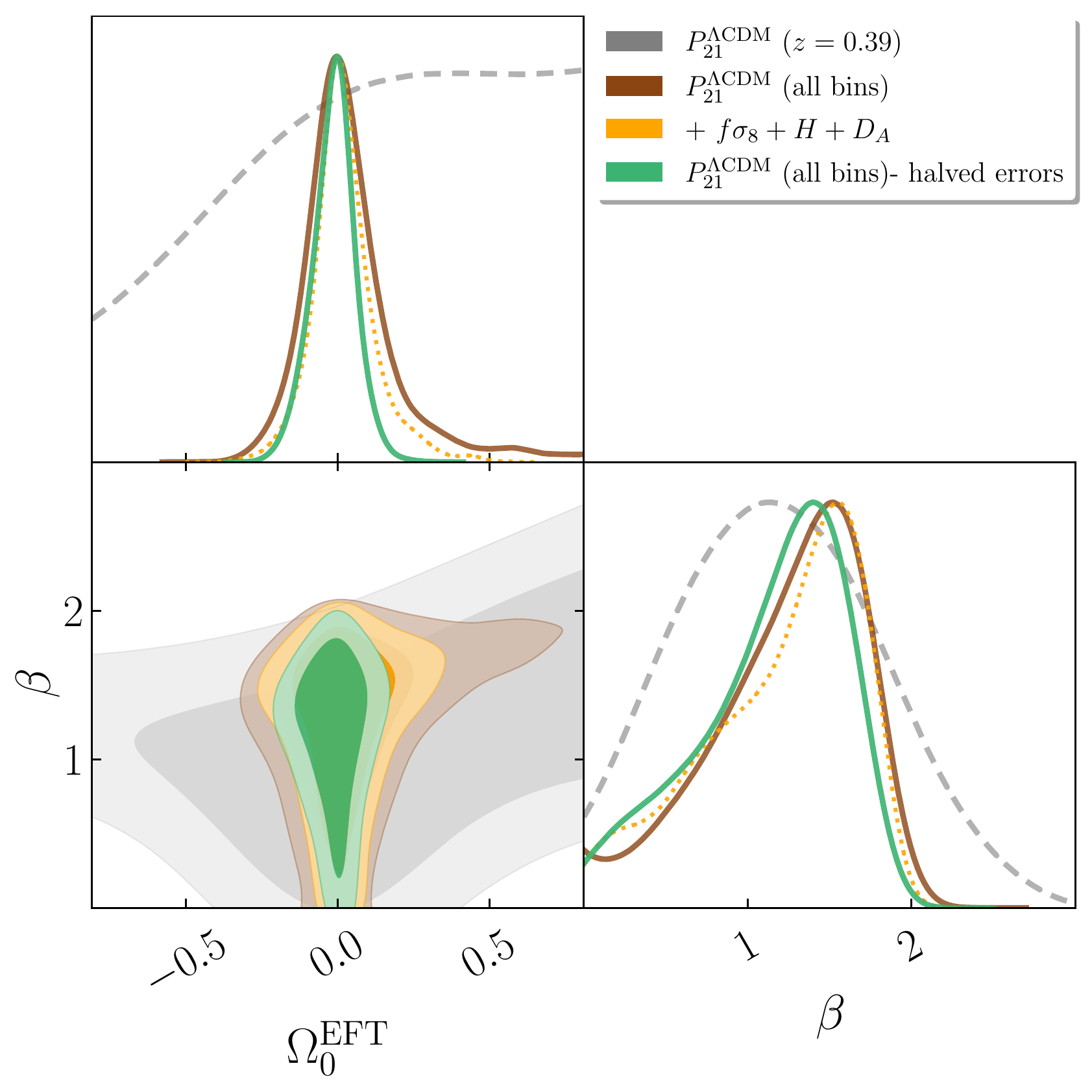}
    \caption{\label{fig:res_tomography_P21}Joint constraints (68\% and 95\% confidence regions) and marginalized posterior distributions on cosmological and EFT parameters. We investigate a \textit{pure} exponential EFT model on a \lcdm background, with negative running of the Planck mass. We test the effects of tomography by means of two new 21cm power spectrum mock data sets: $P_{21}^{\rm EFT}$, constructed around an EFT theory described by $\omegaEFT=-0.178$ and $\beta=1.28$ (left panel), and $P_{21}^{\Lambda\rm CDM}$, constructed around our fiducial cosmology (right panel). We compare the constraining power of a single bin $(z=0.39)$ with the five bins at $z=0,\,0.39,\,0.53,\,0.67,\,2.5$ combined (all bins) and this same set with halved error bars. The label $f\sigma_8 + H + D_A$ represents the additional \textit{background} likelihood described in section~\ref{sec:background}. We vary the nuisance parameter $a_{T_b}$, i.e. the amplitude of the brightness temperature, keeping fixed the HI bias $b_{HI}$.
    }
    \centering
    \includegraphics[width=0.49\textwidth]{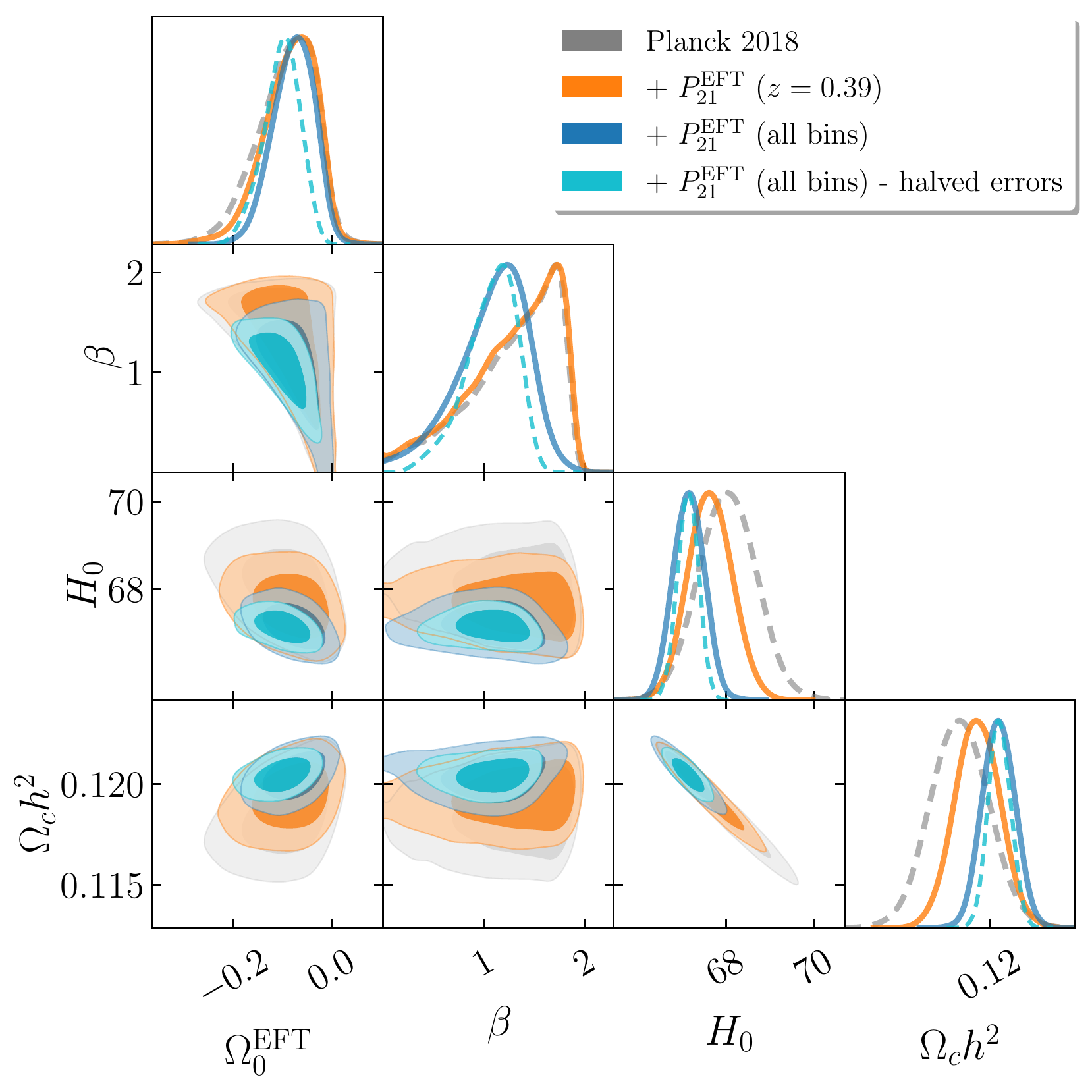}
    \includegraphics[width=0.49\textwidth]{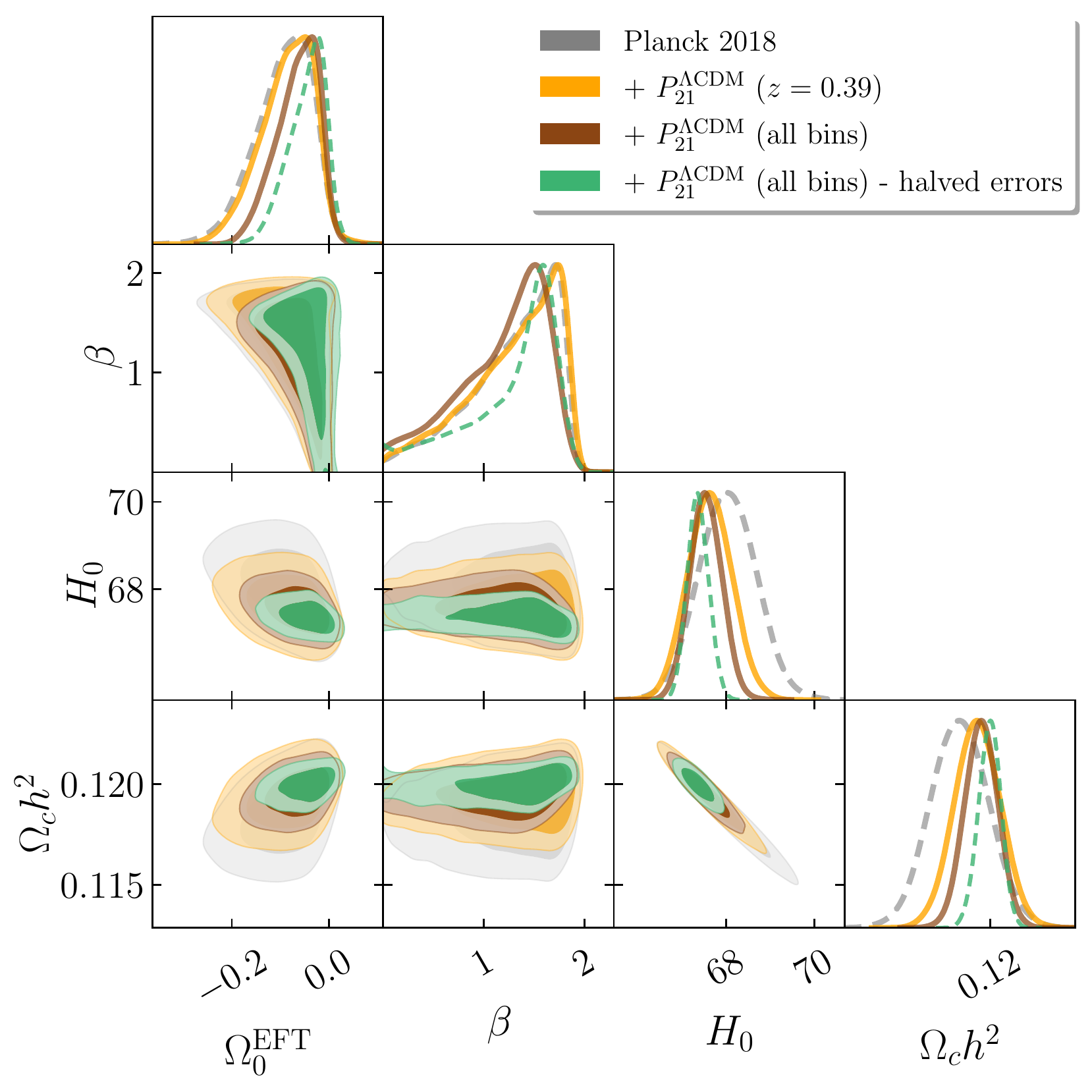}
    \caption{\label{fig:res_tomography_P21_2}Joint constraints (68\% and 95\% confidence regions) and marginalized posterior distributions on cosmological and EFT parameters. We investigate a \textit{pure} exponential EFT model on a \lcdm background, with negative running of the Planck mass. We test the effects of tomography by means of two new 21cm power spectrum mock data sets: $P_{21}^{\rm EFT}$, constructed around an EFT theory described by $\omegaEFT=-0.178$ and $\beta=1.28$ (left panel), and $P_{21}^{\Lambda\rm CDM}$, constructed around our fiducial cosmology (right panel). We compare the constraining power of a single bin $(z=0.39)$ with the five bins at $z=0,\,0.39,\,0.53,\,0.67,\,2.5$ combined (all bins) and this same set with halved error bars. We use the 21cm power spectrum likelihood combined with Planck observations. The label Planck 2018 stands for TT, TE, EE + lowE + lensing. We vary the nuisance parameter $a_{T_b}$, i.e. the amplitude of the brightness temperature, keeping fixed the HI bias $b_{HI}$.
    }
\end{figure}

Central points in the data sets are generated from theory predictions and then randomly displaced, as we did for the data set at $z=0.39$ in section~\ref{sec:mock_data}. For all the five bins, we construct two data sets: $P_{21}^{\Lambda\rm CDM}$ and $P_{21}^{\rm EFT}$. For $P_{21}^{\Lambda\rm CDM}$, the theory we use to generate central points is our \lcdm fiducial cosmology. For $P_{21}^{\rm EFT}$, instead, we consider an exponential \textit{pure} EFT model, in which $\omegaEFT$ is $2\sigma$ away form the Planck constraints, as in section~\ref{sec:res_EFT_lcdm} and figure~\ref{fig:res_exponential_Planck_P21}. I.e., we fix the EFT parameters to $\beta=1.28$ and $\omegaEFT = -0.178$, while we leave the cosmological parameters to their fiducial value.

\begin{figure}[tbp]
    \centering
    \includegraphics[width=0.9\textwidth]{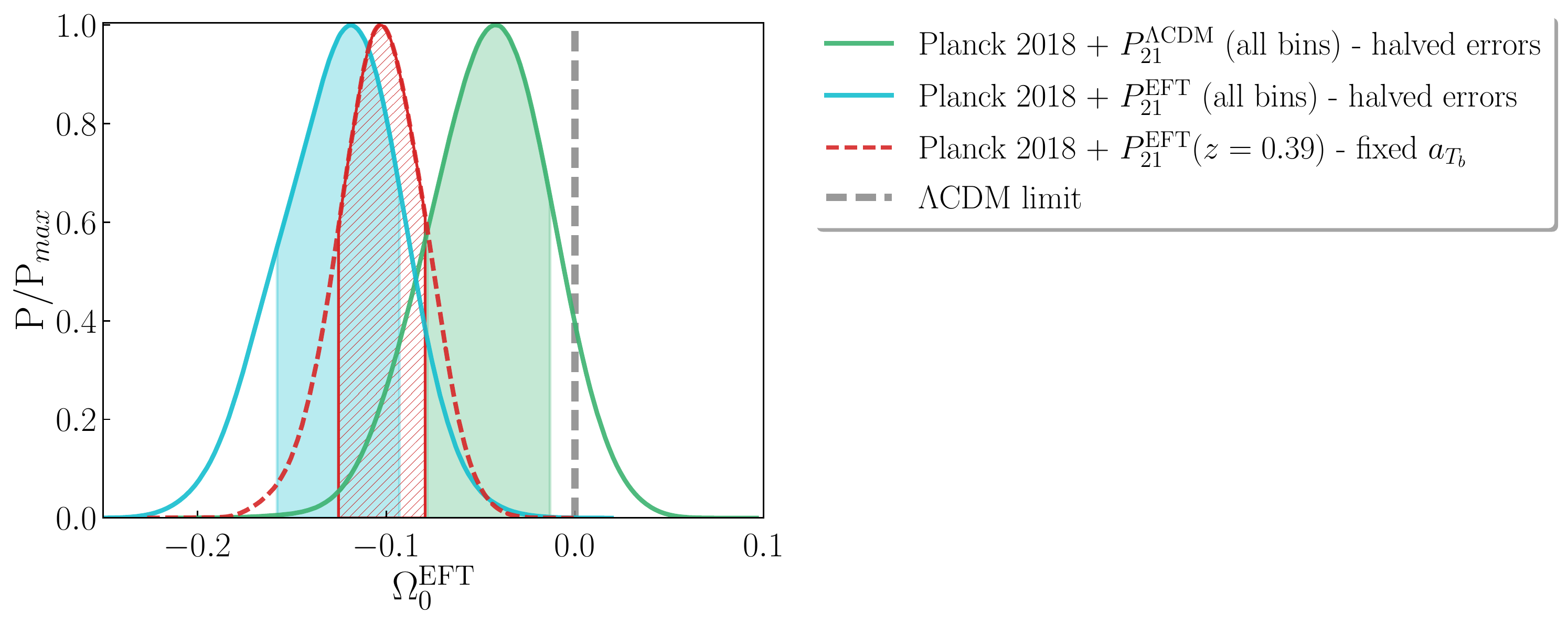}
    \caption{\label{fig:res_tomography_fixed_beta}Marginalized posterior distributions for the EFT parameter $\omegaEFT$, describing a \textit{pure} exponential EFT model on a \lcdm background, with negative running of the Planck mass. We test the effect of adding a stronger hard prior on the parameter $\beta$, which we choose to be $\beta \in [1.2,1.4]$. We consider tomographic data set with five bins and half errors and compare results from $P_{21}^{\rm EFT}$, constructed around an EFT theory described by $\omegaEFT=-0.178$ and $\beta=1.28$ (blue line), and $P_{21}^{\Lambda\rm CDM}$, constructed around our fiducial cosmology (green line). We add also the result for the single bin and with no nuisance parameters. The gray dashed line mark the \lcdm limit for $\omegaEFT$, while the shaded regions correspond to the $1\sigma$ confidence limit. We use the 21cm power spectrum likelihood combined with Planck observations. The label Planck 2018 stands for TT, TE, EE + lowE + lensing.
    }
\end{figure}

We compute constraints from both the $P_{21}^{\Lambda\rm CDM}$ and the $P_{21}^{\rm EFT}$ likelihood, which we use alone and combined with Planck 2018 data. We compare results from the realistic single bin at $z=0.39$ with ideal tomographic data set constructed with all five bins combined and with halved errors. We test a \textit{pure} exponential EFT model with negative running of the Planck mass. The analysis conducted here is similar to the one of section~\ref{sec:res_EFT_lcdm} and figure~\ref{fig:res_exponential_Planck_P21}. However, here we vary the amplitude of the brightness temperature $a_{T_b}$ as nuisance, while in figure~\ref{fig:res_exponential_Planck_P21} $a_{T_b}$ is kept fixed. Results are shown in figures~\ref{fig:res_tomography_P21},~\ref{fig:res_tomography_P21_2},~\ref{fig:res_tomography_fixed_beta} and table~\ref{tab:tomography}.

Most of the constraining power comes from the redshift bins at $z=0,\, 0.39,\, 2.5$. This is because the mock data at the intermediate redshifts are constructed assuming a MeerKAT-like single dish experiment with large error bars due to the $\lambda/D$ scaling of the primary beam.
In figure~\ref{fig:res_tomography_P21}, the constraints from the $\pto$ likelihood alone are shown for both $P_{21}^{\rm EFT}$ (left panel) and $P_{21}^{\Lambda\rm CDM}$ (right panel). We observe that the largest improvement in the EFT constraints for the $\pto$ likelihood is coming from tomography alone, while the inclusion of the \textit{background} (orange dotted lines) has no significant impact on the EFT parameters posteriors. Compared with the very loose constraint from the single bin (gray dashed lines), with tomography the $\omegaEFT$ and $\beta$ parameters are found to be more tightly constrained. The effect is more significant for $\omegaEFT$ and it is maximized when we halve the error bars. 
Observe that tomography influences also the cosmological parameters. E.g., estimates of $H_0$ and $\Omega_ch^2$ from all bins with halved errors are of the same order of the results found above for the single bin combined with Planck.

In figure~\ref{fig:res_tomography_P21_2} constraints from tomographic measures of $\pto$ combined with Planck are presented. As above, compared with the single bin, tomography significantly reduces the errors on both EFT and cosmological parameters (see table~\ref{tab:tomography}). With tomography, deviations in the errors form Planck alone results for $\omegaEFT$ and $\beta$ increase up to the level of $-35\%$, to be compared with the less than $-10\%$ found with the single bin only. The effect is even stronger for cosmological parameters. For $H_0$ and $\Omega_ch^2$ from Planck combined with tomography with halved error bars, we observe a reduction of the error with respect to Planck at the level of $\sim 65\%$. These improvements are comparable with the results from the analysis of sections~\ref{sec:res_lcdm} and~\ref{sec:res_EFT_lcdm}, with no nuisance parameters, that, instead, are varied here. Looking at the 2D contours of figure~\ref{fig:res_tomography_P21_2}, we recognize the characteristic elongated shape in the plane $\beta-\omegaEFT$, that we discussed in section~\ref{sec:res_EFT_lcdm}. Here, we find a much broader confidence region, due to the fact that we vary the nuisance parameter $a_{T_b}$. Adding tomography further reduces the correlation between $H_0$ and $\Omega_ch^2$, resulting in the tight constraints described above. 

Lastly, we observe that constraints from $P_{21}^{\rm EFT}$ (left panels) and $P_{21}^{\Lambda\rm CDM}$ (right panels) are consistent. The two mock data sets produce similar errors with different mean values, as one would expect. By comparing results from the $P_{21}^{\rm EFT}$ and $P_{21}^{\Lambda\rm CDM}$ data sets, we assess if with the $\pto$ likelihood an EFT universe would be distinguishable from a \lcdm one. We recall that the \lcdm limit for the considered EFT model is $\omegaEFT=0$, while $\beta$ is allowed to assume any value. This gives rise to the clustering of the samples along $\omegaEFT=0$ in the $\omegaEFT-\beta$ confidence regions, which, in turn, drives all the contours to be compatible. However, we expect that placing a stronger prior on $\beta$ could remove this effect, as discussed in section~\ref{sec:res_EFT_lcdm}. We test this hypothesis by placing a narrow hard prior on $\beta$, i.e. $\beta \in [1.2,1.4]$, and looking at the modifications in the $\omegaEFT$ marginalized posterior. Results are shown in figure~\ref{fig:res_tomography_fixed_beta}. The significance of the deviation from the \lcdm limit (gray dashed line), is $1.4\sigma$ for $P_{21}^{\Lambda\rm CDM}$ (green line) while it is $3.8\sigma$ for $P_{21}^{\rm EFT}$ (cyan line), where we considered the tomographic data sets with halved errors. The error on $\omegaEFT$ is now the same for both the data sets and the two posteriors are $2.4\sigma$ away from each other. For the single bin $P_{21}^{\rm EFT} (z=0.39)$ with no nuisance parameter (red dashed line), that we presented in figure~\ref{fig:res_exponential_Planck_P21}, the deviation from $\omegaEFT=0$ is even more significant, being $4.5\sigma$. This implies that the constraints on the EFT models obtained using realistic $\pto$ data could greatly benefit from knowing the amplitude of the brightness temperature a priori, even in comparison to a more optimistic (halved errors) tomographic data set. This, in turn, reasserts the utility of the future, tomographic $\pto$ data in detecting beyond \lcdm models.

\section{Summary and conclusions}
\label{sec:res_summary}
In this work we construct a new likelihood function to compute constraints from observations of the 21cm signal power spectrum, measured through intensity mapping techniques. We extend the public available codes \texttt{EFTCAMB}/\texttt{EFTCosmoMC}~\cite{hu:2014,Raveri:2014,num_notes_v3_EFTCAMB} by implementing a new likelihood module fully integrated with the original codes.  

As described in sections~\ref{sec:P21_model} and~\ref{sec:likelihood_implementation}, we model the power spectrum of the 21cm signal $P_{21}(z,k)$ as in refs.~\cite{kaiser1987,Villaescusa-Navarro:2018,Bacon:2018}. Then, we build a realistic mock data set of forecast $P_{21}(z,k)$ measures at redshift $z=0.39$. We model such data set mimicking the MeerKAT~\cite{Santos:2015,Santos:2017,Wang:2021} telescope observations, for which real data will be available in the near future.

We complement our mock $\pto$ with an additional likelihood function to include external 21cm intensity mapping forecasts on background quantities. We use results in ref.~\cite{Obuljen:2017}, where forecasts on $f\sigma_8$, $H$ and $D_A$ are given for an hypothetical HIRAX measure in the redshift range $2.5 <z <5$, as outlined in section~\ref{sec:background}. We refer to this likelihood as \textit{background} likelihood. We further expand the $\pto$ likelihood to include observations in multiple redshift bins.

We test the effect of the $\pto$ likelihood first in a \lcdm framework, then on several beyond \lcdm models that we study within the Effective Field Theory (EFT) formalism. Exploiting the strength of \texttt{EFTCAMB}/\texttt{EFTCosmoMC}, we analyze \textit{pure} EFT models on a \lcdm and a $w$CDM background. We consider two different parametrizations of the $\Omega^{\rm EFT}(a)$ function, i.e. a linear and an exponential evolution in time.

We conduct a MCMC statistical analysis to obtain marginalized confidence levels and regions on cosmological and EFT parameters. Constraints are computed from both $\pto$ data set alone and combined with Planck 2018 CMB observations~\cite{planck:2018}. We test also the effect of adding to $\pto$ alone the \textit{background} likelihood. 

Our main goal is to forecast the state-of-the-art constraining power of the $\pto$ observable on beyond \lcdm theories. This is why we mainly focus on the results obtained with the realistic mock data set at redshift $z=0.39$. However, we believe that the tomographic nature of 21cm signal intensity mapping could be pivotal to constrain DE. Thus, to verify this claim, we complete our analysis by testing the effects of tomography and constructing an ideal mock data set of observations at five different redshifts bins.

As discussed in section~\ref{sec:results}, we find that:
\begin{itemize}
    \item the mock data set $\pto$ at $z=0.39$ alone displays a mild constraining power on the cosmological parameters, which is remarkably improved when we add the \textit{background} data set, by forecast intensity mapping observations at intermediate redshift;
    \item in a \lcdm scenario adding $\pto$ to Planck 2018 CMB data significantly improve the constraints on $\Omega_c h^2$ and $H_0$, at the level of $61\%$ and $69\%$ respectively, for fixed nuisance parameters;
    \item for \textit{pure} EFT models we observe a mild reduction on the constraints when combining CMB data with $\pto$, in the range of $1-18\%$; 
    \item state-of-the-art forecast data for $\pto$ alone are still not able to significantly constrain \textit{pure} EFT models. Forecasts in other redshift bins are expected to improve on the constraining power of $\pto$ on such models;
    \item for the \textit{pure} linear EFT model on a \lcdm background we find $\omegaEFT$ constrained to be $\omegaEFT < 0.031$ ($95\%$ CL) for Planck 2018 + $\pto$ data. This constraint improves on previous results in literature~\cite{Raveri:2014,planck:2015de-mg};
    \item for the \textit{pure} exponential EFT model on a \lcdm background for Planck 2018 + $\pto$ we obtain $\omegaEFT < 0.034$ ($95\%$ CL), $\beta = 0.97^{+0.53}_{-0.19}$ and $\alpha_M^0 < 0.044$ ($95\%$ CL). As above, we improve on previous results in ref.~\cite{planck:2015de-mg};
    \item for the \textit{pure} exponential EFT model on a \lcdm background with a negative running of the Planck mass we obtain from Planck 2018 + $\pto$ the constraints $\omegaEFT = 0.075^{+0.063}_{-0.042}$, $\beta = 1.41^{+0.48}_{-0.15}$ and $\alpha_M^0 = 0.117^{+0.12}_{-0.052}$;
    \item adding one nuisance parameter to the $\pto$ likelihood, e.g. the amplitude of the brightness temperature $a_{T_b}$, reduces the constraining power compared with the case with no nuisances. Deviations with respect to Planck 2018 results for $\Omega_c h^2$ and $H_0$ are at the level of $\sim 15\%$, to be compared with the $\sim 65\%$ found with no nuisances;
    \item repeating the analysis on a $w$CDM background and with a nuisance parameter produces comparable results. The major effect is obtained on $w_0$, $H_0$ and $\sigma_8$ when $\pto$ data are combined with Planck. We obtain an improvement on the errors of these parameters at the level of $40-50\%$. For the EFT models we find that the $\pto$ likelihood has mild constraining power on the EFT parameters, when used both alone and combined with CMB data;
    \item tomography significantly improves the constraining power of the $\pto$ likelihood on \textit{pure} EFT models. With five bins combined in the redshift range from $z=0$ to $z=2.5$, we are able to constrain EFT parameters for a \textit{pure} exponential EFT model with negative running of the Planck mass by means of the $\pto$ likelihood alone;
    \item for the same model, when we combine the tomographic data sets with Planck 2018 CMB data, we find an improvement on the constraints on $\omegaEFT$ and $\beta$ with respect to Planck results at the level of the $\sim25\%$ and $\sim 10\%$, respectively. In the ideal case of halved errors, these values both go up to the $\sim 35\%$, to be compared with the less than $10\%$ found with only one redshift bin.
\end{itemize}

Overall, in our study we find that, at the current state of observations, the 21cm signal power spectrum, measured at one redshift, shows a mild constraining power on the EFT functions that we tested. We expect, and we verified, that $\pto$ is sensitive to EFT parameters. Nevertheless, the one realistic mock data set at redshift $z=0.39$ is still too retained to be able to constrain EFT theories alone. When combined with CMB data, constraints provided by CMB on EFT parameters are too stringent and current $\pto$ mock data do not have a substantial impact. On cosmological parameters, instead, the $\pto$ likelihood proves to be effective in reducing the correlation between $\Omega_ch^2$ and $H_0$ when combined with CMB, thus significantly improving the constraints on cosmological parameters. 

We find that the constraining power of $\pto$ on the considered EFT functions is significantly improved by tomography. With five redshift bins, we are able to constrain the EFT functions, with $\pto$ alone, and to improve Planck results, with $\pto$ and CMB together. For tomographic observations of the 21cm signal, such those that are modelled here, real observations will hopefully be available in the future. However, our test, although preliminary, seems to confirm that tomographic 21cm signal detections will help to expand our knowledge of DE. 

In this work we test simple EFT scenarios and validate the new $\pto$ likelihood code. This analysis could be the starting point to study the constraining power of  21cm signal observations on more complex MG/DE models.  

\appendix  
\section{Effect of the nuisance parameter $a_{T_b}$}
\label{app:nuisance_par}
\begin{figure}[!htb]
    \centering
    \includegraphics[width=0.35\textwidth]{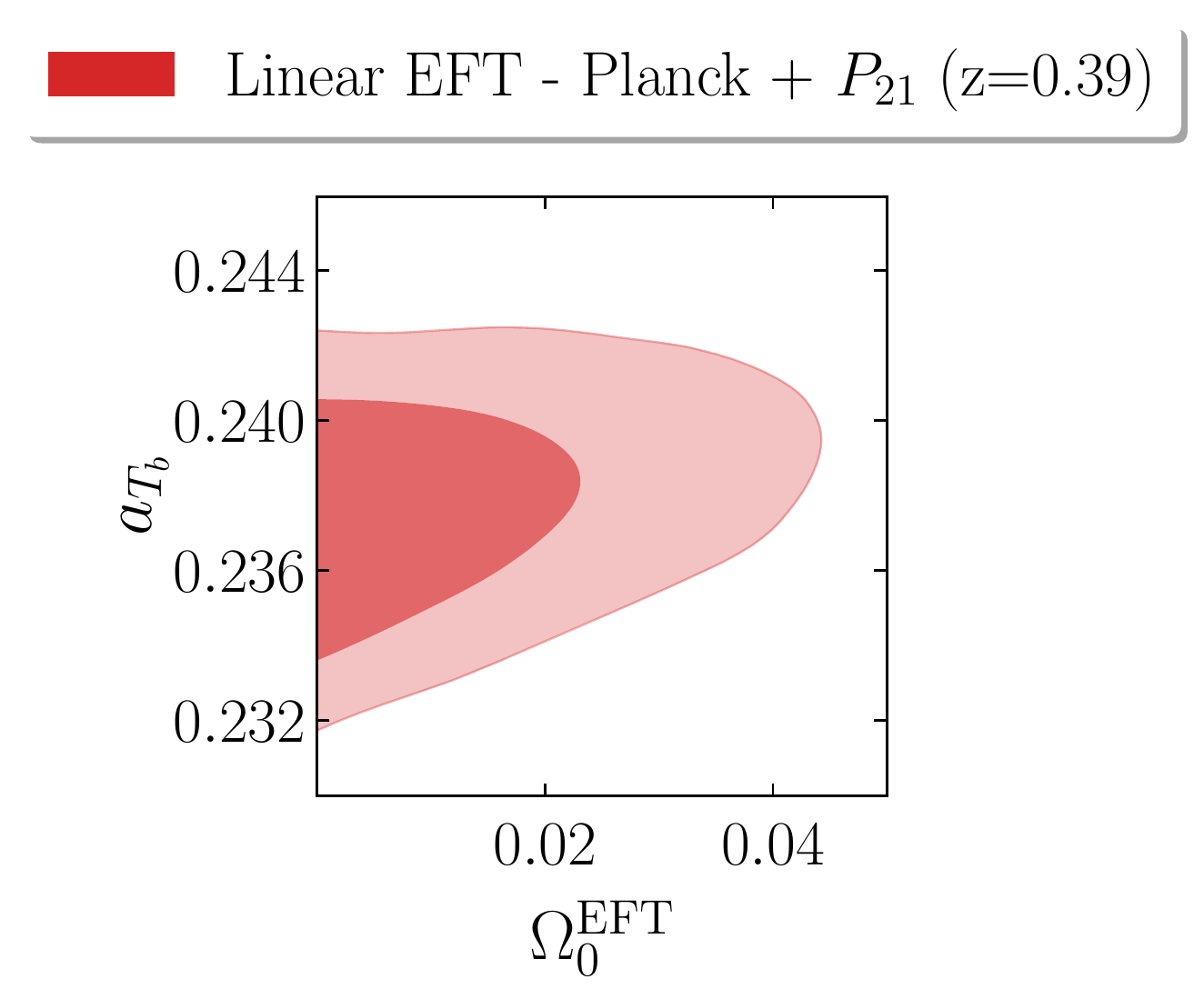} \\
    \hspace*{-1.2cm}
    \includegraphics[width=0.65\textwidth]{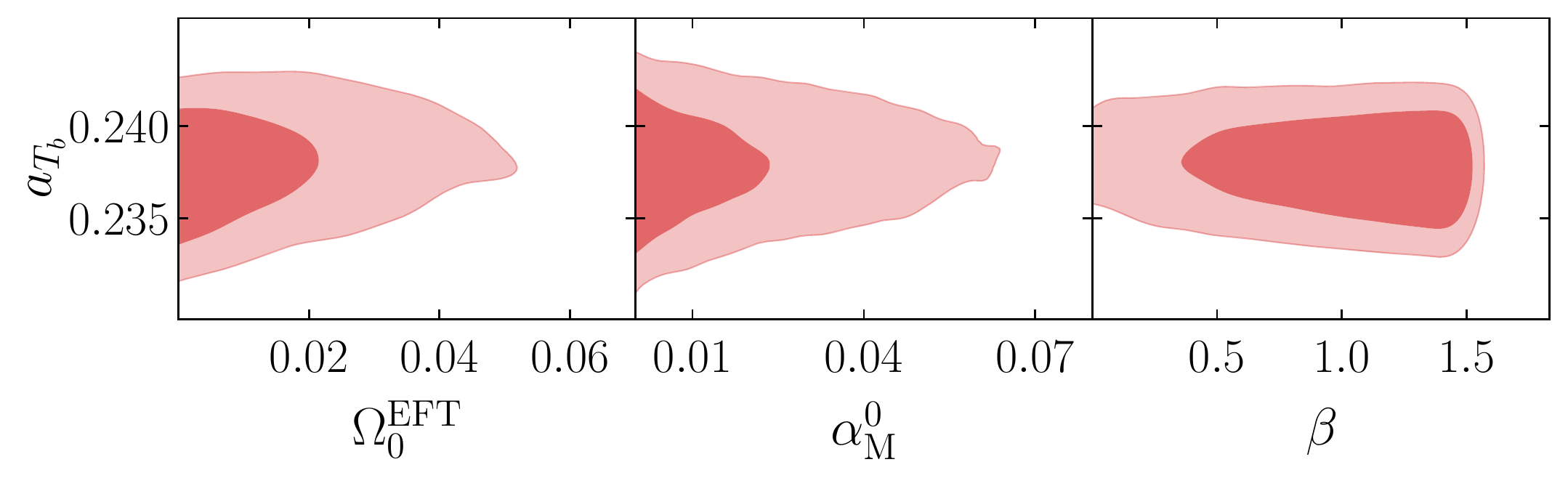} \\
    \hspace*{-1.13cm}
    \includegraphics[width=0.64\textwidth]{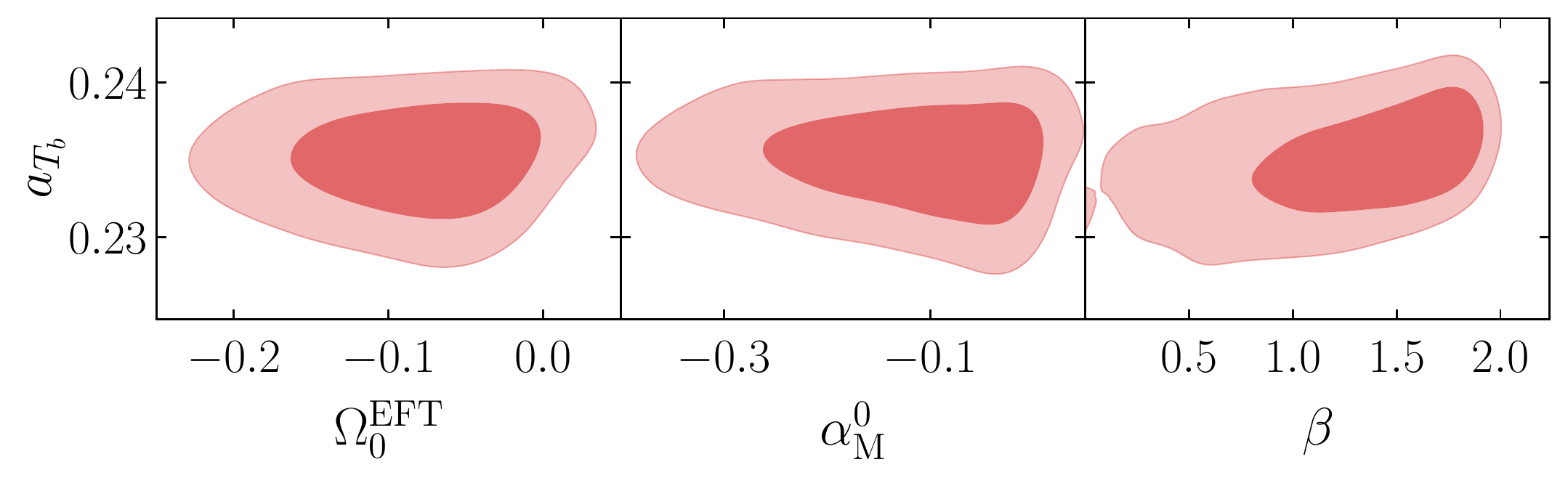}
    \caption{\label{fig:res_nuisance_EFT_2} Joint constraints (68\% and 95\% confidence regions) on EFT and nuisance parameters for a \textit{pure}  linear (upper row), an exponential (second row) and an exponential with negative running of the Planck mass (third row) EFT model, on a $\Lambda$CDM background. Here the label Planck 2018 stands for TT, TE, EE + lowE + lensing while the label $P_{21} (z=0.39)$ stands for the 21cm power spectrum likelihood at redshift $z=0.39$.  We test the effects of varying the nuisance parameter $a_{T_b}$, i.e. the amplitude of the brightness temperature, keeping fixed the HI bias $b_{HI}$. We observe that $a_{T_b}$ is midly correlated with the EFT parameters.}
\end{figure}
\clearpage
\section{Constraints tables}
\label{app:constr_tables}

In this section we present marginalized mean values and 68\% confidence levels on the full set of cosmological, EFT and nuisance parameters. In all the following tables the label Planck 2018 stands for TT, TE, EE + lowE + lensing while the label $P_{21}$ stands for the $\pto$ likelihood at redshift $z=0.39$. The additional \textit{background} likelihood, i.e. $f\sigma_8 + H + D_A$, described in section~\ref{sec:background} and computed at the same redshift $z=0.39$ is labeled as \textit{background}. Percentage deviation in error with respect to Planck 2018 results are shown between brackets. Deviation for asymmetrical constraints are computed with respect to the mean of the left and right errors.  The sign "---" indicates that no significant constraint was found on the parameter, while "$\times$" means that the parameter is absent for the specific set of likelihoods or kept fixed. If only an upper or lower limit is found we show the $95\%$ confidence levels. We highlight it by adding a $^\dag$ sign.
\vfill
\begin{table}[h]
    \centering
    \footnotesize
    \begin{tabular}{lScScScSc} 
    \hline\hline
    \multicolumn{5}{Sc}{\normalsize $\Lambda$\textsc{cdm model - fixed $a_{T_b}$}} \\ \hline
{\textsc{Parameter}} &  Planck 2018 & Planck 2018 + $P_{21}$ & $P_{21}$ & $P_{21}$ + background   \\ \hline
$ \Omega_b h^2 $ \dotfill & $0.02237\pm 0.00014$ & $0.02236\pm 0.00011$ $(-24\%)$  & $0.038\pm 0.015$ & $0.0226\pm 0.0035$ \\
$ \Omega_c h^2 $ \dotfill & $0.1201\pm 0.0012$ & $0.12004\pm 0.00046$ $(-61\%)$ & $0.162^{+0.050}_{-0.033}$ & $0.1227\pm 0.0081$  \\
$ 100\theta_{MC} $ \dotfill & $1.04090\pm 0.00030$ & $1.04089\pm 0.00026$ $(-13\%)$ & $1.061^{+0.036}_{-0.013}$ & $1.042\pm 0.012$  \\
$ \tau $ \dotfill & $0.0543\pm 0.0075$ & $0.0543\pm 0.0056$ $(-25\%)$  & $\times$ & $\times$ \\
$ {\rm{ln}}(10^{10} A_s) $ \dotfill & $3.045\pm 0.014$ & $3.044\pm 0.012$ $(-19\%)$ & $2.87^{+0.10}_{-0.26}$ & $3.037^{+0.083}_{-0.099}$  \\
$ n_s $ \dotfill & $0.9650\pm 0.0041$ & $0.9651\pm 0.0031$ $(-25\%)$ & $< 1.10^\dag$ & $0.951^{+0.072}_{-0.085}$  \\ \hline
$ H_0 $ \dotfill & $67.32\pm 0.53$ & $67.32\pm 0.16$ $(-69\%)$  & $> 57.8^\dag$ & $67.1\pm 1.3$ \\
$ \sigma_8 $ \dotfill & $0.8115\pm 0.0060$ & $0.8113\pm 0.0058$ $(-3\%)$  & $0.815^{+0.015}_{-0.022}$ & $0.812^{+0.018}_{-0.023}$ \\
\hline\hline
\end{tabular}
    \caption{Constraints tables for a $\Lambda$CDM model, where we fix the amplitude of the brightness temperature $a_{T_b}$. Marginalized mean values and 68\% confidence levels.}
    \label{tab:LCDM_fixed_aTb}
\end{table}
\begin{table}[h]
    \centering
    \footnotesize
    \begin{tabular}{lScScSc} 
    \hline\hline
    \multicolumn{4}{Sc}{\normalsize \textsc{eft models - fixed $a_{T_b}$}} \\ \hline
    \multicolumn{4}{Sc}{\footnotesize Planck 2018 + $P_{21}$} \\ \hline
 &   & & Exponential EFT\\ 
{\textsc{Parameter}}& Linear EFT & Exponential EFT  & negative running \\ \hline
$ \Omega_b h^2 $ \dotfill &  $0.02237\pm 0.00011$ $(-26\%)$ & $0.02237\pm 0.00011$ $(-23\%)$     &  $0.02236\pm 0.00011$ $(-32\%)$    \\
$ \Omega_c h^2 $ \dotfill & $0.12003\pm 0.00046$ $(-63\%)$ &  $0.12002\pm 0.00046$ $(-63\%)$   &  $0.11995\pm 0.00047$ $(-67\%)$  \\
$ 100\theta_{MC} $ \dotfill & $1.04090\pm 0.00027$ $(-11\%)$ & $1.04090\pm 0.00027$ $(-12\%)$     &  $1.04090\pm 0.00028$ $(-12\%)$  \\
$ \tau $ \dotfill & $0.0563\pm 0.0059$ $(-22\%)$ &  $0.0561\pm 0.0059$ $(-20\%)$    &  $0.0470\pm 0.0076$ $(-10\%)$  \\
$ {\rm{ln}}(10^{10} A_s) $ \dotfill &  $3.049\pm 0.012$ $(-14\%)$&  $3.048\pm 0.012$ $(-14\%)$   &  $3.027\pm 0.016$ $(-6\%)$ \\
$ n_s $ \dotfill &  $0.9652\pm 0.0031$ $(-24\%)$ & $0.9652\pm 0.0030$ $(-26\%)$     & $0.9653\pm 0.0032$ $(-30\%)$  \\ \hline
$ \omegaEFT $ \dotfill &  $< 0.0309^\dag$ $(-11\%)$ & $< 0.0343^\dag$ $(-5\%)$     &  $-0.075^{+0.063}_{-0.042}$ $(-7\%)$  \\
$ \beta $ \dotfill & $\times$ &  $0.97^{+0.53}_{-0.19}$ $(-1\%)$  &   $1.41^{+0.48}_{-0.15}$ $(-18\%)$ \\
$ \alpha_M^0 $ \dotfill  & $\times$ &  $< 0.0441^\dag$ $(-3\%)$   &  $-0.117^{+0.12}_{-0.052}$ $(-3\%)$  \\
$ H_0 $ \dotfill& $67.33\pm 0.17$ $(-70\%)$ & $67.34\pm 0.17$ $(-69\%)$    & $67.36\pm 0.18$ $(-73\%)$ \\
$ \sigma_8 $ \dotfill& $0.8105\pm 0.0059$ $(-3\%)$ & $0.8104\pm 0.0060$ $(-0\%)$    & $0.8116\pm 0.0065$ $(-19\%)$  \\ 
\hline\hline
\end{tabular}
    \caption{Constraints tables for different EFT models on a \lcdm background, where we fix the amplitude of the brightness temperature $a_{T_b}$. Marginalized mean values and 68\% confidence levels.}
    \label{tab:EFT_fixed_aTb}
\end{table}
\vfill
\begin{table}[h]
    \centering
    \footnotesize
    \begin{tabular}{lScScScSc} 
    \hline\hline
    \multicolumn{5}{Sc}{\normalsize $\Lambda$\textsc{cdm model}} \\ \hline
{\textsc{Parameter}} &  Planck 2018 & Planck 2018 + $P_{21}$ & $P_{21}$ & $P_{21}$ + \textit{background}   \\ \hline
$ \Omega_b h^2 $ \dotfill & $0.02237\pm 0.00014$ & $0.02236\pm 0.00013$ $(-7\%)$  & $0.037^{+0.018}_{-0.016}$ & $0.0224\pm 0.0035$ \\
$ \Omega_c h^2 $ \dotfill & $0.1201\pm 0.0012$ &  $0.1202\pm 0.0010$ $(-17\%)$   & $0.160^{+0.054}_{-0.032}$ & $0.1228\pm 0.0079$  \\
$ 100\theta_{MC} $ \dotfill & $1.04090\pm 0.00030$ &  $1.04088\pm 0.00029$ $(-3\%)$  & $1.057^{+0.041}_{-0.011}$ & $1.042\pm 0.012$    \\
$ \tau $ \dotfill & $0.0543\pm 0.0075$ &  $0.0538\pm 0.0072$ $(-4\%)$   & $\times$ & $\times$ \\
$ {\rm{ln}}(10^{10} A_s) $ \dotfill & $3.045\pm 0.014$ &  $3.043\pm 0.014$ $(-0\%)$ & $2.89^{+0.16}_{-0.32}$ & $3.03\pm 0.13$ \\
$ n_s $ \dotfill & $0.9650\pm 0.0041$ &   $0.9648\pm 0.0039$ $(-5\%)$ & $< 1.12^\dag$ & $0.947^{+0.071}_{-0.087}$ \\ \hline
$ H_0 $ \dotfill & $67.32\pm 0.53$ & $67.27\pm 0.44$ $(-17\%)$ &$> 55.7^\dag$ & $67.0\pm 1.3$  \\
$ \sigma_8 $ \dotfill & $0.8115\pm 0.0060$ &  $0.8114\pm 0.0058$ $(-3\%)$ & $0.817^{+0.047}_{-0.064}$ & $0.810^{+0.041}_{-0.050}$   \\ \hline
$a_{T_b}$ \dotfill & $\times$ &  $0.2371\pm 0.0019$ & --- & $0.238^{+0.015}_{-0.013}$ \\ 
\hline\hline
 \\ 
 \hline\hline
 \multicolumn{5}{Sc}{\normalsize w\textsc{cdm model}} \\ \hline
{\textsc{Parameter}} &  Planck 2018 & Planck 2018 + $P_{21}$ & $P_{21}$ & $P_{21}$ + \textit{background}   \\ \hline
$ \Omega_b h^2 $ \dotfill & $0.02234\pm 0.00015$ & $0.02238\pm 0.00014$ $(-8\%)$ & $0.039^{+0.017}_{-0.014}$ & $0.0199\pm 0.0037$   \\
$ \Omega_c h^2 $ \dotfill & $0.1204\pm 0.0012$ & $0.1199\pm 0.0010$ $(-15\%)$ & $0.167^{+0.046}_{-0.032}$ & $0.112\pm 0.010$   \\
$ 100\theta_{MC} $ \dotfill & $1.04087\pm 0.00032$ & $1.04092\pm 0.00030$ $(-6\%)$ & $1.122^{+0.049}_{-0.059}$ & $1.052^{+0.014}_{-0.016}$   \\
$ \tau $ \dotfill & $0.0547\pm 0.0076$ & $0.0553\pm 0.0074$ $(-2\%)$ & $\times$ & $\times$ \\
$ {\rm{ln}}(10^{10} A_s) $ \dotfill & $3.046\pm 0.015$ & $3.046\pm 0.015$ $(-1\%)$ & $3.24\pm 0.33$ & $3.33^{+0.21}_{-0.32}$   \\
$ n_s $ \dotfill & $0.9644\pm 0.0042$ & $0.9653\pm 0.0039$ $(-7\%)$ & $< 1.06^\dag$ & $0.978\pm 0.081$   \\\hline
$ w_0 $ \dotfill & $< -0.763^\dag$ & $< -0.890^\dag$ $(-53\%)$ & $-0.65^{+0.25}_{-0.20}$ & $< -0.543^\dag$   \\
$ H_0 $ \dotfill & $64.6^{+2.7}_{-1.1}$ & $66.1^{+1.2}_{-0.69}$ $(-50\%)$ & $> 60.0^\dag$ & $63.9^{+2.7}_{-2.3} $   \\
$ \sigma_8 $ \dotfill & $0.787^{+0.025}_{-0.011}$ & $0.799^{+0.013}_{-0.0085}$ $(-40\%)$ & $0.837^{+0.052}_{-0.059}$ & $0.837\pm 0.048$   \\\hline
$a_{T_b}$ \dotfill & $\times$ &  $0.2428^{+0.0032}_{-0.0058}$ & --- & $0.2343^{+0.0092}_{-0.018}$\\ 
\hline\hline
\end{tabular}
    \caption{Constraints tables for a $\Lambda$CDM and a $w$CDM model. Marginalized mean values and 68\% confidence levels.
    }
    \label{tab:LCDM_$w$CDM}
\end{table}
\vfill
\clearpage
\begin{table}[tb]
    \centering
    \footnotesize
    \begin{tabular}{lScScScSc} \hline
    \multicolumn{5}{|Sc|}{\large \textsc{Linear} \textit{pure} \textsc{eft model}} \\\hline\hline
     \multicolumn{5}{Sc}{\normalsize $\Lambda$\textsc{cdm background}} \\ \hline
     {\textsc{Parameter}} &  Planck 2018 & Planck 2018 + $P_{21}$ & $P_{21}$ & $P_{21}$ + \textit{background}   \\ \hline
     $ \Omega_b h^2 $ \dotfill & $0.02234\pm 0.00015$ & $0.02234\pm 0.00014$ $(-6 \%)$       & $0.039^{+0.018}_{-0.016}$ & $0.0232\pm 0.0036$   \\
$ \Omega_c h^2 $ \dotfill & $0.1203\pm 0.0012$ &  $0.1204\pm 0.0010$   $(-16 \%)$  & $0.164^{+0.052}_{-0.032}$ & $0.1230\pm 0.0081$   \\
$ 100\theta_{MC} $ \dotfill & $1.04087\pm 0.00030$ &  $1.04086\pm 0.00030$  $(-0 \%)$   & $1.060^{+0.037}_{-0.013}$ & $1.040\pm 0.012$   \\
$ \tau $ \dotfill & $0.0550\pm 0.0075$ &  $0.0549\pm 0.0073$  $(- 3\%)$   & $\times$ & $\times$   \\
$ {\rm{ln}}(10^{10} A_s) $ \dotfill & $3.047\pm 0.015$ & $3.046\pm 0.014$ $(-7 \%)$    & $3.15^{+0.24}_{-0.34}$ & $3.35\pm 0.22$   \\
$ n_s $ \dotfill & $0.9644\pm 0.0041$ & $0.9643\pm 0.0039$ $(- 5\%)$    & $< 0.964$ & $0.956\pm 0.077$   \\\hline
$ \Omega_0^{\rm EFT} $ \dotfill & $< 0.0349^\dag$ & $< 0.035^\dag$  $(-0\%)$      & --- & ---   \\
$ H_0 $ \dotfill & $67.21\pm 0.55$ & $67.18\pm 0.47$  $(- 15\%)$  & $> 57.7^\dag$ & $67.0\pm 1.3$   \\
$ \sigma_8 $ \dotfill & $0.8101\pm 0.0061$ &  $0.8101\pm 0.0058$   $(- 5\%)$  & $0.831^{+0.047}_{-0.063}$ & $0.844\pm 0.049$   \\\hline
    $a_{T_b}$ \dotfill &   $\times$ &  $0.2378\pm 0.0020$ & --- & $< 0.239$\\
    \hline\hline
\\\hline\hline
    \multicolumn{5}{Sc}{\normalsize $w$\textsc{cdm background}} \\ \hline
{\textsc{Parameter}} &  Planck 2018 & Planck 2018 + $P_{21}$ & $P_{21}$ & $P_{21}$ + \textit{background}   \\ \hline
$ \Omega_b h^2 $ \dotfill & $0.02232\pm 0.00015$ & $0.02236\pm 0.00014$ $(-7\%)$ & $0.039\pm 0.014$ & $0.0219\pm 0.0040$   \\
$ \Omega_c h^2 $ \dotfill & $0.1206\pm 0.0012$ & $0.1201\pm 0.0011$ $(-15\%)$ & $0.167^{+0.042}_{-0.029}$ & $0.119\pm 0.011$   \\
$ 100\theta_{MC} $ \dotfill & $1.04083\pm 0.00030$ & $1.04090\pm 0.00030$ $(-2\%)$ & $1.101^{+0.041}_{-0.048}$ & $1.045\pm 0.014$   \\
$ \tau $ \dotfill & $0.0554^{+0.0069}_{-0.0078}$ & $0.0561^{+0.0070}_{-0.0078}$ $(0\%)$ & $\times$ & $\times$   \\
$ {\rm{ln}}(10^{10} A_s) $ \dotfill & $3.048^{+0.014}_{-0.015}$ & $3.048\pm 0.015$ $(0\%)$ & $3.34^{+0.33}_{-0.28}$ & $3.47\pm 0.22$   \\
$ n_s $ \dotfill & $0.9638\pm 0.0042$ & $0.9647\pm 0.0039$ $(-6\%)$ & $0.921^{+0.049}_{-0.098}$ & $0.965\pm 0.079$   \\\hline
$ w_0 $ \dotfill & $-0.925^{+0.020}_{-0.079}$ & $-0.960^{+0.015}_{-0.043}$ $(-41\%)$ & $-0.77^{+0.25}_{-0.18}$ & $-0.91^{+0.13}_{-0.17}$   \\
$ \omegaEFT$ \dotfill & $< 0.0339^\dag$ & $< 0.0342^\dag$ $(+1\%)$ & --- & $< 0.612$   \\
$ H_0 $ \dotfill & $64.9^{+2.4}_{-0.97}$ & $66.1^{+1.2}_{-0.70}$ $(-44\%)$ &  $> 61.7^\dag$& $65.6^{+2.7}_{-2.4}$   \\
$ \sigma_8 $ \dotfill & $0.790^{+0.023}_{-0.0094}$ & $0.799^{+0.013}_{-0.0083}$ $(-41\%)$ & $0.843^{+0.049}_{-0.060}$ & $0.855^{+0.054}_{-0.045}$   \\\hline
$a_{T_b}$ \dotfill & $\times$ & $0.2430^{+0.0031}_{-0.0058}$ & --- & --- \\
\hline\hline
\end{tabular}
    \caption{Constraints tables for \textit{pure} EFT models on a $w$CDM background. Marginalized mean values and 68\% confidence levels.}
    \label{tab:EFT_linear}
\end{table}

\clearpage
\begin{table}[tb]
    \centering
    \footnotesize
    \begin{tabular}{lScScScSc} \hline
    \multicolumn{5}{|Sc|}{\large \textsc{Exponential} \textit{pure} \textsc{eft model} } \\\hline\hline
    \multicolumn{5}{Sc}{\normalsize $\Lambda$\textsc{cdm background}} \\ \hline
{\textsc{Parameter}} &  Planck 2018 & Planck 2018 + $P_{21}$ & $P_{21}$ & $P_{21}$ + \textit{background}   \\ \hline
$ \Omega_b h^2 $ \dotfill & $0.02235\pm 0.00014$ &  $0.02234\pm 0.00014$  $(-0 \%)$       & $0.041^{+0.016}_{-0.013}$ & $0.0228\pm 0.0035$   \\
$ \Omega_c h^2 $ \dotfill & $0.1203\pm 0.0012$ &    $0.1204\pm 0.0010$  $(- 16\%)$     & $0.172^{+0.042}_{-0.029}$ & $0.1232\pm 0.0080$   \\
$ 100\theta_{MC} $ \dotfill & $1.04087\pm 0.00031$ &   $1.04086\pm 0.00030$  $(-3 \%)$      & $1.070^{+0.025}_{-0.015}$ & $1.042\pm 0.012$   \\
$ \tau $ \dotfill & $0.0548\pm 0.0074$ &    $0.0548^{+0.0068}_{-0.0076}$  $(-3 \%)$     & $\times$ & $\times$   \\
$ {\rm{ln}}(10^{10} A_s) $ \dotfill & $3.046^{+0.013}_{-0.015}$ &  $3.046^{+0.013}_{-0.015}$    $(- 0\%)$     & $3.02^{+0.22}_{-0.34}$ & $3.29^{+0.20}_{-0.29}$   \\
$ n_s $ \dotfill & $0.9643\pm 0.0041$ &   $0.9642\pm 0.0040$   $(- 2\%)$     & $< 1.05^\dag$ & $0.950\pm 0.076$   \\\hline
$ \Omega_0^{\rm EFT} $ \dotfill & $< 0.0361^\dag$ &  $< 0.038^\dag$ $(+5\%)$        & --- & ---   \\
$ \beta $ \dotfill & $0.96^{+0.54}_{-0.19}$ &   $0.96^{+0.54}_{-0.20}$ $(-0\%)$        & $1.30\pm 0.57$ & $1.20^{+0.58}_{-0.47}$   \\
$ \alpha_M^0 $ \dotfill & $< 0.0455^\dag$ &  $< 0.048^\dag$ $(+6\%)$         & $< 1.89^\dag$ & $< 1.72^\dag$   \\
$ H_0 $ \dotfill & $67.20\pm 0.55$ &     $67.17\pm 0.46$  $(- 16 \%)$     & $> 64.3^\dag$ & $66.9\pm 1.3$   \\
$ \sigma_8 $ \dotfill & $0.8101\pm 0.0060$ &  $0.8100\pm 0.0060$    $(- 0\%)$     & $0.829^{+0.050}_{-0.061}$ & $0.835\pm 0.050$   \\\hline
$a_{T_b}$  \dotfill & $\times$ &  $0.2378\pm 0.0021$     & --- & $0.2343^{+0.0076}_{-0.020}$  \\
\hline\hline
 \\ \hline\hline
\multicolumn{5}{Sc}{\normalsize $w$\textsc{cdm background}} \\ \hline
{\textsc{Parameter}} &  Planck 2018 & Planck 2018 + $P_{21}$ & $P_{21}$ & $P_{21}$ + \textit{background}   \\ \hline
$ \Omega_b h^2 $ \dotfill & $0.02233\pm 0.00015$ & $0.02237\pm 0.00013$ $(-8\%)$ & $0.040\pm 0.014$ & $0.0212\pm 0.0040$   \\
$ \Omega_c h^2 $ \dotfill & $0.1205\pm 0.0012$ & $0.1200\pm 0.0010$ $(-16\%)$ & $0.168^{+0.045}_{-0.032}$ & $0.117^{+0.011}_{-0.013}$   \\
$ 100\theta_{MC} $ \dotfill & $1.04084\pm 0.00031$ & $1.04090\pm 0.00030$ $(-1\%)$ & $1.115\pm 0.049$ & $1.048\pm 0.014$   \\
$ \tau $ \dotfill & $0.0555\pm 0.0075$ & $0.0556^{+0.0067}_{-0.0077}$ $(-4\%)$ & $\times$ & $\times$   \\
$ {\rm{ln}}(10^{10} A_s) $ \dotfill & $3.048\pm 0.015$ & $3.047\pm 0.014$ $(-1\%)$ & $3.31^{+0.38}_{-0.28}$ & $3.43\pm 0.24$   \\
$ n_s $ \dotfill & $0.9641\pm 0.0041$ & $0.9650\pm 0.0039$ $(-5\%)$ & $< 1.07^\dag$& $0.969\pm 0.081$   \\\hline
$ w_0 $ \dotfill & $-0.911^{+0.028}_{-0.084}$ & $-0.953^{+0.018}_{-0.041}$ $(-47\%)$ & $-0.68^{+0.23}_{-0.15}$ & $-0.87^{+0.20}_{-0.15}$   \\
$ \Omega_0^{\rm EFT} $ \dotfill & $< 0.0505^\dag$ & $< 0.0470^\dag$ $(-7\%)$ & --- & ---   \\
$ \beta $ \dotfill & --- & $1.79^{+1.1}_{-0.47}$ & --- & $1.50^{+0.66}_{-0.77}$   \\
$ \alpha_M^0 $ \dotfill &  $< 0.115^\dag$ & $< 0.100^\dag$ $(-13\%)$ & $< 2.25^\dag$ & $< 2.01^\dag$ \\
$ H_0 $ \dotfill & $64.6^{+2.5}_{-1.1}$ & $66.0^{+1.1}_{-0.73}$ $(-49\%)$ &$> 61.3^\dag$ & $65.1^{+2.4}_{-3.2}$   \\
$ \sigma_8 $ \dotfill & $0.787^{+0.023}_{-0.011}$ & $0.797^{+0.012}_{-0.0086}$ $(-39\%)$ & $0.844^{+0.048}_{-0.063}$ & $0.848\pm 0.050$   \\\hline
$a_{T_b}$ \dotfill & $\times$ & $0.2437^{+0.0033}_{-0.0055}$ & --- & --- \\
\hline\hline
\end{tabular}
    \caption{Constraints tables for \textit{pure} EFT models on a $w$CDM background. Marginalized mean values and 68\% confidence levels.}
    \label{tab:EFT_exponential}
\end{table}

\clearpage
\begin{table}[tb]
    \centering
    \footnotesize
    \begin{tabular}{lScScScSc} \hline
    \multicolumn{5}{|Sc|}{\large \textsc{Exponential} \textit{pure} \textsc{eft model} - negative Planck mass running } \\\hline\hline
\multicolumn{5}{Sc}{\normalsize $\Lambda$\textsc{cdm background}} \\ \hline
{\textsc{Parameter}} &  Planck 2018 & Planck 2018 + $P_{21}$ & $P_{21}$ & $P_{21}$ + \textit{background}   \\ \hline
$ \Omega_b h^2 $ \dotfill & $0.02249\pm 0.00016$ & $0.02243\pm 0.00014$  $(-12 \%)$   & $0.037\pm 0.015$ & $0.0226\pm 0.0038$   \\
$ \Omega_c h^2 $ \dotfill & $0.1185\pm 0.0014$ &  $0.1192\pm 0.0011$   $(- 21\%)$  & $0.159^{+0.047}_{-0.036}$ & $0.1229\pm 0.0079$   \\
$ 100\theta_{MC} $ \dotfill & $1.04106\pm 0.00032$ & $1.04096\pm 0.00031$ $(-3 \%)$      & $1.058^{+0.037}_{-0.012}$ & $1.042\pm 0.012$   \\
$ \tau $ \dotfill & $0.0490^{+0.0086}_{-0.0075}$ &  $0.0487\pm 0.0082$ $(+ 1\%)$    & $\times$ & $\times$   \\
$ {\rm{ln}}(10^{10} A_s) $ \dotfill & $3.029^{+0.018}_{-0.015}$ &$3.029\pm 0.017$ $(+6 \%)$       & $2.90\pm 0.43$ & $3.07\pm 0.29$   \\
$ n_s $ \dotfill & $0.9687\pm 0.0045$ & $0.9671\pm 0.0041$ $(- 7\%)$    & $0.929^{+0.039}_{-0.12}$ & $0.948\pm 0.075$   \\\hline
$ \Omega_0^{\rm EFT} $ \dotfill & $-0.094^{+0.071}_{-0.042}$ &$-0.083^{+0.065}_{-0.040} $  $(-7\%)$    & --- & $0.19^{+0.69}_{-0.34}$   \\
$ \beta $ \dotfill & $1.28^{+0.56}_{-0.21}$ & $1.33^{+0.54}_{-0.20}$ $(-4\%)$    & $1.21^{+0.57}_{-0.70}$ & $1.38\pm 0.55$   \\
$ \alpha_{\rm M}^0 $ \dotfill & $-0.133^{+0.13}_{-0.048}$ &$-0.122^{+0.12}_{-0.047}$  $(-6\%)$   & $0.31^{+0.84}_{-0.70}$ & $0.36\pm 0.80$   \\
$ H_0 $ \dotfill & $68.03\pm 0.66$ &   $67.69\pm 0.51$  $(- 23\%)$ &$> 56.3^\dag$ & $67.0\pm 1.3$   \\
$ \sigma_8 $ \dotfill & $0.8130\pm 0.0080$ & $0.8131\pm 0.0073$  $(- 9\%)$   & $0.819^{+0.052}_{-0.061}$ & $0.817^{+0.049}_{-0.058}$   \\\hline
$a_{T_b}$ \dotfill & $\times$ & $0.2351^{+0.0026}_{-0.0023}$    & --- & $0.236\pm 0.013$ \\
\hline\hline
 \\ \hline\hline
\multicolumn{5}{Sc}{\normalsize $w$\textsc{cdm background}} \\ \hline
{\textsc{Parameter}} &  Planck 2018 & Planck 2018 + $P_{21}$ & $P_{21}$ & $P_{21}$ + \textit{background}   \\ \hline
$ \Omega_b h^2 $ \dotfill & $0.02246\pm 0.00016$ & $0.02243\pm 0.00016$ $(-4\%)$ & $0.040^{+0.016}_{-0.014}$ & $0.0212^{+0.0042}_{-0.0049}$   \\
$ \Omega_c h^2 $ \dotfill & $0.1188\pm 0.0014$ & $0.1192\pm 0.0014$ $(-5\%)$ & $0.169^{+0.040}_{-0.030}$ & $0.118^{+0.013}_{-0.015}$   \\
$ 100\theta_{MC} $ \dotfill & $1.04103\pm 0.00032$ & $1.04100\pm 0.00032$ $(-0\%)$ & $1.125\pm 0.061$ & $1.049^{+0.014}_{-0.019}$   \\
$ \tau $ \dotfill & $0.0501\pm 0.0081$ & $0.0490\pm 0.0085$ $(5\%)$ & $\times$ & $\times$   \\
$ {\rm{ln}}(10^{10} A_s) $ \dotfill & $3.031\pm 0.016$ & $3.030\pm 0.018$ $(8\%)$ & $3.19^{+0.54}_{-0.32}$ & $3.07^{+0.47}_{-0.42}$   \\
$ n_s $ \dotfill & $0.9677\pm 0.0046$ & $0.9671\pm 0.0044$ $(-3\%)$ & $0.915^{+0.036}_{-0.11}$ & $0.964\pm 0.078$   \\\hline
$ w_0 $ \dotfill & $< -0.535^\dag$ & $-0.980^{+0.065}_{-0.079}$ & $-0.69^{+0.34}_{-0.14}$ & $-0.91\pm 0.24$   \\
$ \Omega_0^{\rm EFT} $ \dotfill & $-0.144^{+0.13}_{-0.042}$ & $-0.126^{+0.10}_{-0.046}$ $(-15\%)$ & --- & $-0.05^{+0.41}_{-0.76}$   \\
$ \beta $ \dotfill & $1.89^{+1.1}_{-0.35}$ & $> 0.625^\dag$ & $> 0.587^\dag$ & $1.56^{+0.70}_{-0.87}$   \\
$ \alpha_{\rm M}^0 $ \dotfill & $-0.305^{+0.31}_{-0.062}$ & $-0.250^{+0.25}_{-0.022}$ $(-27\%)$ & $0.20^{+0.88}_{-1.2}$ & $0.03^{+0.60}_{-1.1}$   \\
$ H_0 $ \dotfill & $72\pm 9$ & $67.1^{+2.1}_{-1.6}$ $(-79\%)$ & $> 62.4^\dag$ & $65.6\pm 3.5$   \\
$ \sigma_8 $ \dotfill & $0.849^{+0.097}_{-0.036}$ & $0.814\pm 0.018$ $(-76\%)$ & $0.84\pm 0.10$ & $0.813^{+0.055}_{-0.063}$   \\\hline
$a_{T_b}$  \dotfill & $\times$ & $0.2357^{+0.0070}_{-0.0078}$ & --- & --- \\
\hline\hline
\end{tabular}
    \caption{Constraints tables for \textit{pure} EFT models on a $w$CDM background. Marginalized mean values and 68\% confidence levels.}
    \label{tab:EFT_exp_neg}
\end{table}

\acknowledgments
The authors would like to thank José Fonseca for his valuable help in the construction of the MeerKAT-like $\pto$ mock data. We thank Marco Raveri for useful discussions and technical help. MB, MS, BSH and MV are supported by the INFN INDARK grant.

\bibliographystyle{unsrtnat}  
\bibliography{Bibliography}

\end{document}